\documentclass[final,11pt,3p,doublespacing]{elsarticle}
\usepackage{sectsty}
\sectionfont{\color{black}\large\MakeUppercase}
\subsectionfont{\color{black}\large}
\subsubsectionfont{\color{black}\large\itshape}
\usepackage{mathptmx} 
\usepackage{times}
\emergencystretch=\maxdimen
\biboptions{sort&compress}
\usepackage[colorlinks]{hyperref}
\AtBeginDocument{%
	\hypersetup{
		plainpages = false,
		bookmarksopen = true,
		bookmarksnumbered = true,
		breaklinks = true,
		linktocpage,
		linkcolor = purple,
		urlcolor  = black,
		citecolor = Magenta,
}}
\usepackage{multicol}
\usepackage{multirow}
\usepackage{imakeidx}
\usepackage{easyReview}
\makeindex
\usepackage{nomencl}
\makenomenclature
\usepackage[xindy]{glossaries}
\makeglossaries
\usepackage{graphics}
\usepackage{graphicx}
\usepackage{epsf,psfrag}
\usepackage{epstopdf}
\usepackage[para]{threeparttable}
\usepackage{subfigure}
\usepackage{epsfig}
\usepackage{pdflscape}
\usepackage{rotating}
\usepackage{lscape}
\usepackage{float}
\usepackage{stfloats}
\usepackage{textgreek}
\usepackage{longtable}
\usepackage{lscape}
\usepackage{comment}
\usepackage{tablefootnote}
\usepackage{footnote}
\usepackage{caption}
\usepackage[autopunct=true]{csquotes}
\usepackage{rotating}
\usepackage[normalem]{ulem}
\usepackage{resizegather}
\usepackage{siunitx}
\usepackage[nodisplayskipstretch]{setspace}
\usepackage{color,soul}
\usepackage[usenames,dvipsnames]{xcolor}
\usepackage{amsfonts,amsmath,amssymb,gensymb,mathtools}
\usepackage{latexsym,array}
\usepackage{textcomp}

\newcommand{\RomanNumeralCaps}[1]
\linenumbers
\makesavenoteenv{tabular}
\makesavenoteenv{table}
\usepackage[left,displaymath, mathlines]{lineno}
\providecommand\bnabla{\mathbf{\nabla}}

\DeclareMathAlphabet{\mathpzc}{OT1}{pzc}{m}{it}
\def\fig{Fig.~}
\def\figs{Figs.~}
\def\eqn{Eq.~}
\def\eqns{Eqs.~}
\def\tab{Table~}

\def\micro{\textmu}
\providecommand\bnabla{\mathbf{\nabla}}

\providecommand\p{{\partial}}

\newcommand\cac{Ca_{\text{c}}}
\newcommand\qr{Q_{\text{r}}}
\def\tsc#1{\csdef{#1}{\textsc{\lowercase{#1}}\xspace}}
\tsc{WGM}
\tsc{QE}
\tsc{EP}
\tsc{PMS}
\tsc{BEC}
\tsc{DE}
\usepackage{enumitem}
\usepackage[makeroom]{cancel}
\usepackage{easyReview} 
\newcommand{\removeEq}[1]{\ifistoreview{\expandafter\removeColor{\xcancel{#1}}\expandafter} \else {}\fi}
\newcommand{\removeBox}[1]{\ifistoreview{\removeColor{\setitemize{itemsep=0pt, parsep=0pt, topsep=0pt}{#1}}\expandafter} \else {\nonumber} \fi}
\graphicspath{{../}{Figures/}}
\DeclareGraphicsExtensions{.png, .PNG,.pdf,.PDF,.jpg,.JPG,.eps,.EPS}
\newif\ifblind
\blindfalse 
\begin{document}
\setcounter{page}{1}
\begin{frontmatter} 
\title{Experimental and Computational Analysis of the Hydrodynamics of Droplet Generation in a Cylindrical Microfluidic Device}
%
\ifblind
\else
\author[labela]{Pratibha {Dogra}}
\author[labela]{Ram Prakash {Bharti}\corref{coradd}}\emailauthor{rpbharti@iitr.ac.in}{R.P. Bharti}
\author[labelc]{Gaurav {Sharma}}
\address[labela]{Complex Fluid Dynamics and Microfluidics (CFDM) Lab, Department of Chemical Engineering, Indian Institute of Technology Roorkee, Roorkee - 247667, Uttarakhand, India}
%
%
\address[labelc]{Fluid and Instability Research (FAIR) Lab, Department of Chemical Engineering, Indian Institute of Technology Roorkee, Roorkee - 247667, Uttarakhand, India}
\fi
\cortext[coradd]{\textit{Corresponding author. }}
%
\begin{abstract}
\fontsize{10}{14pt}\selectfont	
\noindent
Droplet-based microfluidic systems underpin a wide range of emerging chemical and biochemical technologies, from precision emulsification and materials synthesis to drug delivery and LOC platforms, additive manufacturing, and biomedical applications.  
This study has investigated the hydrodynamics of the droplet formation in a T-shaped cylindrical microfluidic device using \micro-PIV experiments and CFD simulations. The microfluidic devices of 150 $\mu$m internal diameter are fabricated from PDMS using a highly economical embedded template method. The flow visualization experiments are performed using silicone oil and deionized water, which are immiscible; thus, water-in-oil (W/O) droplets forms. The mathematical model coupling Navier-Stokes and conservative level set equations has been solved using finite element method. The detailed flow fields  (velocity, pressure, and phase composition) are obtained over a wide range of flow-rate ratios ($0.1\le\qr\le 10$) and capillary numbers ($10^{-3}\le\cac\le 0.1$) in order to investigate the hydrodynamics governing droplet formation. The phase profiles reveal droplet breakup stages (lag, filling, necking, and pinch-off stages), alongside several distinct regimes (squeezing, dripping, sausage flow, and parallel flow with tip streaming). The flow regimes map has subsequently been developed which delineate `droplet' and `non-droplet' regimes based on the flow conditions. The droplet features such as radii, length and internal flow profiles during droplet formation have shown complex dependence on the flow parameters ($\cac$,  $\qr$). The scaling analysis further indicated that droplet size exhibits a linear dependence, and  droplet curvature remains nearly independent, of $\qr$ in squeezing (interfacial tension-dominated) regime, however, both droplet size and curvature depicts a power-law dependence on $\cac$ and $\qr$ in dripping (viscosity-dominated) regime. Furthermore, velocity field inside the forming droplet exhibits a complex dependence on the governing parameters such that velocity profiles inside droplets are laminar and parabolic in the central region. While symmetric plugs at both the front and rear sides display fully developed velocity profiles in the squeezing regime, although the central velocity profile remains parabolic, the front and rear sections are still in the developing stage in dripping regime. The proposed correlations for droplet length, curvature, and film thickness enable predictive modeling within the studied parameter range. In addition, a novel empirical correlation is proposed to determine the thin film thickness accounting the visco-inertial and capillary effects. Overall, these findings advance the fundamental understanding of droplet dynamics in confined microchannels and provide valuable guidelines for design and optimization of droplet-based microfluidic systems, with applications in emulsification, drug delivery, and LOC technologies.
\end{abstract}
\end{frontmatter}
%
%
\section{Motivation and Novelty}
Droplet-based microfluidic systems underpin a wide range of emerging chemical and biochemical technologies, from precision emulsification and materials synthesis to drug delivery and lab-on-a-chip (LOC) platforms, additive manufacturing, and biomedical applications. Despite their widespread use, the ability to predict and control droplet formation across operating regimes remains limited, particularly when geometrical confinement, interfacial forces, and viscous stresses interact in a strongly coupled and nonlinear manner. This limitation hampers the rational design and scaling of droplet-based processes, which are increasingly required to operate robustly under varying flow conditions and material properties.
\newline
Most existing studies focus on planar microchannels or rely on empirical correlations with limited transferability across geometries and flow regimes. In contrast, cylindrical microchannels, which are directly relevant to capillaries, fibers, and industrial microstructured reactors, introduce distinct hydrodynamic and interfacial effects that are not yet fully understood. As a result, a mechanistic and quantitatively predictive framework that links interfacial-scale physics to macroscopic droplet characteristics in cylindrical confinement remains lacking.
\newline
In this work, we address this gap by combining high-resolution experiments with physics-based computational modeling to elucidate droplet generation in a T-shaped cylindrical microfluidic device. By systematically mapping flow regimes and extracting scaling relations for droplet size, curvature, and breakup dynamics across capillary and flow-rate ratios, we establish generalizable design principles that extend beyond a specific device geometry. In addition, a novel empirical correlation is proposed to determine the thin film thickness accounting the visco-inertial and capillary effects. These findings advance the development of predictive chemical engineering frameworks for multiphase flows in confined geometries and support the rational design and scale-up of droplet-based chemical and materials processing technologies.
\section{Background}
Over the recent decades, droplet microfluidics has emerged  \citep{Anna2003,Christopher_2007,Eggers2008,Abate2012,Glawdel2012a,Glawdel2012b,Glawdel2012c,Zhu2017,Doufene2019,Totlani_2020} as a transformative technology, enabling the precise manipulation of extremely small fluid volumes within microchannels forming diverse geometries. Recent reviews \citep{Chen_2022,Lathia_2023,Cao_2024,Han_2024,Harispe_2024,Li_2024,Mora_2024,Nan_2024} have summarized key advancements, including droplet formation mechanisms, cell encapsulation strategies, and droplet-to-microgel transformations, while also outlining future directions for biomedical applications such as 3D cell culture, single-cell analysis, and in-vitro organ and disease modeling. Within this framework, microdroplets have revolutionized chemical, biomedical, diagnostic, and additive manufacturing applications by serving as discrete reaction vessels that enable high-throughput screening of chemical and biological samples \citep{Shang_2017,Jain_2024,matula2020single,Jing_2024}. They have also been widely employed in single-cell analysis \citep{mazutis2013single,Li_2021,Liang_2024}, facilitating the investigation of cellular heterogeneity with unprecedented resolution. Furthermore, microdroplets provide a controlled environment for chemical synthesis and reaction engineering \citep{wang2017droplet,wu2023microdroplet}, and have been proposed as a promising platform for nanoparticles synthesis \citep{Gimondi_2023}, enabling precise tuning of reaction parameters and the fabrication of well-defined materials.

\noindent
Microdroplets form due to interfacial instabilities between the fluids, arising from a complex interplay of interfacial, inertial, and viscous forces governing the flow. The droplet breakup, arising from a pressure difference across the interface together with inertial effects, is governed by the Rayleigh-Plateau instability \citep{Rykner_2024}. An increase in interfacial tension accelerates the deformation of the interface, enhancing the pressure gradient according to the Young-Laplace equation, and thereby expediting droplet formation.
In order to understand these complex dynamics, \citet{Thorsen2001} employed a cross-flow microfluidic device, which has emerged as a promising platforms for generating monodisperse and stable droplets, offering unique advantages in terms of scalability and controllability.  Subsequent studies \citep{Anna2003,Garstecki2006} laid the foundation for elucidating droplet formation mechanisms in T-junction microchannels. Subsequently, the concept of ``flow focusing'' was  introduced to emphasize the critical role of capillary forces in governing droplet generation.

\noindent
Recent studies \citep{venkateshwarlu2021effects,Venkateshwarlu_2022,Venkateshwarlu_2023,Venkateshwarlu_2024,Dogra_icfd22,Dogra_2024,Maurya_2023,Chen_2022,Nan_2024,Shen_2024,Pandey_2025} have comprehensively reviewed the literature on droplet generation in cross-flow microfluidic devices; here, it is only briefly summarized to ensure completeness while avoiding redundancy. 
For instance, \citet{Garstecki2006} demonstrated droplet generation in a T-junction microfluidic channels of square cross-section. They identified two distinct flow regimes, squeezing and dripping, each characterized by different mechanisms responsible for the abrupt breakup of the dispersed phase into droplets. In the squeezing regime, droplet formation is primarily governed by interfacial forces (surface tension), whereas in the dripping regime, it is dominated by the viscous stresses of the continuous phase. Their simple scaling laws further indicate that the droplet size ($L$) and flow rate ratio ($Q_\text{r}$) exhibits a linear dependence  in the squeezing (interfacial tension-dominated) regime, and a sub-linear dependence in the dripping (viscosity-dominated) regime. \citet{Christopher2008} carried out a systematic experimental investigation demonstrating that the droplet-breakup process in microfluidic T-junctions, particularly near the transition from squeezing-dominated pinch-off to dripping, where viscous shear stress becomes increasingly influential. They reported that under these conditions, highly uniform droplets could be generated, with polydispersity below 2\%, across a wide range of capillary numbers and flow-rate ratios spanning the transition region.
\citet{zhao2010effect} investigated the effects of surface modification and hybrid wall coatings on droplet formation in T-junction microchannels, enhancing understanding of wettability impact  \citep{mousavi2022effect,Venkateshwarlu_2023,Venkateshwarlu_2024} on droplet dynamics. \citet{wang2015fluid} numerically quantified internal circulation rates governing mixing within droplets in a serpentine microchannel using the volume-of-fluid method, and found that the the presence of the winding section improved the effectiveness of mixing due to the occurrence of uneven circulation.

\noindent
Similarly, \citet{li2014computational} computationally examined mixing efficiency inside liquid slug formed in T-type microchannel. They reported that increasing velocity of the continuous phase ($u_\text{c}$) had a restricted impact on mixing performance for constant $Q_\text{r}$, as the density of symmetrical vortices reduces with increasing $u_\text{c}$, which constrained the achievable mixing efficiency.
\citet{li2017experimental} experimentally and numerically investigated liquid plug hydrodynamics in circular microchannels (0.2 and 0.5 mm diameters) and found that the liquid film thickness increases non-linearly with the capillary number ($\cac = 0.0224 - 0.299$), and the front and rear menisci shapes deform significantly at higher $\cac$. \citet{Nekouei2017} numerically investigated the effects of viscosity ratio ($\mu_{\text{r}}=0.01-15$) and capillary number ($\cac=0.001-0.5$) on droplet formation in a rectangular T-junction microchannel. They found that the reduction in droplet size with increasing $\cac$ is more pronounced for low viscosity ratios ($\mu_{\text{r}}<1$) than for high viscosity ratios ($\mu_{\text{r}}>1$). Moreover, at a given $\cac$, the droplet size increases with $\mu_{\text{r}}>1$, while it remains comparatively less sensitive to variations in $\mu_{\text{r}}<1$.
\citet{kovalev2018flow,kovalev2021viscosity} experimentally investigated the influence of viscosity ratio ($\mu_{\text{r}}=0.001-0.67$) on two-phase flow ($Q_{\text{c}}=0.28-27.8$ $\mu$L/min; $Q_{\text{d}}=0.14-556$ $\mu$L/min) in T-shaped rectangular microchannels with non-uniform cross-sections (inlet: $w \times w$; outlet: $w \times 2w$; $w=120$ $\mu$m \citep{kovalev2018flow}, $w=200$ $\mu$m \citep{kovalev2021viscosity}). At very low viscosity ratio ($\mu_{\text{r}}=0.001$), various flow regimes (i.e., plug, droplet, slug, throat-annular, and parallel flow) were identified, primarily as a function of the dispersed-phase Reynolds number \citep{kovalev2018flow}.
Furthermore, three dominant flow regimes (plug, droplet, and parallel flow) were reported \citep{kovalev2021viscosity} for broader flow conditions ($Q_{\text{c}}$, $Q_{\text{d}}$, $\mu_{\text{r}}$).
These findings \citep{kovalev2018flow,kovalev2021viscosity} were further validated by the experimental study \citep{ma2022effect}, who examined viscosity effects ($\mu_{\text{r}}=0.016-1.423$) on liquid–liquid slug flow ($Q_{\text{d}}=5-50$ $\mu$L/min, $Q_{\text{c}}=5-1000$ $\mu$L/min) in a step T-junction microchannel. They reported shifts in flow transitions with capillary and Reynolds numbers, along with distinct plug morphologies at higher $\mu_{\text{r}}$. In addition, the internal circulation within droplets was found to be strongly influenced by variations in flow parameters. Such circulation greatly enhances mixing efficiency compared to single-phase flows.

\noindent
Recently, \citet{jena2023effect} experimentally investigated the influence of T-junction rectangular microchannel geometry ($w_{\text{c}}=250-440$ $\mu$m, $w_{\text{d}}=250 - 460$ $\mu$m, $h=200$ $\mu$m) and flow conditions ($\mu_{\text{r}}=0.0146$, 0.4167; $\rho_{\text{r}}=0.773$, 0.960; $\qr=0.25-1$; $\cac=0.00015-0.22$) on droplet size and generation frequency. Their findings offer valuable guidelines for optimizing T-junction designs for specific droplet microfluidics applications, further supported by a other study \citep{He_2023}. \citet{Jafari_2023} numerically investigated the effects of viscosity ratio ($\mu_{\text{r}}$), flow rate ratio ($\qr$), and outlet channel geometry on droplet formation in a square cross-junction microfluidic device. Their study demonstrated that modifying the outlet design can shift the transition boundary between the squeezing and jetting regimes and alter droplet generation frequency under otherwise identical operating conditions.  

\noindent
Subsequent recent numerical contributions \citep{venkateshwarlu2021effects,Venkateshwarlu_2022,Dogra_icfd22,Venkateshwarlu_2023,Venkateshwarlu_2024} have systematically investigated droplet generation in rectangular T-junction microfluidic systems, emphasizing the effects of capillary number ($\cac = 10^{-4} - 1$), flow-rate ratio ($\qr = 0.1 - 10$), viscosity ratio ($\mu_{\text{r}}=7.143\times 10^{-3} - 0.7143$), and wettability ($\theta=120^{\circ} - 180^{\circ}$) while fixing other parameters ($w_{\text{c}}=100~\micro$m, $w_{\text{r}}=1$, $Q_{\text{d}}=0.14~\mu$L/s, $\rho_{\text{c}}=1000$ kg/m$^{3}$, $\rho_{\text{r}}=1$, $\mu_{\text{d}}=0.001$ Pa·s, $Re_{\text{c}}=0.1$). These studies thoroughly analyzed instantaneous phase profiles, droplet formation stages, flow regimes, interface and pressure evolution, interface-to-neck ratio (INR), and droplet size and frequency as functions of key parameters. Their regime maps show that droplet size decreases with increasing $\cac$, with stronger sensitivity at low $\mu_{\text{r}}$ while scaling laws confirm linear ($\propto \qr$) dependence in the squeezing regime and weaker (sub-linear, $\propto\qr^n$, $n < 1$) trends in dripping/jetting.  Time-resolved simulations reveal distinct formation stages (filling, necking, pinch-off, relaxation), pressure fluctuations, and INR as useful descriptors. Wettability shifts regime boundaries and modulates droplet size and frequency, whereas geometric variations (non-uniform or slit-type junctions) alter confinement and shear fields, thereby tuning breakup dynamics. Internal recirculation within droplets (governed by $\cac$, $\qr$, and confinement) was linked to enhanced mixing.  Predictive correlations for time for each stage of droplet formation, interfacial pressure drop, and droplet length and frequency  as a function of governing parameters were also developed to depict the complex nature of the physics.
A complementary limited study \citep{Dogra_2024} explored cylindrical T-junctions, highlighting the influence of flow rate ratio ($\qr=0.667-1$), contact angle ($\theta=120^{\circ}-135^{\circ}$), and channel diameter ($D=140-280~\micro$m) under other fixed conditions  ($\rho_{\text{r}}=0.998$, $\mu_{\text{r}}=3.5$, $\cac=10^{4}$) on droplet hydrodynamics, and confirmed the critical roles of governing parameters ($\qr$, $\theta$ and channel geometry) in tuning breakup hydrodynamics.
More recently, \citet{Datar_2025} have performed high-fidelity measurements of liquid film thickness in an air-water adiabatic slug flow in a T-type circular ($R=250$ \micro m) glass microchannel for wide ranges of the bubble length ($8\le L/R\le 90$), bubble velocity ($0.5 - 1.4$ m/s) and capillary numbers ($0.0062\le \cac\le 0.0185$).
They reported that bubble length increases with a given bubble velocity, while the film thickness asymptotically approaches a maximum value, consistent with the semi-infinite bubble approximation previously described in the literature \citep{Bretherton_1961,aussillous2000quick,Han_2009,Klaseboer_2014,Youn_2021}.
Altogether, these efforts \citep{venkateshwarlu2021effects,Venkateshwarlu_2022,Dogra_icfd22,Venkateshwarlu_2023,Venkateshwarlu_2024,Dogra_2024,Datar_2025} establish a coherent framework linking force balances, interface dynamics, and confinement effects, providing practical guidelines for controlled, monodisperse droplet generation in microfluidics.

\noindent
In summary, the existing literature provides extensive insights into the hydrodynamics and mechanisms of droplet generation in T-type `rectangular' microfluidic devices across a broad range of flow-governing ($\cac$, $\qr$, $\mu_{\text{r}}$, $\rho_{\text{r}}$, $\theta$) and geometrical ($w_{\text{c}}$, $w_{\text{d}}$, $h$) parameters. In contrast, knowledge of droplet generation in `cylindrical' capillary microfluidic devices remains scarcely limited, despite their widespread applications in reactors, biochemical and biological analyses, cell sorting, vascular diagnostics, surgical investigations, and additive manufacturing. 

\noindent
Therefore, Addressing this gap in the literature underpins the novelty of the present work. In the present study, combined experimental and numerical investigations are carried out to elucidate the hydrodynamics and mechanisms of droplet generation in cross-flow T-type cylindrical microfluidic devices. The experimental results, obtained using micro-particle image velocimetry (\micro-PIV), are complemented by extensive three-dimensional (3D) computational fluid dynamics (CFD) simulations performed with the finite element method (FEM) based COMSOL Multiphysics solver. The microfluidic devices are fabricated through an `in-house' developed technique based on the embedded template approach. This study establishes a novel understanding of the complex interplay between hydrodynamic characteristics (phase evolution, pressure and velocity fields, flow regimes, droplet size, and generation frequency, film thickness) and the governing parameters (microchannel dimensions, physical properties, flow rates of both phases, and contact angle).

\section{Physical Description}
\label{sec:problem}
\begin{figure}[!b]
	\centering\includegraphics[width=1\linewidth]{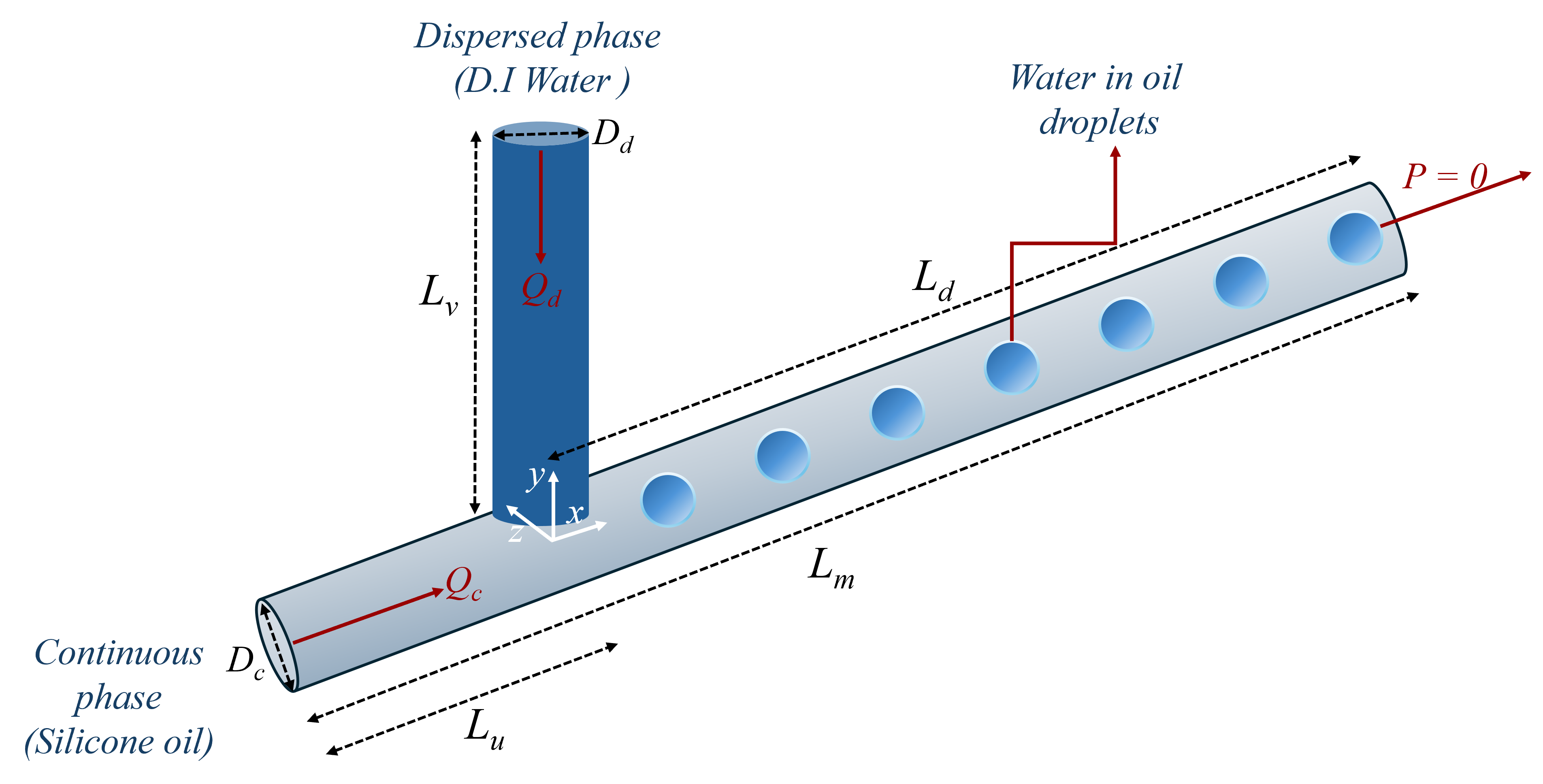}
	\caption{Schematic representation of the droplet formation in the two-phase cross-flow through a T-junction cylindrical microfluidic device.}
	\label{fig:1}
\end{figure}

Consider the two-phase immiscible liquid–liquid (LL) flow through a T-shaped cylindrical microfluidic device, as illustrated in \fig\ref{fig:1}. While the underlying mechanisms of droplet breakup are similar in both rectangular and cylindrical geometries, the cylindrical confinement introduces distinct interfacial curvatures and wettability effects, which can significantly influence the hydrodynamics of droplet formation. It is noteworthy that the fundamental hydrodynamic description of droplet generation in two-phase flows through T-junction microchannels with rectangular cross-sections \citep{venkateshwarlu2021effects,Venkateshwarlu_2022,Venkateshwarlu_2023,Venkateshwarlu_2024} remains broadly consistent across different cross-sectional geometries and is briefly outlined here.

\noindent
In a typical cross-flow configuration, two immiscible fluids converge at the T-junction, where their interaction leads to droplet formation downstream in the microchannel. 
The microfluidic device (\fig\ref{fig:1}) is configured using a main horizontal cylindrical channel (length $L_{\text{m}}$; diameter $D_{\text{c}}=2R_{\text{c}}$) and a side vertical cylindrical branch (length $L_{\text{v}}$; diameter $D_{\text{d}}=2R_{\text{d}}$). The side channel intersects the main channel perpendicularly at a position defined by the upstream length ($L_{\text{u}}$) and downstream length ($L_{\text{d}}$). Here, $L_{\text{u}}$ and $L_{\text{d}}$ are measured from the vertical centerline of the side channel to the inlet and outlet of the main channel, respectively, as shown in \fig\ref{fig:1}. Accordingly, the total length of the main channel is given as $L_{\text{m}} = (L_{\text{u}} + L_{\text{d}}$). In the present study, both channels are considered to possess uniform and equal cross-sectional areas, such that their diameters are identical ($D_{\text{d}} = D_{\text{c}}$) in the cylindrical configuration.

\noindent
The continuous-phase (CP, denoted by subscript c) liquid enters the main channel and the dispersed-phase (DP, denoted by subscript d) liquid enters the side channel with volumetric flow rates  (in \micro L/min) of $Q_{\text{c}}$ and $Q_{\text{d}}$, respectively. The two phases meet at the T-junction and subsequently flow together downstream in the main channel, which is open to the ambient environment, corresponding to an outlet pressure of $P=0$. Subsequently, different flow regimes, including droplet and non-droplet regimes \citep{venkateshwarlu2021effects}, emerge from the perturbations induced by the complex interplay of interfacial, inertial, and viscous forces.

\noindent
In this study, the physical properties of both immiscible Newtonian liquids, such as density ($\rho$, kg/m$^{3}$), viscosity ($\mu$, Pa·s), and interfacial tension ($\sigma$, mN/m), are assumed constant, i.e., independent of temperature, pressure, time, and position, in both experiments and simulations. The diameter of the main channel ($D_{\text{c}}$) is taken as the characteristic length for scaling the geometric parameters. The relevant dimensionless governing and influencing flow parameters are defined as follows.
\begin{gather}
X_{\text{r}}	= \frac{X_{\text{d}}}{X_{\text{c}}}; \qquad
Re_{\text{c}}	= \frac{D_{\text{c}} u_c \rho_c}{\mu_c}; \qquad
\cac	= \frac{u_c \mu_c}{\sigma}
\label{eq:1}
\end{gather}
where $X = (D, Q, \rho, \mu)$ denotes the set of governing parameters, and $Re$ and $\cac$ represent the Reynolds and Capillary numbers, respectively. 
\section{Research Methodology}
This study has performed flow visualization experiments using micro-particle image velocimetry (\micro-PIV) alongside three-dimensional (3D) computational fluid dynamics (CFD) simulations performed with the finite element method (FEM) in COMSOL Multiphysics. The microfluidic devices have been fabricated in-house using a template-embedding technique. The following sections detail the experimental and computational methodologies used in this work.
\subsection{Experimental methodology}
This section presents the details of the experimental materials, the fabrication of the microfluidic device, the \micro-PIV experimental setup, and the data analysis approaches used in this study.
\subsubsection{Materials}
This study has used silicone oil (procured from Sigma Aldrich) as the continuous phase (CP) fluid, and deionized (DI) water as the dispersed phase (DP) fluid to generate the water-in-oil (W/O) droplets.  The interfacial tension ($\sigma$) is measured using a tensiometer, the rheological properties ($\mu$) of fluids are measured using a rheometer (MCR 702, Anton Paar, Germany), and the contact angle ($\theta$) is determined by the sessile drop method using goniometer (Kruss). The physical properties of the liquid materials used are summarized in \tab\ref{tab:1}.
\begin{table}[!b]
	\renewcommand{\arraystretch}{1.5}
		\caption{Physical properties (density $\rho$, viscosity $\mu$, interfacial tension $\sigma$, contact angle $\theta$) of liquids and flow parameters (flow rate $Q$, capillary number $Ca$) used in this study.}\label{tab:1}
		\resizebox{\textwidth}{!}{
			\begin{tabular}{|l|l|c|c|c|c|c|c|}
				\hline
				Phase   & Experimental Fluid   &
				$\rho$ (kg/m$^3$) & $\mu$ (mPa.s)  &  $\sigma$ (mN/m) & {$\theta$} (\degree)& $Q$ (mL/h) &$Ca$ (-) \\
				\hline
				Continuous phase (CP) &  Silicone oil & 966 &    $305$  &56  & 150 &$0.03 - 5$& $10^{-3} - 10^{-1}$\\ \hline
				Dispersed phase (DP)&  Deionized (DI) water & 1000 & $0.90$    & -- & -- &$0.03 - 5$& --\\ \hline
				\multicolumn{2}{|r|}{$X_{\text{r}}$ (\eqn\ref{eq:1}) $\rightarrow$}  & $1.035$ &$0.00295$    & -- & -- &$0.1 - 10$& -- \\
				\hline
			\end{tabular}
		}
\end{table}

\noindent
Further, the microfluidic device has been fabricated using the widely adopted elastomeric material \citep{Nge_2013,Ren_2013,Nielsen_2019}, polydimethylsiloxane (PDMS) Sylgard 184 (procured from Sigma Aldrich). Nylon wires are used as a raw template, glass slides are used as a substrate, solvents (chloroform) are used for swelling of the cross-linked polymer,  and silicone oil (CP fluid) is used to mitigate swelling effects resulting from solvents. The silicon tubing and other connections assemblies are used to enable the fluid supply and distribution throughout the microfluidic device setup.

\noindent
For flow visualization, fluorescent seeding particles (polystyrene, 1.05 g/cm$^3$ density, 1 \micro m mean diameter, 3\% monodispersity, green absorption, red emission, refractive index of 1.59 at 589 nm and 25{\degree}C) are suspended in the working fluid (DP), enabling illumination and tracking of particle displacements, thereby facilitating reconstruction of the flow field.
\subsubsection{Fabrication of T-junction cylindrical microfluidic device}
In the present study, the microfluidic devices are fabricated using the \textit{embedded template method} \citep{Verma_2006}, which employs  templates to shape structures and form channels within a polymeric material.
%
\begin{figure}[!b]%
	\centering
	{\includegraphics[width=1\linewidth]{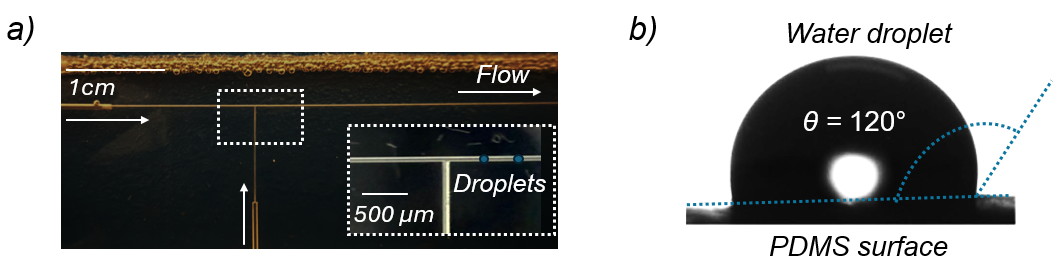}}
	\caption{(a) Fabricated sample of T-junction cylindrical microchannel inside a PDMS block. (b) Measured contact angle of water on PDMS block.}%
	\label{fig2a}
\end{figure}
In this process, a nylon wire of specified diameter (corresponding to the desired inner diameter of the microchannel, e.g., 150 \micro m in this study) is first thermally treated (at 120\degree C) to impart rigidity and then shaped into the required T-configuration followed by  affixing to a glass substrate.
Subsequently, a liquid precursor, consisting of PDMS and a curing agent in a $10\mathbin{:}1$ by weight, is poured onto the nylon template fixed on the glass substrate. The entire assembly, with the liquid precursor, is then cured in an oven at 90 {\degree}C for 12 hours. After curing, the nylon wire is carefully removed by solvent-assisted swelling and extraction, leaving behind a cylindrical microchannel of the desired T-shaped geometry within the PDMS block. A  snapshot of a fabricated T-junction cylindrical microchannel is shown in \fig\ref{fig2a}. The prepared device is then bonded to a glass substrate, followed by integration with inlet-outlet tubing assemblies for fluid delivery. To prevent any swelling effects during flow visualization experiments, the channel is primed with the continuous-phase (CP) liquid at a fixed flow rate ($0.5$ mL/h) for sufficiently long duration ($3 - 4$ hours).
\newline
In this work, the T-junction cylindrical microfluidic devices with uniform and equal channel diameters ($D_\text{c}=D_\text{d}$ or $D_\text{r} = 1$) have been in-house fabricated and employed in \micro-PIV setup (\fig\ref{fig2b}) to investigate the hydrodynamics of droplet generation.
\subsubsection{Experimental setup}\label{sec:exset}
In this study, the hydrodynamics of droplet generation in microfluidic devices has been examined using the two-dimensional (2D) micro-particle image velocimetry (\micro-PIV) technique \citep{Adrian_1991,Raffel_2018}. 
\begin{figure}[!b]%
	\centering
	{\includegraphics[width=0.9\linewidth]{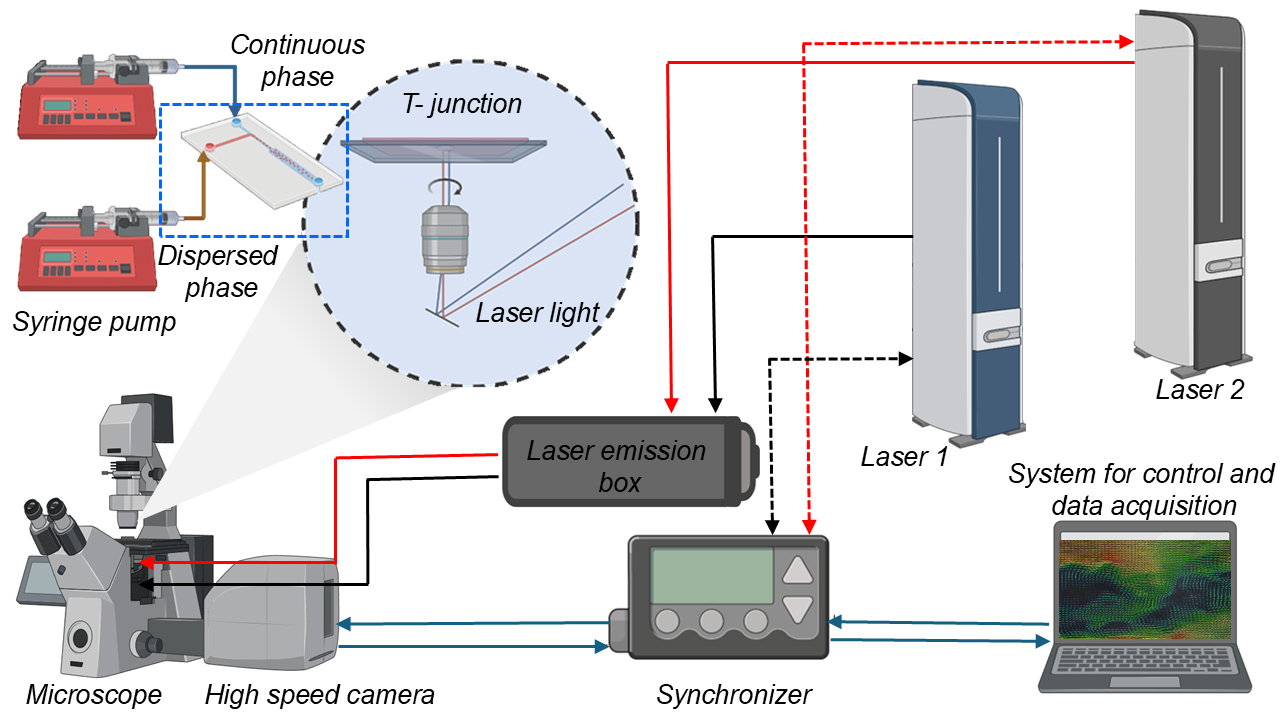}}%
	\caption{{Schematic representation of} experimental $\mu$-PIV setup. }%
	\label{fig2b}
\end{figure}
The experimental setup (\micro-PIV; procured from TSI Instruments India Private Limited), as illustrated in \fig\ref{fig2b}, consists of the following major components:
\begin{itemize}
	\item A high-speed camera (Phantom VEO E340, CMOS sensor, resolution $2560 \times 1600$ pixels $\approx 4$ MP at 800 fps, 36 GB RAM, 10 {\micro m} pixel pitch, 12-bit pixel depth, 2 {\micro s} frame straddle time, $532 \pm 5$ nm optical camera filter, ISO 6400 sensitivity);
	\item A dual-cavity Nd:YAG laser (wavelength 532 nm for each head, 100 Hz frequency per cavity, pulse energy $2 \times 50$ mJ, pulse width $\leq 10$ ns at 1064 nm, integrated laser beam emission box);
	\item An inverted fluorescence microscope (Olympus, high-contrast optics with both laser and halogen light sources, F-mount camera support);
	\item A high-resolution synchronization device (computer-controlled programmable laser pulse synchronizer with delay range 0 -- 5000 s, pulse width 10 ns -- 1000 s, and time resolution 0.5 ns);
	\item A data acquisition, storage, and processing unit (Intel Xeon Silver 4110 processor, 32 MB cache, 2.2 GHz base frequency, 128 GB RAM, 16 GB dedicated GPU); 
	\item Two syringe pumps (New Era Pump systems. Inc, USA) to independently introduce the continuous and dispersed phases into the microchannel through fine silicone tubing at flow-rates of $Q_{\text{c}}$ and $Q_{\text{d}}$, respectively.
\end{itemize}
\subsubsection{Experimental and analysis procedures}\label{sec:exsetp}
The experiments were conducted using T-junction cylindrical microfluidic devices with uniform cross-section and equal channel diameters ($D_\text{r} = 1$; $D_\text{c} = D_\text{d} = 150$ \micro m). Both the continuous phase (CP) and dispersed phase (DP) flow-rates (in mL/h) were varied as $0.03\le Q_\text{c},\ Q_\text{c}\le 5$, while maintaining the flow-rate ratio ($0.1\le Q_\text{r}\le 10$), refer to \tab\ref{tab:1}. Both phases (CP and DP) were continuously injected, using syringe pumps, through the inlets of the main and side channels, respectively, as shown in \fig\ref{fig:1} to ensure uniformity. The measurements were taken after allowing sufficient time for the flow to become fully developed and at an adequate distance from the channel entrance to minimize entrance effects.  
 Following each round of experimental runs, the microchannel was sequentially flushed with acetone and pressurized air. After cleaning, the microchannel was further primed with silicone oil to mitigate any swelling effects that may have been caused by the acetone treatment.

\noindent
In order to examine the physical behavior of phase distributions in the form of droplets and plugs, such as their length, radius, and film thickness, the experiments were conducted using a high-speed camera operated at a variable frame rate (800 --1400 fps), depending on the flow rates ($Q_\text{c}$, $Q_\text{d}$).
The raw images captured during the experiments were analyzed using Insight 4G (Global Image Acquisition, Analysis and Display) Software (TSI Incorporated) and ImageJ (an open-source software) for image processing and scientific analysis \citep{Schneider_2012}.
Further, the experimental setup was switched from the `camera-only' mode to the `PIV' mode for flow visualization and subsequent determination of the flow field. For the PIV experiments, the dispersed phase (DP) was seeded with a small fraction (<1\%) of fluorescent tracer particles. These tracer particles were illuminated at each laser pulse, and the corresponding images were recorded using a high-speed camera. 
\newline
The image processing and analysis procedures for PIV, extensively discussed in the literature \citep{Adrian_1991,Keane_1992,Shi_2015,Raffel_2018,Scharnowski_2020,Kumar2018,Kumar2021}, have been briefly outlined here as follows. For instance, each image is discretized into small interrogation windows. The spatial displacement ($\Delta \mathbf{X}$) of tracer particles is determined by cross-correlating \citep{Keane_1992,Shi_2015} two successive images recorded at time instants $t_0$ and $t_0+\Delta t$, where $\Delta t$ is the time delay between two subsequent laser pulses. The displacement ($\Delta \mathbf{X}$) that statistically produces the maximum cross-correlation ($R_\text{max}$) corresponds to the most probable average particle displacement ($\Delta \mathbf{X}_\text{avg}$) within each interrogation window. Repeating this process yields the particle displacement as a function of time ($\Delta \mathbf{X}$ vs. $\Delta t$), which in turn provides the local velocity field ($\mathbf{V} = \Delta \mathbf{X} / \Delta t$).

\noindent
In this study, images were captured at pulse delay times ($\Delta t$ \micro s) ranging from $10$ to $500$, and the velocity field was computed using the Insight 4G software. As a pre-processing step, the gray-scale values of the raw images were inverted to obtain bright particle signals on a dark background, as required by the software. For enhanced visualization of low-intensity images, pseudo-coloring was also applied.
Subsequently, the images were processed using a median filter to effectively remove/suppress any extraneous background noise, followed by normalization with a minimum-intensity mask to strongly enhance the contrast between tracer particles and the background, thereby improving the accuracy of velocity field estimation. 
\newline
The displacement of tracer particles between two consecutive frames was determined using square interrogation windows ($32\times 32$ pixels) with 50\% overlap, ensuring compliance with the Nyquist sampling criterion. The Fast Fourier Transform (FFT) algorithm was employed for determining the correlation, and the correlation peak position was refined by fitting a Gaussian curve to the highest-intensity pixel and its four neighboring pixels. The velocity vectors were then generated and validated using the median test and the peak-to-noise ratio (PNR). The spurious vectors were replaced with the median value of adjacent vectors to ensure robustness of the velocity field.

\noindent
A post-processing algorithm embedded in the Insight 4G software was employed to compute the ensemble average of approximately 50 consecutive image frames, yielding instantaneous velocity vector fields. To investigate the hydrodynamics of droplet generation, nearly 500 image frames were recorded and categorized according to their respective process stages. The sorted images were then ensemble-averaged to obtain representative instantaneous velocity vector fields for each stage.
Furthermore, a minimum of three experiments ($N_\text{ex} \geq 3$) were performed for each set of flow conditions ($Q_{\text{c}}$, $Q_{\text{d}}$, $\cac$) to ensure repeatability. The quantitative values reported in this study represent the average of three consistent experiments ($N_\text{ex} = 3$) performed for each set of flow conditions.
%
\subsection{Numerical methodology}
This section presents the numerical modeling approach, detailing the governing equations, boundary conditions, computational methodology, and the choice of simulation parameters adopted to analyze the problem under study.
\subsubsection{Mathematical model}\label{sec:gde}
The considered flow problem (refer to Section \ref{sec:problem}) is mathematically described \citep{venkateshwarlu2021effects,Venkateshwarlu_2022,Venkateshwarlu_2023,Venkateshwarlu_2024,Dogra_2024} using the Navier-Stokes equations (NSE) and the conservative level set method (CLSM), which represent the conservation of mass (\eqn\ref{eqn:l}), momentum (\eqn\ref{eqn:2}), and phase (\eqn\ref{eqn:lsm}) fields, respectively.
\begin{gather}
	\nabla \cdot \mathbf{{V}} = 0, 
	\label{eqn:l} \\
	\rho \left[\dfrac{\p \mathbf{{V}}}{\p t}+\mathbf{{V}} \cdot \bnabla\mathbf{{V}}\right] = (-\bnabla P+\bnabla \cdot \mathbf{\tau}+\mathbf{F}_{\sigma}), 
	\label{eqn:2} \\
	\dfrac{\p \phi}{\p t}+\mathbf{{V}} \cdot {\bnabla} \phi = \gamma\ {\bnabla} \cdot \left[\epsilon_{\text{ls}}{\bnabla} \phi- \phi (1-\phi)\mathbf{n}\right],  
	\label{eqn:lsm} 
\end{gather}
Here, $\mathbf{V}$,  $P$ and $\phi$ represent the velocity vector,  pressure and phase fields, respectively.
The deviatoric stress tensor ($\mathbf{\tau} = 2\mu \mathbf{D}$) is expressed in terms of the rate-of-deformation tensor $\mathbf{D}=(1/2)\left[\nabla\mathbf{{V}} + (\nabla\mathbf{{V}})^T \right]$. Both the density ($\rho$) and dynamic viscosity ($\mu$) are assumed to vary linearly with the level set function ($\phi$), which spans from $\phi = 0$ (pure CP) to $\phi = 1$ (pure DP), with $\phi = 0.5$ representing the liquid-liquid interface.
The two critical parameters for CLSM, namely the re-initialization parameter ($\gamma$) and the interface thickness regulator ($\epsilon_{\text{ls}}$), as used in \eqn(\ref{eqn:lsm}), effectively enable the accurate tracking of instantaneous topological changes of the liquid-liquid interface.
\newline 
The interfacial force ($\mathbf{F}_{\sigma}$) in \eqn\eqref{eqn:2} is modeled using the continuum surface force (CSF) formulation \citep{Brackbill_1992}, which represents it as a volumetric force localized near the liquid-liquid interface, as follows. 
\begin{gather}
		\mathbf{F}_{\sigma} = \sigma \kappa \delta \mathbf{n}, \qquad
	\kappa = -(\bnabla\cdot \mathbf{n}),\qquad \delta =6\phi(1-\phi){|\mathbf{\bnabla \phi}|},\qquad
	\mathbf{n} = \frac{\mathbf{\bnabla \phi}}{|\mathbf{\bnabla \phi}|}, 
	\label{eqn:lsmp}
\end{gather}
Here, $\kappa$ denotes the mean interface curvature ($\kappa = 1/R_{ic}$, where $R_{ic}$ is the radius of curvature of the interface), $\delta$ is the Dirac delta function, and $\sigma$ is the interfacial tension.

\noindent
The above-discussed fully coupled partial differential equations (PDEs), \eqn\eqref{eqn:l} to \eqref{eqn:lsm}, are solved subject to the following boundary conditions (BCs): (a) constant flow-rates ($Q_\text{c}$ and $Q_\text{d}$) for the CP and DP at the inlets of the main and side channels, respectively; (b) fully developed conditions at the outlet ($x=L_{\text{m}}$) of the main channel, which is open to ambient pressure (i.e., $P=0$, and $\partial f/\partial \mathbf{n} = 0$,  with $f=\mathbf{V}, \phi$); and (c) no-slip and wetted wall conditions (WWBC) on the surface of the cylindrical channels.

\noindent
The wetted wall boundary condition (WWBC) has been imposed to account for the motion of the liquid-liquid (LL) interface along the solid surface \citep{venkateshwarlu2021effects,Venkateshwarlu_2022,Venkateshwarlu_2023,Venkateshwarlu_2024,Dogra_2024}. 
%
%
In laminar microfluidic flows, the WWBC enforces no penetration normal to the wall ($\mathbf{V}\cdot \mathbf{n}_{w}=0$) and incorporates the tangential stress  ($\mathbf{K}_{\mathbf{nt}}$) as follows. %
\begin{gather}
	\mathbf{K}_{\mathbf{nt}} = \mu (\mathbf{V}/\beta), \qquad
	\mathbf{K}_{\mathbf{nt}} = \mathbf{K}_{\mathbf{n}} - (\mathbf{K}_{\mathbf{n}}\cdot \mathbf{n}_{w})\mathbf{n}_{w}, \qquad
	\mathbf{K}_{\mathbf{n}} = (\mathbf{K}\ \mathbf{n}_{w}), \qquad
	F_{\theta}= \sigma \delta (\mathbf{n}_{w}\cdot \mathbf{n} - \cos\theta)\mathbf{n}
	\label{eqn:06}
\end{gather}
 where, $\beta$ is the Navier slip coefficient, $\mathbf{K}$ is the viscous stress tensor,  and $\mathbf{n}_{w}$ is the unit vector normal to the wall.  Physically, $\beta \to 0$ corresponds to the classical no-slip condition, whereas $\beta \to \infty$ represents perfect slip at the wall. 
%
%
Furthermore, in the presence of interfacial forces ($\mathbf{F}_{\sigma}$, \eqn\ref{eqn:2}), the WWBC incorporates the wetting force ($F_{\theta}$), i.e., boundary force that enforces the contact angle, expressed as follow.
\begin{gather}
	F_{\theta}= \sigma \delta (\mathbf{n}_{w}\cdot \mathbf{n} - \cos\theta)\mathbf{n}
	\label{eqn:07}
\end{gather}
where $\theta$ is the static contact angle between the liquid-liquid interface and the solid wall.

\noindent
The numerical solution of the above-discussed mathematical model provides detailed information on the flow and phase fields, thereby offering valuable insights into the hydrodynamics of droplet generation.
\subsubsection{Simulation approach and parameters}
The numerical solution of the mathematical model (refer to Section \ref{sec:gde}), which couples the Navier-Stokes equations (NSE) with the conservative level set method (CLSM) under physically consistent boundary conditions, is obtained through three-dimensional (3D) simulations performed using COMSOL Multiphysics, a finite element method (FEM) based computational fluid dynamics (CFD) software. The computational domain (see \fig\ref{fig:1}) is discretized using non-uniform linear tetrahedral mesh elements.
\newline
The mathematical model is implemented in COMSOL Multiphysics using the three-dimensional (3D) laminar flow (\texttt{spf}), level set (\texttt{ls}), two-phase flow  (\texttt{tpf}), and wetted wall (\texttt{ww}) modules. The governing partial differential equations (PDEs) are discretized using the finite element method (FEM) on a non-uniform grid, with temporal discretization with an implicit backward differentiation formula (BDF). The detailed discretization and solution procedures are available in recent studies \citep{venkateshwarlu2021effects,Venkateshwarlu_2022,Venkateshwarlu_2023,Venkateshwarlu_2024,Dogra_2024}, and not repeated here.

\noindent
The segregated solution of the algebraic equations is obtained using the linear solver, PARDISO (Parallel Direct Solver), which is particularly efficient for large-scale, sparse systems of equations. The non-linear system is solved using Newton’s method with a sufficiently small time step ($\Delta t = 10^{-4}$ s) to accurately resolve the coupled flow ($\mathbf{V}$, $P$) and phase ($\phi$) fields. The level set parameters ($\gamma = 0.5$ m/s, $\epsilon_{\text{ls}} = h_{\text{max}}/2 = 4.5$ \micro m) and slip length ($\beta = 5$ \micro m), required for the wetted wall boundary condition (WWBC), are selected based on their independence studies \citep{Venkateshwarlu_2024}.  The fully converged solution at each time step is ensured by enforcing a relative tolerance of $10^{-3}$ and absolute tolerance determined automatically using a tolerance factor of 0.05 for all dependent variables, with a maximum of 50 nonlinear iterations per time step.
\subsubsection{Computational domain and mesh selection}
The reliability and accuracy of the numerical solution depend critically on the appropriate refinement of the mesh and the optimized size of the computational domain. In this study, the cross-sections of both the main and side channels are cylindrical and identical ($D_{\text{c}} = D_{\text{d}} = 150$ \micro m), consistent with the fabricated T-junction geometry employed in the experiments (refer to Section \ref{sec:exset}). An adequately long computational domain is necessary to minimize end effects, while a highly refined mesh (with element size approaching zero) is required to reduce discretization errors and ensure accurate results. However, both strategies significantly increase computational costs in terms of CPU time and memory usage. To achieve an optimal balance between numerical accuracy and computational efficiency, systematic independence studies with respect to domain length ($L_{\text{u}}$, $L_{\text{d}}$, and $L_{\text{v}}$) and mesh size have been performed.

\begin{figure}[!b]
	\centering
	\subfigure[Downstream length independence study $P^{*}$$(t)^{*}$ vs $L_{d}^{*}$ ]{
		\includegraphics[width=0.48\linewidth]{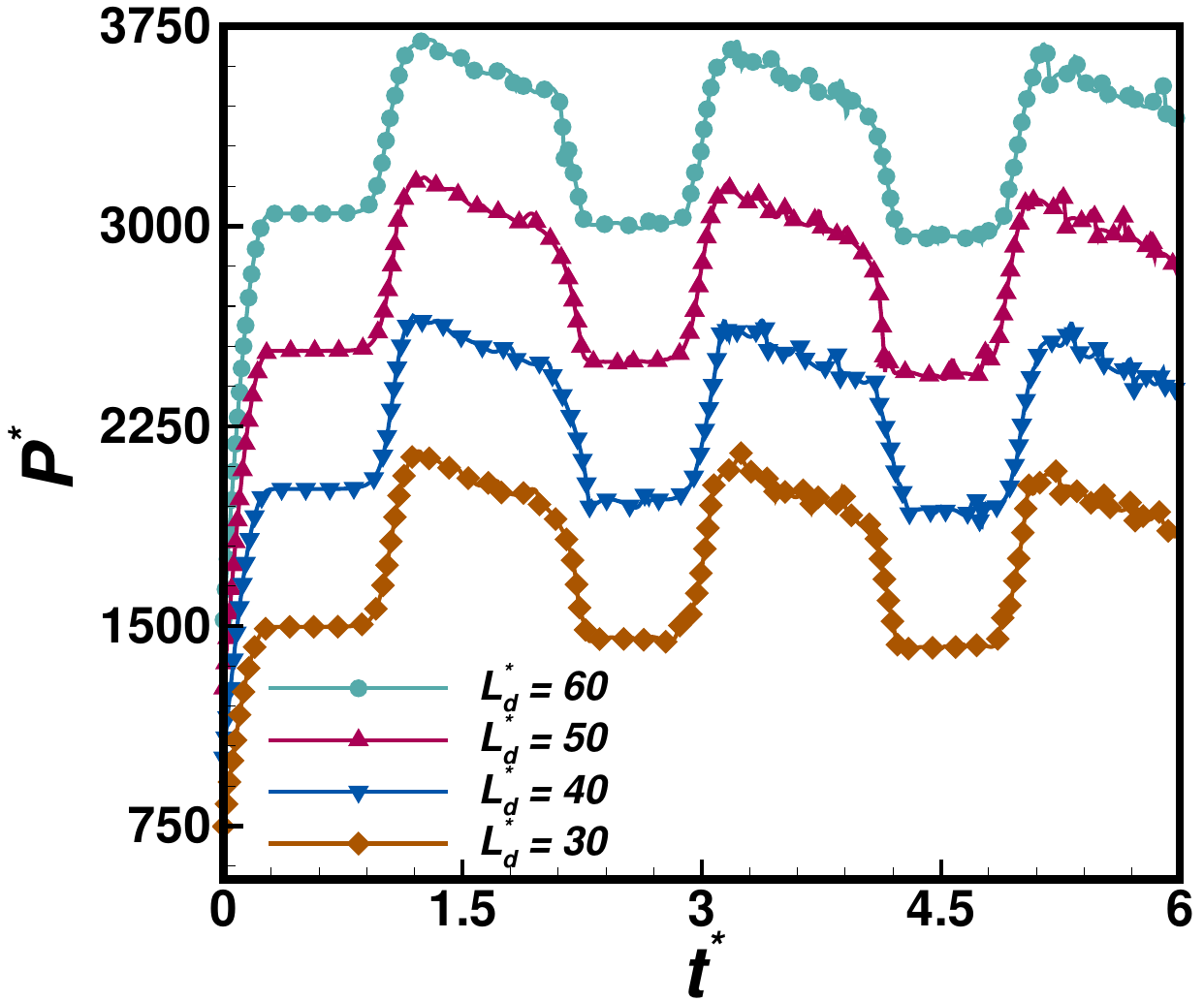}\label{fig:down-1}
	}
	\subfigure[Downstream length independence study $p^{*}$$(t)^{*}$ vs $L_{d}^{*}$ ]{
		\includegraphics[width=0.48\linewidth]{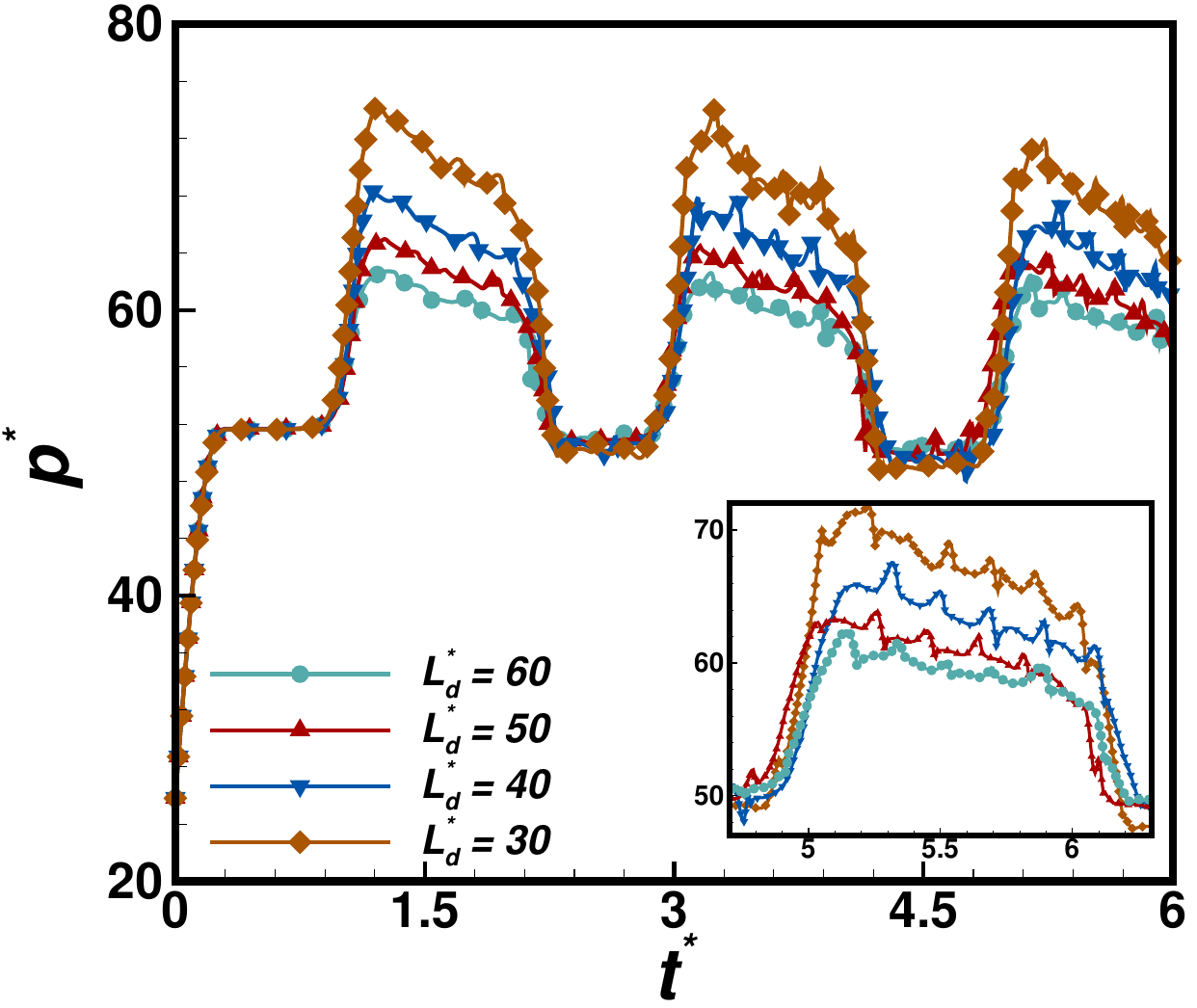}\label{fig:down-2}
	}\\ 
	\subfigure[Upstream length independence study $P^{*}$$(t)^{*}$ vs $L_{u}^{*}$ ]{
		\includegraphics[width=0.48\linewidth]{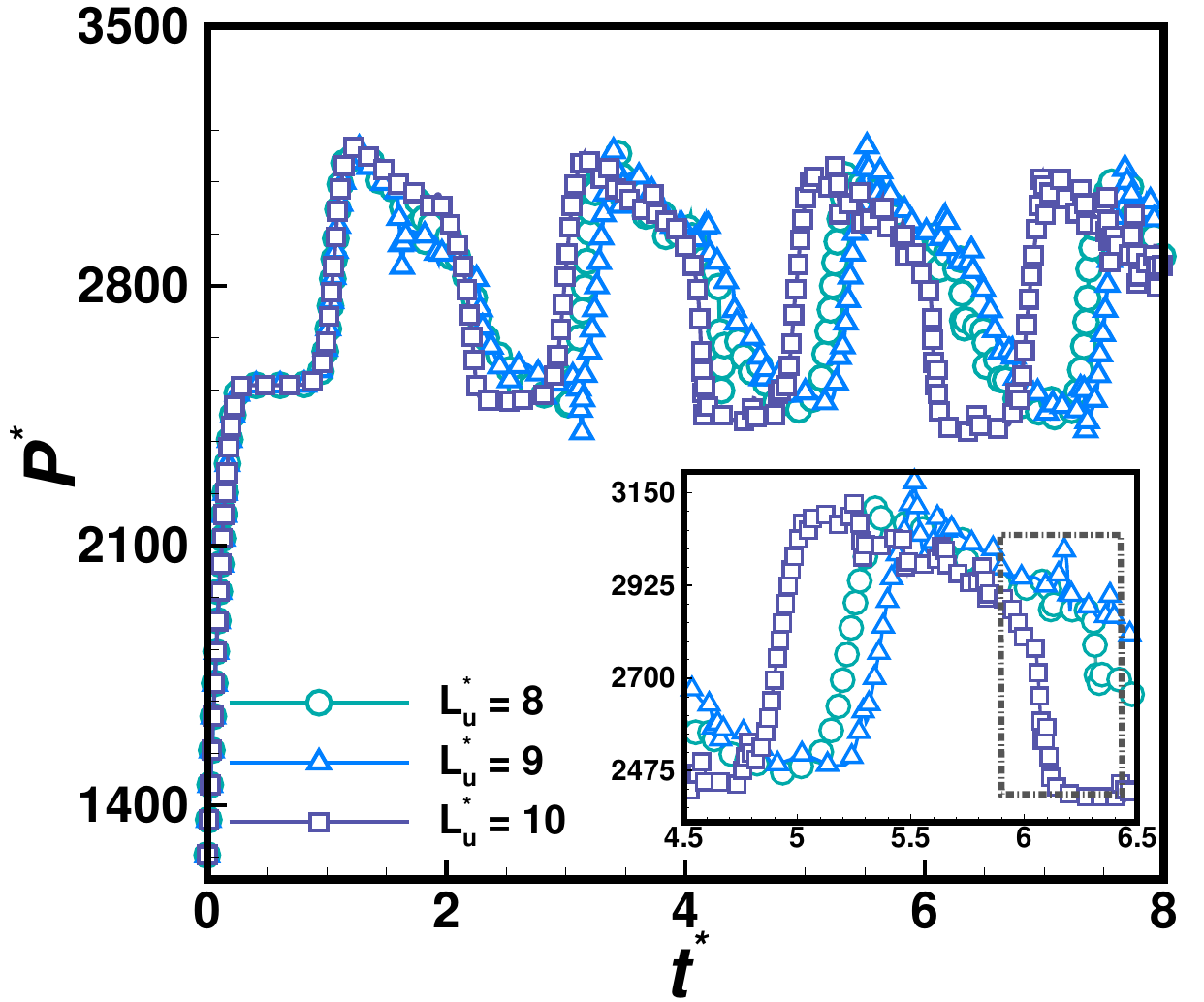}\label{fig:up-1}
	}
	\subfigure[Mesh independence study $P^{*}$$(t)^{*}$ vs $\lambda_\text{m}$ ]{
		\includegraphics[width=0.48\linewidth]{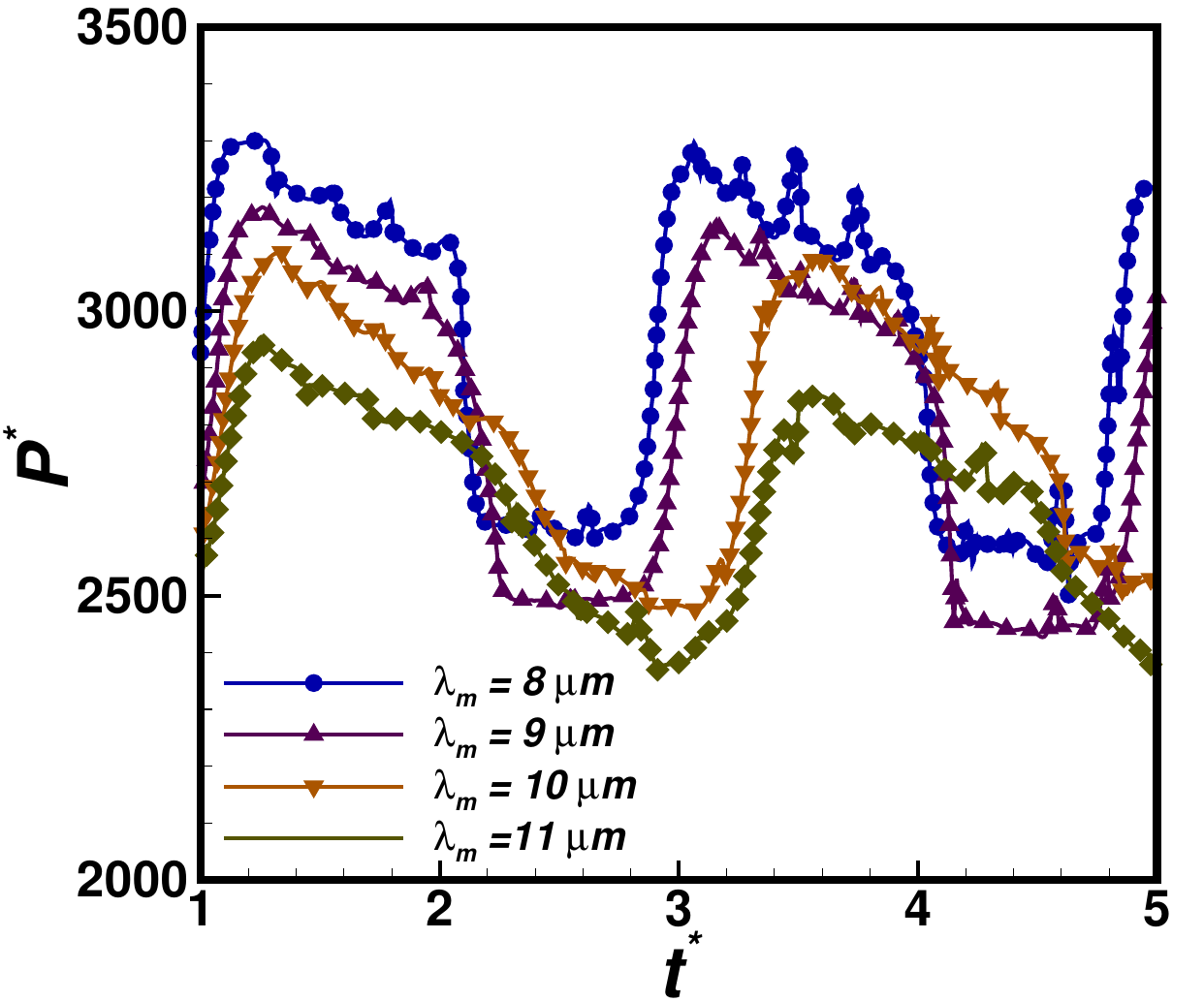}\label{fig:mesh-1}
	}
	\caption{The dimensionless instantaneous pressure, $P^{\ast}(t^{\ast})$, profiles at a probe point ($L_{\text{u}} + D_{\text{c}}, 0$) as a function of domain length ($L_{\text{u}}$, $L_{\text{d}}$, and $L_{\text{v}}$) and mesh size ($\lambda_\text{m}= h_\text{max}$) at the fixed flow rates ($Q_{\text{c}}=\mbox{0.03}$ mL/h, $Q_{\text{d}}=\mbox{0.03}$ mL/h).}
		\label{fig:domain_mesh}
\end{figure}

\noindent
\fig\ref{fig:domain_mesh} and \tab\ref{tab:2} present the results of the domain and mesh independence studies, showing the instantaneous local pressure profiles ($P^{\ast}$ and $p^{\ast}$, as defined in \eqn\ref{eq:scalePt}) at a probe location ($L_{\text{u}} + D_{\text{c}}, 0$) in the main channel, and the droplet length ($L^{\ast}$) as functions of computational domain size and mesh resolution. 
\begin{equation}
	P^{\ast} = \frac{P}{(\mu_{\text{c}}V_{\text{c}})/D_{\text{c}}}; \qquad
	p^{\ast} =  \frac{P^{\ast}}{L^{\ast}_\text{m}};
	\qquad
    t^{\ast} = \frac{t}{(D_{\text{c}}/V_{\text{c}})}
	\label{eq:scalePt}
\end{equation}
where $V_{\text{c}}$ is the average inlet velocity of CP.
In this study, the domain length independence analysis is performed by fixing the maximum mesh element size at $\lambda_\text{m}=10$ \micro m, and the corresponding results are shown in \fig\ref{fig:domain_mesh}(a–c).
To determine the sufficiency of the downstream length ($L_{\text{d}}$), \fig\ref{fig:down-1} shows the instantaneous pressure distribution, $P^{\ast}$, at the probe point ($L_{\text{u}} + D_{\text{c}}, 0$) for four downstream lengths ($L_\text{d}^{\ast} = L_\text{d}/D_{\text{c}} = 30$, 40, 50, 60), while keeping the fixed upstream length  ($L_\text{u}^{\ast} = L_\text{u}/D_{\text{c}} = 10$) and the side-channel length ($L_\text{v}^{\ast} = L_\text{v}/D_{\text{c}} = 10$). The profiles are obtained over a time duration up to $t^{\ast} = 7$, which is sufficient to capture three complete cycles of the droplet generation. Qualitatively, the pressure ($P^{\ast}$) profiles remain consistent; as expected, $P^{\ast}$ increases with $L_\text{d}$ due to the increasing total length ($L_\text{m}$) of the main channel.
\newline
Fundamentally, according to the Hagen-Poiseuille equation, the local pressure ($P$) in a cylindrical tube is governed by the pressure gradient ($\Delta P$) along the total length ($L_{m}$) of the tube. For a fixed volumetric flow rate ($Q$), the pressure drop is proportional to the tube length (i.e., $\Delta P \propto L_{m} \equiv \Delta P=C\times L_{m}$). Consequently, the pressure decreases linearly from its maximum value ($P=P_{\text{max}}$) at the inlet to zero ($P=0$) at the outlet. For a given $Q$, the value of $C$ decreases with increasing total length, which in turn raises the local pressure at a fixed location. This trend is qualitatively captured in \fig\ref{fig:down-1}; however, the sufficiency of $L_{\text{d}}$ cannot be determined solely from this observation. Therefore, for quantitative analysis, the pressure ($P^{\ast}$, \fig\ref{fig:down-1}) has been transformed with respect to the  channel length ($L_{m}^{\ast}$), and the normalized pressure ($p^{\ast}$, \eqn\ref{eq:scalePt}) is plotted in \fig\ref{fig:down-2}. The results indicate that $L_{\text{d}}^{\ast} = 50$ is sufficiently long to eliminate downstream end effects.
Further, to determine the sufficiency of the upstream length ($L_{\text{u}}$), \fig\ref{fig:up-1} shows the instantaneous pressure distribution, $P^{\ast}$, for three upstream lengths ($L_\text{u}^{\ast} = 8$, 9, and 10) at a fixed downstream length ($L_\text{d}^{\ast} = 30$) and side channel length ($L_\text{v}^{\ast} = 10$), with other parameters identical to those in \fig\ref{fig:down-1}. The results indicate that $L_\text{u}^{\ast} = 10$ is sufficiently long to eliminate entrance effects, yielding a smooth pressure distribution. A similar analysis (not shown here) for the side channel length confirms that $L_\text{v}^{\ast} = 10$ is also adequate to avoid entrance effects and ensure accurate results.
\begin{table}[!b]
	\begin{center}\renewcommand{\arraystretch}{1.5}
		\caption{Mesh specifications and droplet size}.\label{tab:2}
		\resizebox{0.8\textwidth}{!}{
			\begin{tabular}{|l|c|c|c|c|}
				\hline
				Mesh  &  M1&M2 &M3 & M4   \\ \hline
				Largest mesh element size ($\lambda_\text{m} = h_\text{max}$ \micro m) &  11&10 &9 & 8   \\ \hline
				Total number of elements ($N_{e}$) & 30,88,033 &  38,18,822   &49,20,184  &66,69,141 \\ \hline
				Dimensionless droplet length ($L^{\ast}=L/D_\text{c}$) & 2.2533   &  2.2333 &2.2000 &2.2033 \\
				\hline		\end{tabular}}
	\end{center}
\end{table}
\noindent
Furthermore, the results of the mesh independence study are shown in \fig\ref{fig:mesh-1} and \tab\ref{tab:2}. \fig\ref{fig:mesh-1} shows the instantaneous pressure distribution, $P^{\ast}$, while \tab\ref{tab:2} lists the total number of mesh elements ($N_{e}$) and droplet length ($L^{\ast}$) for four mesh sizes (M1, M2, M3, M4), using the above-optimized computational domain ($L_\text{u}^{\ast} = 10$, $L_\text{v}^{\ast} = 10$, $L_\text{d}^{\ast} = 30$) under the same conditions as in \figs\ref{fig:down-1}-\ref{fig:up-1}. As shown in \fig\ref{fig:mesh-1}, refining the mesh from M3 ($N_{e}=4,920,184$) to M4 ($N_{e}=6,669,141$) results in negligible changes in both droplet size ($L^{\ast}$= 2.200 to 2.203) and pressure ($P^{\ast}$) profile, but incurs a substantial increase in computational cost. Therefore, a mesh M3 is deemed sufficiently refined to capture the local flow features and droplet dynamics with accuracy.

\noindent
Overall, a computational domain of $L_\text{u}^{\ast} = 10$, $L_\text{v}^{\ast} = 10$, and $L_\text{d}^{\ast} = 50$, together with a mesh M3 ($\lambda_\text{m} = 9$ \micro m, $N_{e}=4,920,184$), is adopted for the results presented in the following section.
\section{Results and discussion}
In this section, the experimental and numerical results elucidating on the hydrodynamics of droplet generation in a T-type cylindrical microfluidic device are presented and discussed over a wide range of conditions (see \tab\ref{tab:1}). Before presenting these new results, however, the reliability and accuracy of the present findings are assessed to establish confidence in the experimental and modeling approaches.

\begin{figure}[!b]
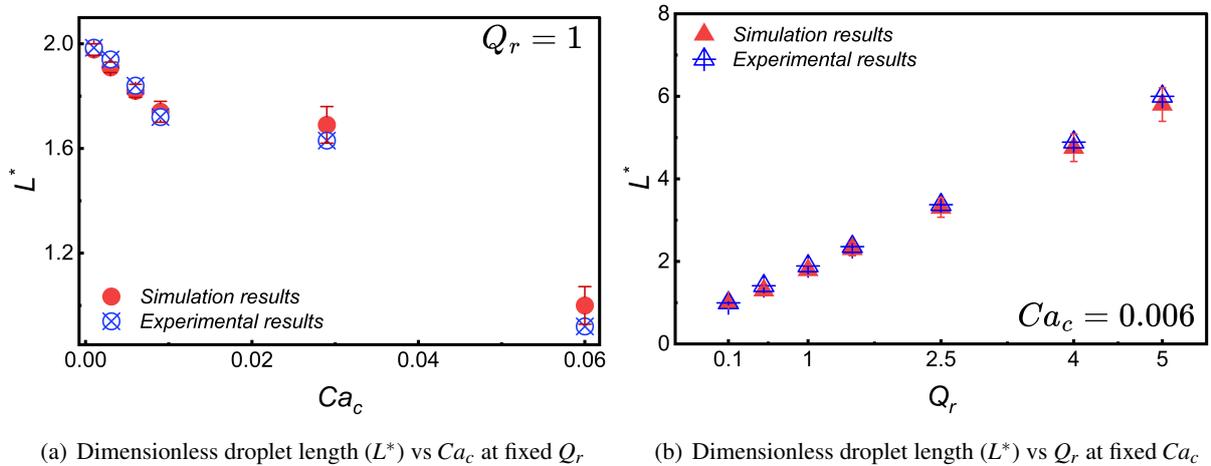
%
	\centering
	\subfigure[Dimensionless droplet length ($L^{*}$) vs $Ca_{c}$ at fixed $Q_{r}$ ]{{\includegraphics[width=0.48\linewidth]{Figure/valiation-Ca} } \label{validation-1}}%
	\subfigure[Dimensionless droplet length ($L^{*}$) vs $Q_{r}$ at fixed $Ca_{c}$ ]{{\includegraphics[width=0.48\linewidth]{Figure/valiation-Qr} } \label{validation-2}}%
	\caption{Comparison of present experiment and simulation results in terms of Dimensionless droplet length ($L^{\ast}=L/D_\text{c}$).}%
\label{validation}%
\end{figure}

\subsection{Validation}
The present CFD modeling and simulation approach has already been thoroughly validated in recent studies by comparing experimental and numerical results for droplet generation in T-junction rectangular microfluidic devices \citep{venkateshwarlu2021effects,Venkateshwarlu_2022,Venkateshwarlu_2023,Venkateshwarlu_2024,Dogra_2024}, and is therefore not repeated here. Building on this prior experience and confidence, the current experimental and computational results for droplet generation in T-junction cylindrical microfluidic devices are compared to further validate the approach, thereby establishing the reliability and accuracy of the present findings as outlined below.
%
\fig\ref{validation} compares the experimental and numerical results for the non-dimensional droplet length ($L^{\ast} = L/D_\text{c}$) as a function of the capillary number of the continuous phase ($\cac$) at a fixed flow-rate ratio ($\qr = 1$) in \fig\ref{validation-1}, and the flow-rate ratio ($\qr$) at a constant capillary number ($\cac = 0.006$) in \fig\ref{validation-2}. Both results show excellent agreement at lower $\cac$ for all values of $\qr$; however, small deviation appear at higher $\cac$ and $\qr$, where the simulations slightly overestimate the experimental data.
\newline
{ It is worth noting that the wetting conditions implemented in the simulations are defined by the equilibrium contact angle of the dispersed phase on PDMS. Experimentally, the static contact angle of water on PDMS is measured as $120\degree$, whereas a value of $150\degree$  is prescribed in the simulations to represent an idealized hydrophobic boundary condition. Nevertheless, the simulated droplet sizes and breakup dynamics remain within an acceptable range, with the present results are considered reliable and accurate to within $\pm 2\%$. Further, the simulated T-junction has a perfectly smooth surface, however, the microfluidic channel used in the experiments have some degree of surface heterogeneity \cite{thai2018formation}. In the recent study \cite{Dogra_2024} under the squeezing regime, the variations in contact angle ($\theta = 120\degree - 150\degree$)  were found to influence the droplet breakup dynamics rather than the final droplet size, which is primarily governed by flow rate ratios ($\qr$). However, for higher capillary numbers ($\cac$),  the study \cite{Liu2009} using a rectangular T- junction ($w = 100\ \mu$m) indicated that increasing hydrophobic conditions ($\theta$ varying from $120\degree$ to $180\degree$) tend to decrease droplet size, with variations of up to $8\ \mu$m in droplet diameter, at $\qr=1$. In the present case, however, the simulated droplet sizes differ by $\sim3\ \mu$m at $\theta = 120\degree$ and $\theta = 150\degree$ for $\cac=0.06$ and  $\qr=1$; with that at $\theta = 150\degree$ remaining in close agreement with the experimental measurements. This minor variation may be attributed to the compensating effect of surface heterogeneity and localized contact line pinning inherent to PDMS substrates as evidenced in the literature \cite{VanderGraaf2006}.}
\subsection{Droplet formation stages}
The stages of droplet formation in a typical T-junction geometry, irrespective of the cross-section, remain the same as those reported in recent studies on rectangular microchannels \citep{Garstecki2006, venkateshwarlu2021effects, Venkateshwarlu_2022, Venkateshwarlu_2023} over a wide range of conditions ($\cac$, $\qr$). The droplet formation process is briefly summarized as follows.
 The two immiscible phases (e.g., oil and water) enter through their respective inlets, meet at the T-junction, and form an interface due to immiscibility. Both phases then flow simultaneously downstream in the main channel, initiating droplet growth. The upstream interface of the dispersed phase advances toward the downstream edge, forming a neck between the main channel and the dispersed stream. This neck elongates and eventually pinches-off, producing either a liquid plug or a droplet depending on $\qr$ and $\cac$. The droplet subsequently flows downstream in the main channel, while the dispersed phase retracts to the inlet tip, and the entire cycle repeats.

\begin{figure}[!b]
	\centering\includegraphics[width=1\linewidth]{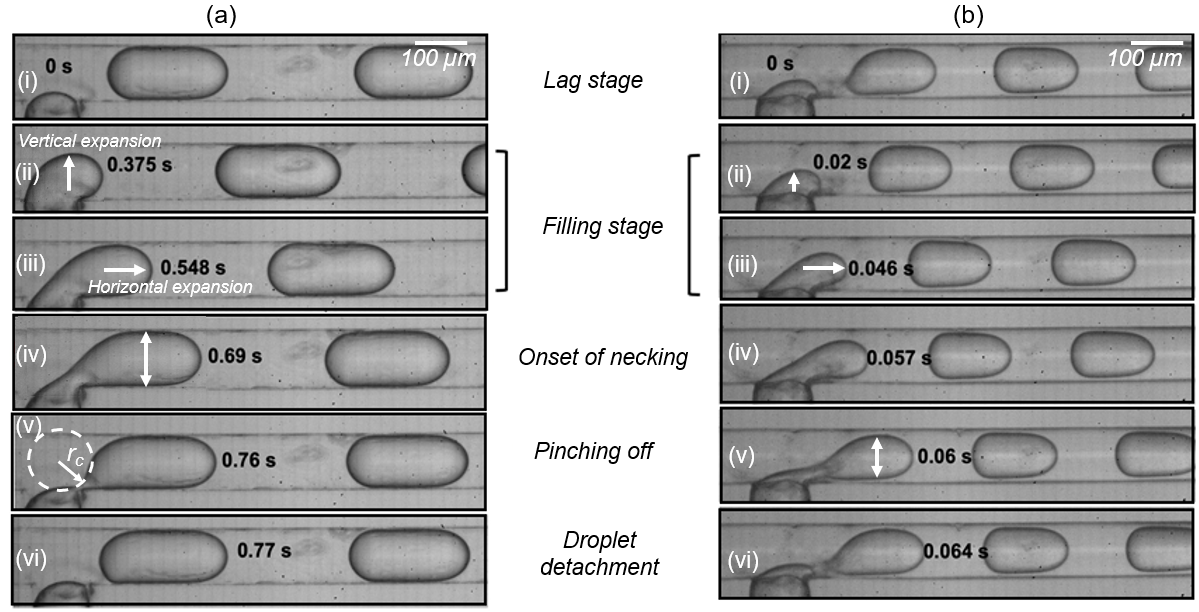}
	\caption{Droplet formation stages at $\qr = 1$ for (a) squeezing regime ($Q_\text{c} = Q_\text{d} = 0.03$ mL/h; $\cac = 0.003$) and (b) dripping regime ($Q_\text{c} = Q_\text{d} = 0.3$ mL/h; $\cac = 0.029$).}
	\label{fig:Expregime}
\end{figure}

\noindent
The stages of the droplet formation cycle \citep{venkateshwarlu2021effects, Venkateshwarlu_2022, Venkateshwarlu_2023} include (i) lag stage, (ii) filling stage, (iii) necking stage, (iv) pinch-off stage, and (v) detachment stage, as illustrated in \fig\ref{fig:Expregime} at low ($Q_\text{c} = Q_\text{d} = 0.03$ mL/h; $\cac = 0.003$) and high ($Q_\text{c} = Q_\text{d} = 0.3$ mL/h; $\cac = 0.029$) flow rates and capillary numbers for fixed flow-rate ratio ($\qr = 1$).
The initial time is set to zero ($t=0$ s) at the instant when a preceding droplet pinches-off from the dispersed phase (DP). The time interval between this pinch-off and the subsequent evolution of the dispersed phase in the main channel during consecutive droplet formation is referred to as the \textit{lag stage}. During this period, both the continuous phase (CP) and dispersed phase (DP) continue flowing through their respective inlets. Once the DP overcomes pressure barrier and enters the main channel, it expands vertically and horizontally with a convex shape as depicted in Fig. \ref{fig:Expregime}a(ii-iii), this marks the \textit{filling stage}. The vertical expansion is more pronounced at low capillary numbers ($\cac$), i.e.,  at  $\cac = 0.003$, compared to $\cac= 0.029$  as shown in Figs. \ref{fig:Expregime}a(iv) and \ref{fig:Expregime}b(v) respectively. Subsequently, the DP neck (i.e., the cross-sectional width of the DP near the junction) gradually shrinks from its maximum to minimum, marking the \textit{necking stage}, as visualized in \figs\ref{fig:Expregime}(a,b)(v). The interface then contracts further with time, adopting a concave shape and reaching a critical (minimum) neck width ($r_c$), marked in \figs\ref{fig:Expregime}a(v), before instant detachment (i.e., \textit{droplet pinch-off}),  as visualized in \figs\ref{fig:Expregime}(a,b)(vi).
\begin{figure}[!b]
	\centering\includegraphics[width=0.95\linewidth]{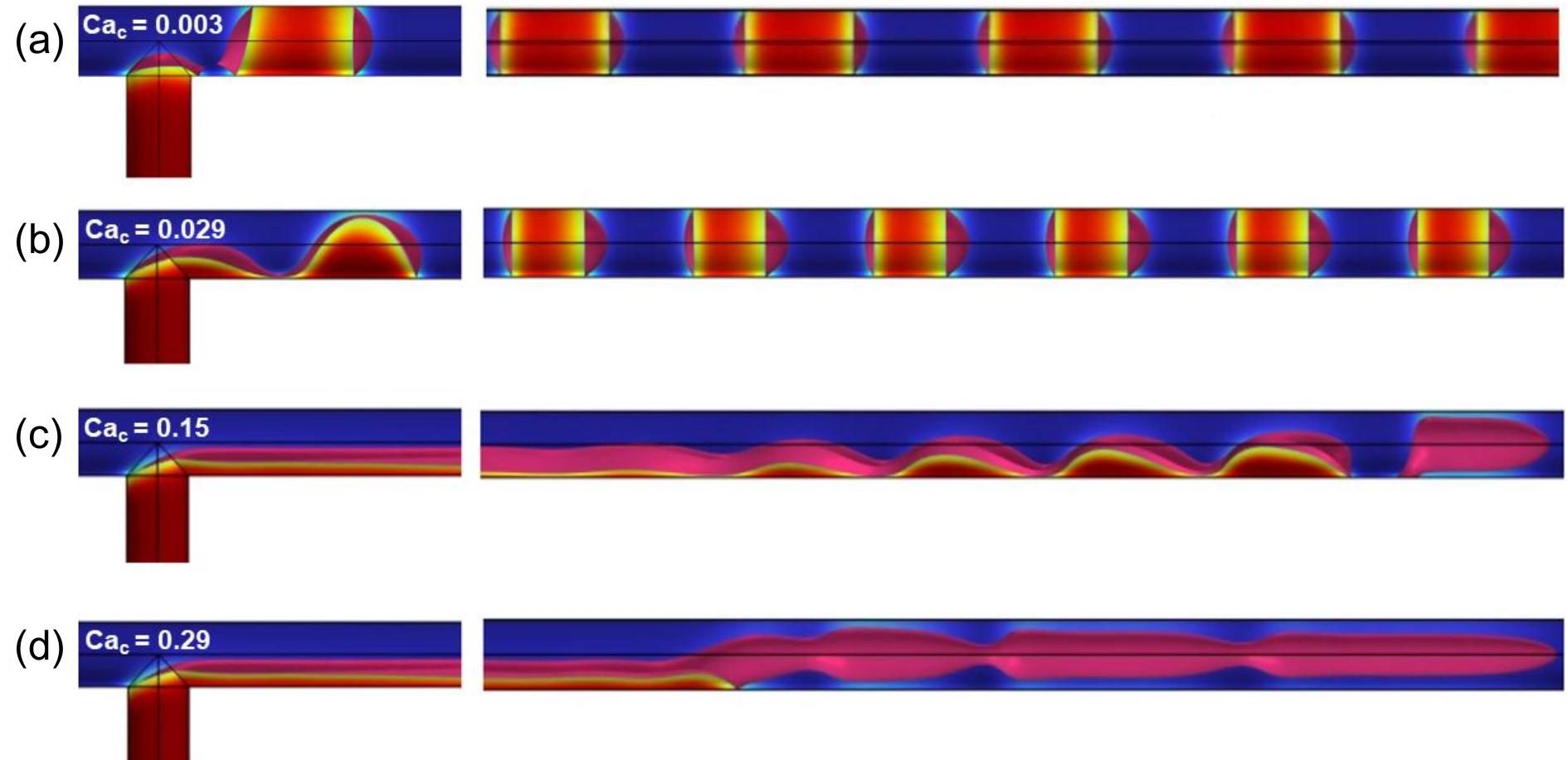}
	\caption{ {Representative phase flow patterns as a function of capillary number ($\cac$) for fixed flow-rate ratio ($\qr = 1$) under (a) squeezing regime (b) dripping regime (c) parallel flow with tip streaming, and (d) sausage flow}.}
	\label{fig:simulationregime}
\end{figure}
\begin{figure}[!b]
	\centering\includegraphics[width=1\linewidth]{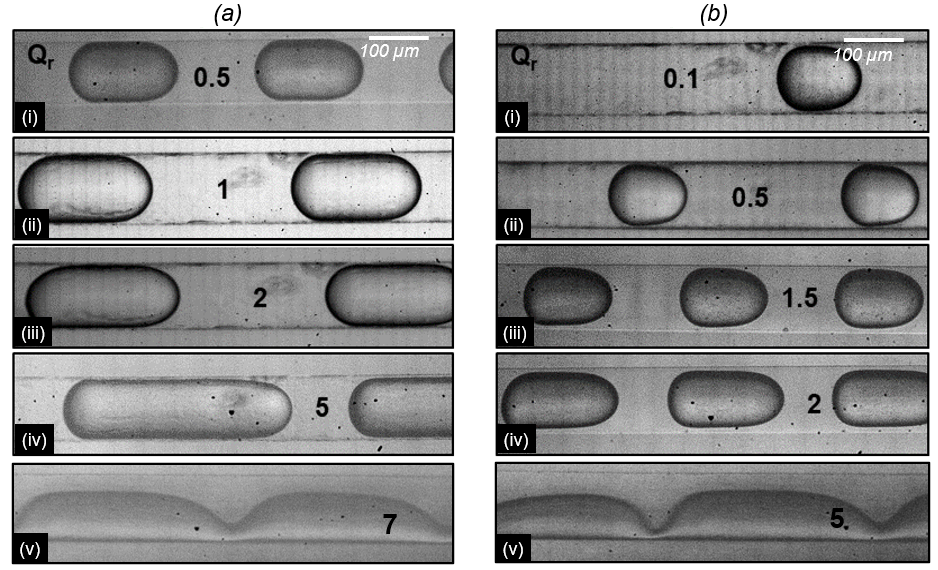}
	\caption{Effect of flow-rate ratio {($\qr$) and capillary number ($\cac$)} on phase flow patterns under (a) squeezing regime at $\cac = 0.001$, and (b) dripping regime at $\cac = 0.029$.}
		\label{fig:flowrate ratio}
\end{figure}
\subsection{Phase profiles and flow regimes}
The distribution of the phase-field composition ($\phi$) in two-phase flow through a T-type rectangular microfluidic device is known 
\citep{Garstecki2006, venkateshwarlu2021effects, Venkateshwarlu_2022, Venkateshwarlu_2023} 
to complexly depend on the flow-rate ratio ($\qr$) and the capillary number ($\cac$). In this section, the corresponding patterns for a T-type cylindrical device are examined to construct the flow regime map. 
The representative phase-flow profiles and corresponding flow regimes obtained computationally as a function of $\cac$ at a fixed $\qr=1$ are shown in  \fig\ref{fig:simulationregime}, whereas the  experimentally observed regimes as a function of $\qr$ are presented in \fig\ref{fig:flowrate ratio}. 
Subsequently, four distinct regimes are identified as (a) squeezing, (b) dripping, (c) parallel flow with tip streaming, and (d) sausage flow. Among these, droplet formation is observed under the first three regimes. 
Notably, the occurrence of \textit{parallel flow with tip streaming}, as shown in \fig\ref{fig:simulationregime}c and \fig\ref{fig:flowrate ratio}v, represents a novel observation of the present study and, to the best of our knowledge, has not been reported previously in cross-flow devices. Owing to the limited field of view in our experiments, the fully developed tip-streaming stage (seen in \fig\ref{fig:simulationregime}c)  could not be captured in \fig\ref{fig:flowrate ratio}v; however, the onset of this regime is clearly visible in the early stage (\fig\ref{fig:flowrate ratio}iv). Due to the experimentally restricted field of view, the present experimental observations could capture the flow only under the squeezing and dripping regimes, as depicted in \fig\ref{fig:flowrate ratio}, over a wide range of the flow-rate ratios ($0.1 \le \qr \le 6$).
\begin{figure}[!b]%
	\centering
	{{\includegraphics[width=0.75\linewidth]{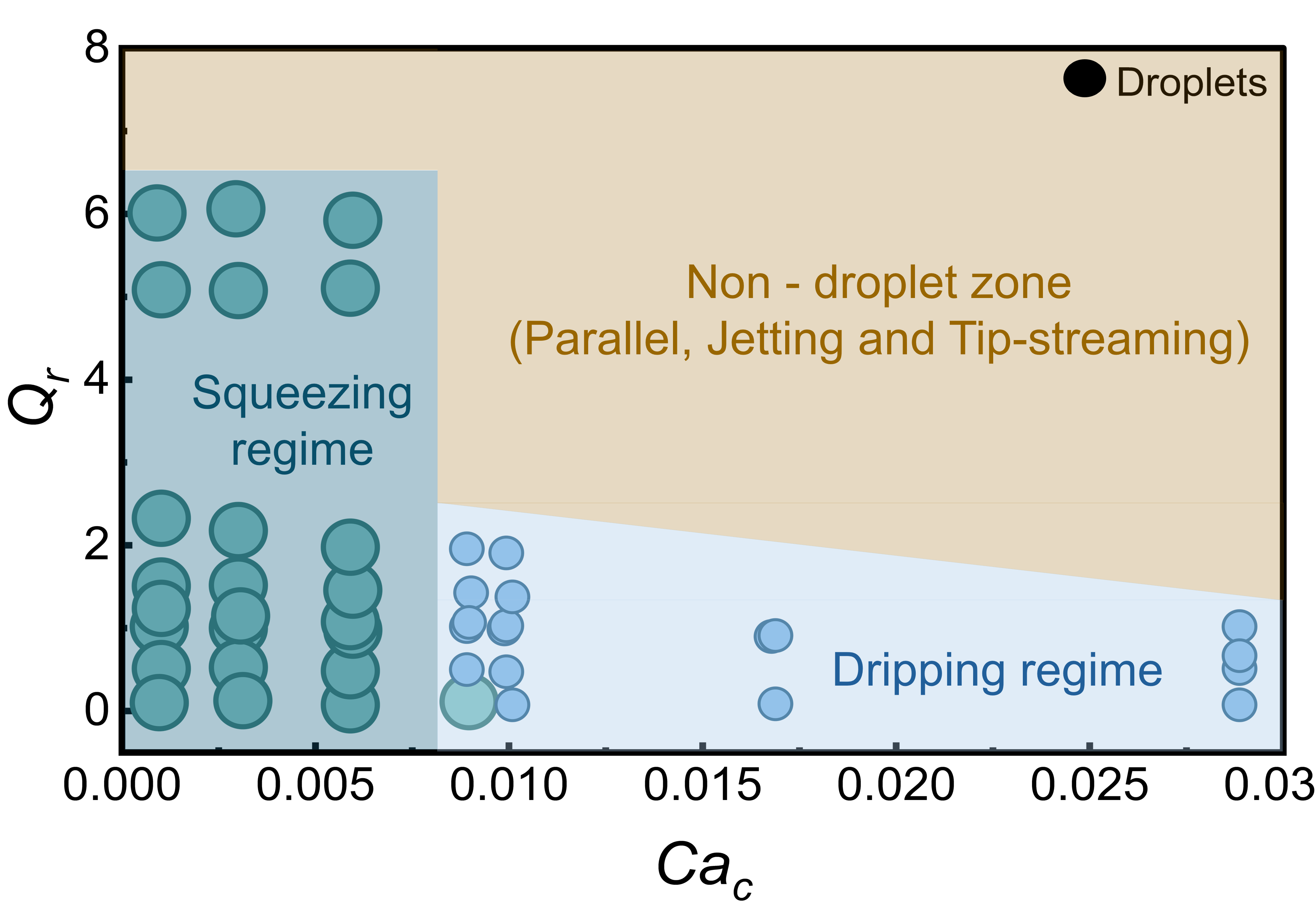}}}%
	\caption{Flow regime map  as a function of {$\cac$ and $\qr$.  } }%
	\label{L/Qr}%
\end{figure}
\noindent
Furthermore, 
as DP evolves into the main channel, it progressively blocks the path for CP flow. This obstruction induces a pressure build-up upstream in CP, thereby exerting an additional squeezing force on DP thread. Small pressure perturbations can also arise from fluctuations in the local flow fields and interface deformation. However, these perturbations are effectively damped by the confining channel walls, so the downstream (frontal) interface remains largely unaffected, while the upstream interface continues to evolve due to the interplay of capillary, viscous, and pressure forces. 
At low $\cac$, the neck of  DP forms exactly at the downstream side of  channel junction, and the droplet detaches  once the pressure from  CP overcomes the interfacial forces from DP inflow. 
Subsequently, the plug shaped droplets with length exceeding the channel cross-section ($L^{\ast} = L/D_\text{c} > 1$) are formed. These plugs exhibit symmetric front (leading) and rear (trailing) interfaces and are surrounded by a very thin continuous-phase film between the droplet and the channel walls, characterizing the squeezing regime, as shown in \figs\ref{fig:simulationregime}(a) and \ref{fig:flowrate ratio}(a). 
At higher $\cac$,  viscous shear starts dominating over interfacial tension. This accelerates neck-thinning and promotes earlier breakup. For example, at the flow rate ratio ($\qr = 1$), droplet pinch-off occurs within 64 ms (at $\cac = 0.029$), compared to 770 ms (at $Ca_c = 0.003$), as shown in \fig\ref{fig:Expregime}(a,b)(vi). These droplets are smaller in size and and exhibit a bullet shaped configuration, with distinct front and rear curvatures; the rear interface tapers under shear stresses, and a relatively thicker continuous-phase film forms around them, as observed in \figs\ref{fig:simulationregime}(b) and \ref{fig:flowrate ratio}(b). Such characteristics are indicative of the dripping regime. 
\newline
The analysis of the phase profiles reveals a complex dependence of the flow regimes on the explored ranges of capillary number ($10^{-3}\le\cac\le 0.1$) and flow-rate ratio ($0.1\le\qr\le 10$), as summarized qualitatively in \fig\ref{L/Qr} and quantitatively in \tab\ref{tab:regimes}.
%
%
\begin{table}[!tb]
	\centering
	\caption{Summary of flow regimes as functions of capillary number ($10^{-3}\le\cac\le 0.1$) and flow-rate ratio ($0.1\le\qr\le 10$), with corresponding dominant force balance.}
		\label{tab:regimes}
	\renewcommand{\arraystretch}{1.3}
	\resizebox{0.8\textwidth}{!}{
	\begin{tabular}{|c|c|p{0.38\linewidth}|p{0.42\linewidth}|}
		\hline
		$\cac$ & $\qr$ & Observed Regime & Dominant Force Balance \\ 		\hline
		$0.001-0.006$ & $0.1-6$  & Squeezing 	& Interfacial tension $\gg$ viscous forces \newline (pressure-driven confinement) \\  		 \cline{2-4}
		 & $>6$  & Non-droplet \newline (jetting, parallel,  or sausage-like flow)  & Increasing dispersed-phase inertia \\ 		\hline
		$0.006 - 0.009$ & $0.1$  & Squeezing  & Interfacial tension dominant \\  		 \cline{2-4}
		 & $0.1-2$ 	& Dripping 	& Viscous shear $\sim$ interfacial tension \\ 		 \cline{2-4}
		 & $>2$  & Non-droplet  \newline (jetting, parallel,  or sausage-like flow) & Inertial effects significant \\ 		\hline
		$0.01-0.029$ & $0.1-2$ 	& Dripping  & Shear-dominated with reduced confinement \\ 		 \cline{2-4}
		 & $>2$ 	& Parallel with tip streaming \newline (or sausage-like flow) 
		& Shear $\gg$ interfacial tension \newline (interface destabilization) \\ 		\hline
		 $> 0.029$ &  all $\qr$ & Non-droplet \newline (jetting, parallel,  or sausage-like flow) & Inertial effects significant \\ \hline
	\end{tabular}}
\end{table}
To the best of our knowledge, studies on cylindrical T-junctions remain scarce in the literature. Therefore, to assess the reliability and accuracy of the demarcated flow regimes (\tab\ref{tab:regimes}), the transitional capillary numbers marking the squeezing-to-dripping regime transition  are compared in \tab\ref{tab:transition2} with reported values for rectangular T-junctions.
Previous studies report a wide range of critical capillary numbers ($\cac$) marking the squeezing-to-dripping transition, reflecting differences in geometry, interfacial tension, fluid properties, and flow conditions. As summarized in Table \ref{tab:transition2}, the reported 
$\cac$ values span approximately 0.002 to 0.081; most studies report transitions in the range 0.005 to 0.02, and higher values ($0.05-0.08$) are less common and associated with specific geometric or viscosity configurations. This wide variation is primarily attributed to channel geometry, since most prior investigations were conducted in rectangular or square T-junctions, where corner-induced recirculation and stronger confinement amplify interfacial shear stresses, thereby promoting transition at lower capillary numbers. The transitional value ($\cac=0.009$) reported in the present study, compared with literature,  can be attributed to the reduced hydrodynamic confinement and more uniform, axisymmetric flow redistribution inherent to the circular junction geometry. 
Further, the key differences between droplets in the dripping regime (large $\cac$, typically for $\qr > 1$) and the squeezing regime (small $\cac$, across all $\qr$) lie in their length, radii, formation frequency, and the film thickness of CP between the droplet and the channel wall. These droplet characteristics and their dependence on flow parameters are discussed in the following sections. 
\begin{table}[!bt]
	\centering
	\caption{Comparison of transitional Capillary numbers ($\cac$) marking the squeezing-to-dripping transition for different T-type geometries.}
	\label{tab:transition2}
	\renewcommand{\arraystretch}{1.3}
	\resizebox{\textwidth}{!}{
		\begin{tabular}{|c|c|l|l|c|c|c|c|} 
			\hline
			{\bfseries Ref.} & {\bfseries Type} & \multicolumn{3}{c|}{\bfseries Geometry Parameters}
			& {\bfseries $\mu_r$} & {\bfseries $\cac$} & {\bfseries Transitional $\cac$} \\ \hline
			\textbf{Present} & N, E & 3-D  &Cylindrical & $d = 150~\mu$m & 0.003 & $0.001- 0.1$ & 0.009 \\ \hline 
			\cite{Abate2012} & E & 3-D  & Square & $w/h = 1$, $w = 25~\mu$m& 1 & $0.01 - 0.23$ & 0.07 \\ \hline
			\cite{Arias2020} & N, E & 3-D  & Rectangular & $w/h \gg 1$, $w = 100~\mu$m & 1 & $(1 \times 10^{-4}) - 0.1$ & 0.005 \\ \hline
			\cite{Oishi2018} & E & 3-D  & Rectangular & $w/h = 1.2$, $w = 100~\mu$m & 0.15 & $(4.89 \times 10^{-4}) - (2.04 \times 10^{-2})$ & 0.008 \\ \hline
			\cite{Demenech2008} & N & 3-D  &Square & $w/h = 1$, $w = 100~\mu$m&$0.125 - 0.1$ & $0.01 - 0.15$ & 0.015 \\ \hline
			\cite{Xu2008} & E & 3-D  & Rectangular & $w/h = 1.4$, $w = 200~\mu$m &$0.68 - 10.84$& $0.001 - 0.3$ & $0.002 - 0.01$ \\ \hline
			\cite{van_2006} & N, E & 3-D  & Square & $w/h = 1.4$, $w = 100~\mu$m &3.44& $0.0054 - 0.1$ & 0.081 \\ \hline
			\cite{venkateshwarlu2021effects} & N & 2-D  & Planar & $w/h = 1$, $w = 100~\mu$m & $0.007143 - 0.7143$ & $10^{-4} - 1$ & 0.01 \\ \hline
			\cite{Liu2009} & N & 2-D  & Planar &  $w/h \gg 1$, $w = 150~\mu$m &$0.0125 - 0.25$ & $(5.6 \times 10^{-4}) - (5.9 \times 10^{-2})$ & 0.01 \\ \hline
			\cite{guo_2009} & N & 2-D  & Planar & $w/h \gg 1$, $w = 200~\mu$m & $0.0001- 0.001$ & $(6.4 \times 10^{-4}) - (1.7 \times 10^{-2})$ & 0.0058 \\ \hline
			\multicolumn{8}{|c|}{\textbf{N} (Numerical); \textbf{E} (Experimental);  $\mu_r$ (Viscosity ratio); $\cac$ (Capillary number); $w$ (cross-sectional width); $h$ (cross-sectional height)} \\ \hline
		\end{tabular}
	}
\end{table}
\begin{figure}[!b]%
	\centering{{\includegraphics[width=1\linewidth]{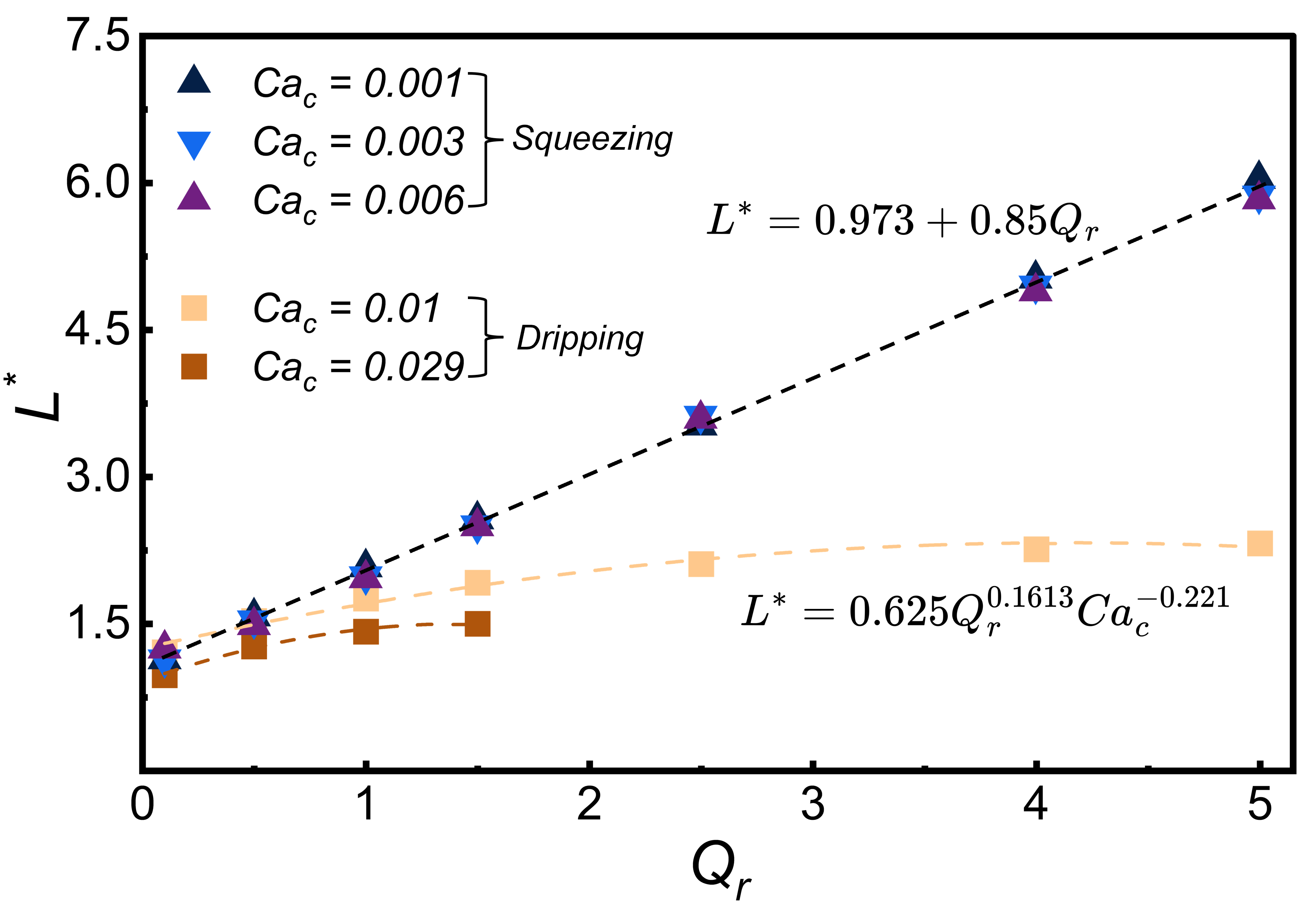}}}%
	\caption{Droplet length ($L^{\ast} =L/D_{\text{c}}$) as a function of governing parameters ($\cac$ and $\qr$).}%
	\label{L/Ca}%
\end{figure}
\subsection{Droplet length}
The phase flow profiles shown in \figs\ref{fig:Expregime}–\ref{fig:flowrate ratio} qualitatively demonstrate a reduction in droplet size with increasing capillary number ($\cac$). \fig\ref{L/Ca} shows the quantitative variation of droplet length ($L^{\ast} = L/D_{\text{c}}$) as a function of the capillary number ($\cac$) and flow rate ratio ($\qr$).
%
%
The results reveal that $L^{\ast}$ decreases with increasing $\cac$, as the higher viscous shear force from CP, relative to to interfacial tension, promotes earlier pinch-off of the dispersed phase. Conversely, $L^{\ast}$ increases with $\qr$, since a greater influx of DP, relative to CP, allows more volume to accumulate before breakup. These trends are consistent with the physical interplay between interfacial tension, viscous shear, and pressure-driven expansion governing droplet formation in microfluidic T-junctions.
\newline
The statistical analysis of the present experimental and numerical data indicates that the droplet length ($L$) exhibits a linear dependence on the flow rate ratio ($\qr$) and the characteristic dimension of the device ($D_{\text{c}}$)  in the squeezing regime ($\cac < 10^{-2}$, for all values of $\qr$). This trend arises because, under low $\cac$, the interfacial tension dominates over viscous shear, and droplet growth is primarily governed by the relative volumetric contributions of the dispersed and continuous phases. Consequently, the empirical correlation (\eqn\ref{eq:l_squeezing}) for the droplet length ($L^{\ast}$) in the squeezing regime is established as follows.
\begin{gather}
	{L^{\ast} }= \alpha_1 + \beta_1 \qr \qquad \text{(squeezing regime)}
	\label{eq:l_squeezing}
	\\ \text{where}\qquad \alpha_1=0.973, \quad\text{and}\quad \beta_1 = 0.85\nonumber
\end{gather}
where $\alpha_1$ and $\beta_1$ are statistical fitting parameters that account for the combined effects of device geometry and interfacial dynamics. 
Notably, the present correlation coefficients ($\alpha_1$, $\beta_1$) are in close agreement with the widely reported values ($\alpha_1 \approx 1$ and $\beta_1 \approx 1$) for droplet-length correlations in rectangular T-type microchannels \citep{Garstecki2006, Christopher2008}, thereby reinforcing and lending further confidence to the accuracy and reliability of the present findings.
\newline
Furthermore, in the dripping regime ($10^{-2} < \cac < 10^{-1}$; $0.1 \leq \qr \leq 2$), the present experimental and numerical data indicate that the droplet length ($L$) exhibits a nonlinear power-law dependence on the flow parameters ($\cac$,  $\qr$), and whereas linear dependence on the characteristic length of the device ($D_{\text{c}}$). 
This observation qualitatively agrees with the comprehensive review which demonstrates that $L^{\ast}$ decreases with increasing $\cac$ and increases with $\qr$ within the dripping regime of rectangular T-type microchannels\citep{Christopher_2007}. Consequently, the empirical correlation (\eqn\ref{eq:l_dripping}) for the droplet length ($L^{\ast}$) in the dripping regime is established as follows.
\begin{gather}
	{L^{\ast}}=\alpha \qr^{m}  \cac^{n}\qquad \text{(dripping regime)}
	\label{eq:l_dripping}
	\\ \text{where}\qquad \alpha=0.625, \quad m = 0.1613, \quad\text{and}\quad m = -0.221\nonumber
\end{gather}
where, $\alpha$, $m$, and $n$  are statistically fitted parameters obtained from regression analysis of the present data in the dripping regime. In \eqn\eqref{eq:l_dripping}, $\alpha$ accounts for geometric confinement and interfacial effects, the positive exponent ($m > 0$) confirms that higher dispersed-phase inflow (larger $\qr$) increases droplet size, and the negative exponent ($n < 0$) reflects the stronger shear force at larger $\cac$, which promotes earlier detachment and hence smaller droplets.

\begin{table}[!bt]
	\caption{Comparison of empirical correlation coefficients for the droplet length ($L^{\ast}$) for different T-type geometries and regimes, squeezing regime: $L^{\ast}= \alpha_1 + \beta_1 \qr$ (\eqn\ref{eq:l_squeezing}), and dripping regime: $L^{\ast}=\alpha \qr^{m}  \cac^{n}$ (\eqn\ref{eq:l_dripping}).}
	\label{tab:Lcoeff}
	
	\begin{center}
		\vspace{-1em}
		\renewcommand{\arraystretch}{1.5}
		\resizebox{\textwidth}{!}{
			\begin{tabular}{|l|l|l|p{0.2\linewidth}|c|c|c|c|c|}
				\hline
				\multirow{2}{*}{Ref.} 
				& \multirow{2}{*}{Type} 
				& \multirow{2}{*}{Geometrical nature}
				& \multirow{2}{*}{Conditions explored}  
				& \multicolumn{2}{c|}{Squeezing regime} 
				& \multicolumn{3}{c|}{Dripping regime} \\  
				\cline{5-9}
				& & & & $\alpha_1$ & $\beta_1$ & $\alpha$ & $m$ & $-n$ \\ 
				\hline
				
				\textbf{Present} 
				& E + N
				& 3D cylindrical 
				& $D_{c}=D_{d} = 100\ \mu$m, $\qr = 0.1-10$, $\mu_{\text{r}}=0.00295$, $\gamma= 36.5$ mN m$^{-1}$ , $\cac = 0.001-0.1$
				& 0.973 & 0.8500 & 0.6250 & 0.1613 & 0.2210 \\ 
				\hline
				
				\citep{Garstecki2006} 
				& E
				& 3D rectangular 
				& $h = 33\ \mu$m, $w = 50-200\ \mu$m, $\qr= 0.01-10$ , $\mu_{\text{r}}=0.01-0.1$, $\gamma = 36.5$ mN m$^{-1}$  
				& $\approx 1$ & $\approx 1$ & - & - & - \\ 
				\hline		
				
				\citep{VanderGraaf2006} 
				& E + N
				& 2D rectangular 
				& $\qr= 0.05-1$, $\cac= 0.003-0.065$, $\theta=135^{\circ}-180^{\circ}$, $\gamma= 1-15$ mN m$^{-1}$ 
				& 1 & 1 & 1 & 1 & 0.25 \\ 
				\hline		
				
				\citep{VanSteijn2007} 
				& E
				& 3D rectangular  
				& $w_d = 800\ \mu$m, $\mu_{\text{r}}= 0.01-0.1$, $\cac < 0.01$  
				& 1.5 & 1.5 & - & $\sim 0.5-0.7$ & $\sim 0.3-0.5$ \\ 
				\hline	
				
				\citep{VanSteijn2010} 
				& E
				& 3D rectangular 
				& $\cac < 0.01$, $\qr= 0.05-5$, $w_d= 65-375\ \mu$m, $h_d = 65-375\ \mu$m, $\mu = 1.2-7.2$ mPa s, $\gamma= 17.9-22.7$ mN m$^{-1}$
				& $\approx 1$ & $\approx 1$ & 1 & - & 0.316 \\ 
				\hline		
				
				\citep{venkateshwarlu2021effects} 
				& N
				& 2D rectangular 
				& $\cac = 10^{-4} - 1$, $\qr = 0.1-10$, $\mu_{\text{r}}=0.007143-0.7143$, $\theta=120^{\circ}-180^{\circ}$, $Re_{\text{c}}=0.1$
				& 1 & 1.7648 & 0.5358 & 0.2307 & 0.3682 \\ 
				\hline	
				
				\citep{Nekouei2017}  
				& N
				& 3D rectangular 
				& $\mu_{\text{r}}=0.01$, \newline $0.001< Ca_c < 0.02$
				& 1.000 & 1.13 $\pm$ 0.16 & - & - & - \\ 
				\cline{4-9}		
				
				& & 
				& $\mu_{\text{r}}=0.1$, \newline $0.001< Ca_c < 0.02$
				& 1.000 & 1.67 $\pm$ 0.41 & - & - & - \\ 
				\hline		
				\multicolumn{9}{|c|}{\textbf{N} (Numerical); \textbf{E} (Experimental);  $\gamma$ (Interfacial tension) ; $Q_r$ ( Flow rate ratio) ; d (dispersed phase), c (continuous phase)} \\ \hline
				
		\end{tabular}}
	\end{center}
\end{table}

%

\noindent
The present empirical correlations (\eqns\ref{eq:l_squeezing} and \ref{eq:l_dripping}) are qualitatively consistent, as compared in \tab \ref{tab:Lcoeff}, with several experimental and computational studies on droplet generation. 
In the squeezing regime, the present values ($\alpha_1 = 0.973$, $\beta_1 = 0.85$) are close to those reported \citep{Garstecki2006, VanSteijn2010} for rectangular channel ($w/h >> 1$). The slightly smaller value of $\beta_1$ in the present study indicates a weaker dependence of $L^{\ast}$ on $Q_r$, which can be attributed to the reduced hydrodynamic confinement and more uniform flow redistribution in the 3D cylindrical T-junction compared to rectangular geometries. In the dripping regime, the obtained coefficients ($\alpha = 0.625$, $m = 0.1613$, $n = 0.221$) align well with the ranges reported \citep{VanderGraaf2006, venkateshwarlu2021effects} for rectangular channel (2D planar configurations), whereas they deviate more noticeably from correlations based on 3D rectangular geometries. This agreement is attributed to the inherent characteristics of 3D rectangular geometries, where corner-induced recirculation and liquid entrapment within the gutters enhance confinement and modify the interfacial shear field, thereby strengthening sensitivity to $\qr$ compared to 2D planar and cylindrical geometries. Overall, the present correlations reaffirm classical scaling trends while highlighting the pronounced influence of device geometry on droplet size.
\subsection{Droplet radii}
\begin{figure}[!bt]
	\centering
	\hspace*{0.7cm}
		\subfigure[Illustration of droplet radii measurement for droplets under (i) Squeezing regime (ii) Dripping regime]
		{\includegraphics[width=0.94\linewidth]{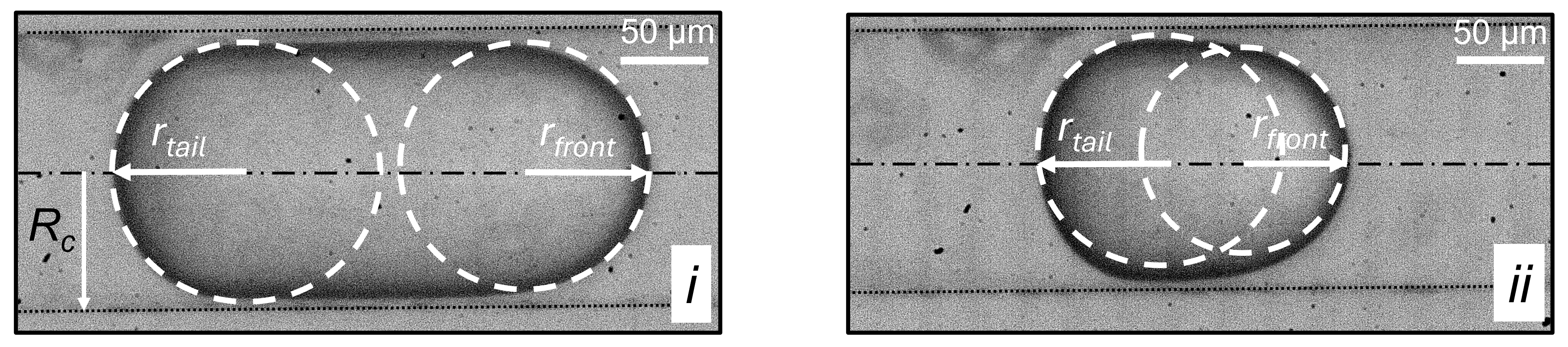}\label{radii-1}}\\
	\subfigure[$r^{\ast}= f (\cac,\qr)$] {\includegraphics[width=0.48\linewidth]{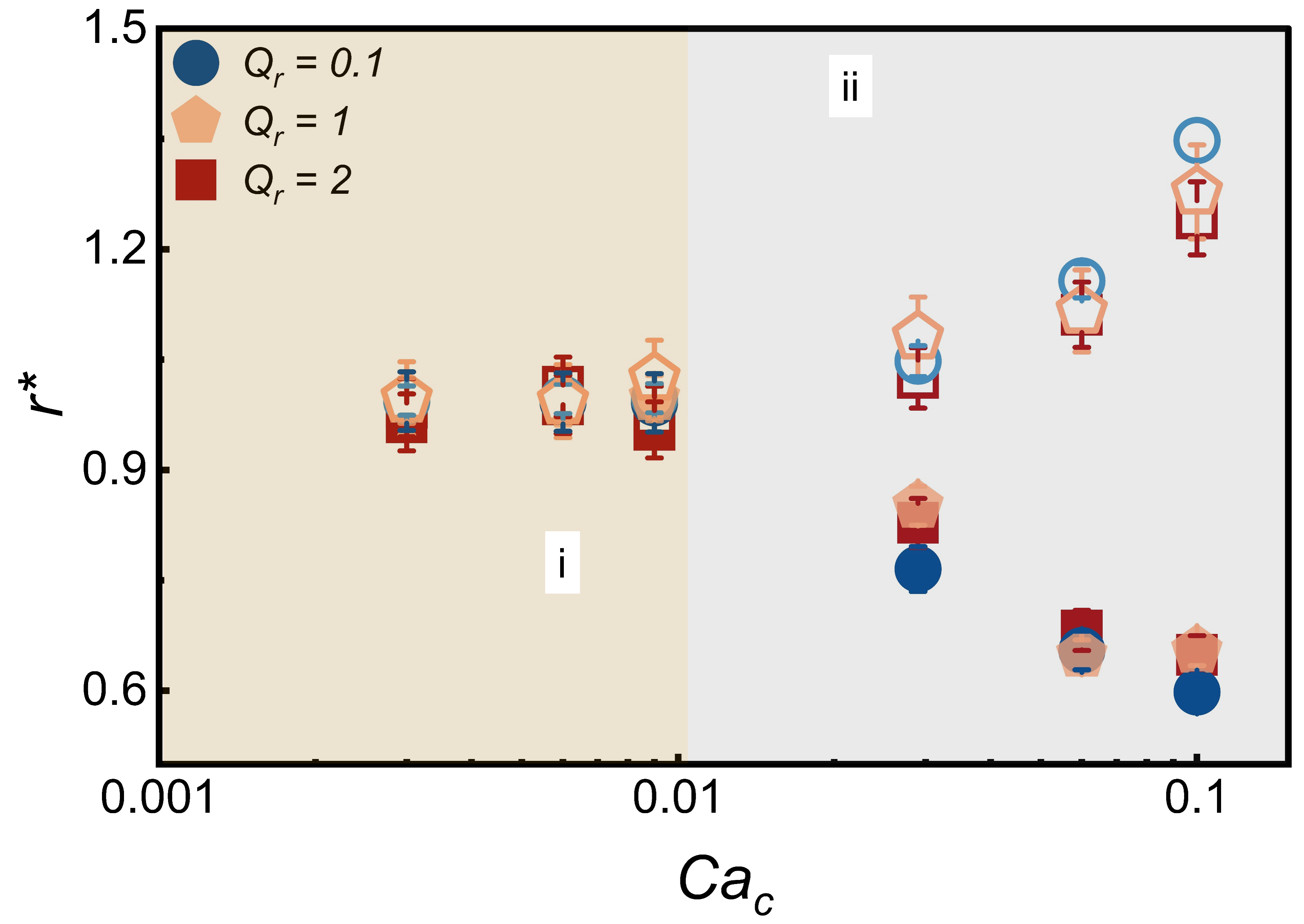}\label{radii-2}}
	\subfigure[$\Delta r^{\ast}= f (\cac, \qr$)] {\includegraphics[width=0.49\linewidth]{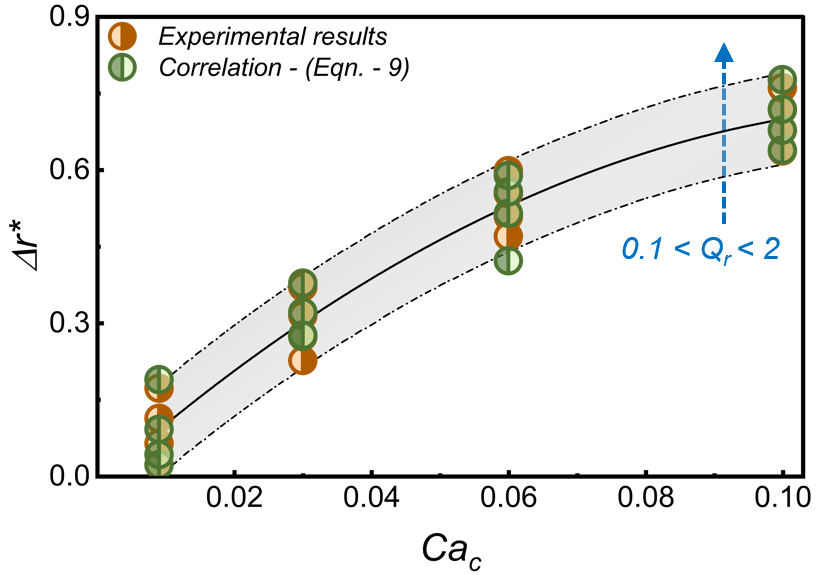}\label{dradii}}
	\caption{(a) Illustration of droplet radii measurement, (b) Droplet  radii ($r^{\ast}=r/R_\text{c}$)  as a function of flow parameters ($\cac$, $\qr$); Filled and unfilled symbols represent the front ($ r^{\ast}_\text{front}$) and tail ($ r^{\ast}_\text{tail}$) radii respectively. (c) Droplet asymmetry ($\Delta r^{\ast} = r^{\ast}_\text{tail} -  r^{\ast}_\text{front}$) as a function of flow parameters ($\cac$, $\qr$).}
	\label{radii}
\end{figure}

An interplay among viscous, inertial, and interfacial tension forces influences not only the droplet length ($L$) but also the droplet topology, particularly the characteristic radii ($r$). \fig\ref{radii-1} schematically illustrates the front and rear radii  of the droplet, denoted as $r_\text{front}$  and $r_\text{tail}$, under the squeezing and dripping regimes. Qualitatively, these topological features are strongly governed by both the capillary number ($\cac$) and the flow rate ratio ($\qr$).
%
\noindent
As discussed in the preceding sections, droplets in the squeezing regime ($\cac \le 10^{-2}$) exhibit a plug-like shape, characterized by a length-to-diameter ratio greater than unity ($L^{\ast} > 1$). \fig\ref{radii-2} illustrates the relationship between the droplet radii ($r^{\ast} = r/R_\text{c}$, where $R_\text{c}=D_\text{c}/2$) at the front ($r^{\ast}_\text{front}$) and rear ($r^{\ast}_\text{tail}$) sides of the droplet, the capillary number ($\cac$), and the flow rate ratio ($\qr$). In addition, \fig\ref{dradii} presents the variation in  asymmetry between the front and rear sides curvature of the droplet, defined as $\Delta r^{\ast} = (r^{\ast}_\text{tail} - r^{\ast}_\text{front})$, as a function of $\cac$ and $\qr$, providing further insights into the topological evolution of the droplet.

\noindent
The analysis of \figs\ref{radii-2} and \ref{dradii} reveals that droplets exhibit nearly symmetric radii ($\Delta r^{\ast} \approx 0$) at low $\cac$ because interfacial tension dominates over viscous stresses, resisting deformation. As $\cac$ increases, viscous shear stresses stretch the dispersed phase, leading to $r^{\ast}_\text{tail} > r^{\ast}_\text{front}$, thereby increasing asymmetry ($\Delta r^{\ast} > 0$). This asymmetry is further amplified with increasing $\qr$, since the relative volumetric flux of the DP over CP intensifies the pressure and velocity gradients across the interface. Thus, the droplet asymmetry ($\Delta r^{\ast}$) serves as a sensitive indicator of the transition from symmetric plug-like droplets in the squeezing regime to elongated asymmetric droplets in the dripping regime.

\noindent
Furthermore, in the squeezing regime ($\cac \leq 10^{-2}$), the droplet length varies linearly with $\qr$ but remains independent of $\cac$ (\eqn\ref{eq:l_squeezing}). The interfacial tension dominates over viscous forces, producing stable plug-shaped droplets with negligible sensitivity of radii to either $\qr$ or $\cac$.
While the droplet expands (or shrinks) with $\qr$, its overall shape remains symmetric ($r^{\ast}_\text{front} \approx r^{\ast}_\text{tail}$) about both the horizontal and vertical centerlines, as depicted in  \fig\ref{radii-1}(i), owing to the negligible influence of $\cac$ on the stability of the droplet after formation. Consequently, the front and rear radii are nearly identical ($r^{\ast}_\text{front} \approx r^{\ast}_\text{tail} \approx 1$) representing symmetrical droplet ($\Delta r^{\ast} \approx 0$) and show minimal dependence on either $\qr$ or $\cac$, as shown in \fig\ref{radii-2} (Regime - i). This trend is consistently observed in the squeezing regime ($\cac \leq 10^{-2}$) across all flow rate ratio ($\qr$) in \figs\ref{radii-2}.

\noindent
As the capillary number ($\cac$) increases, irrespective of $\qr$, the radius at the front end of the droplet ($r^{\ast}_\text{front}$) decreases, resulting in higher curvature, while the rear end flattens ($r^{\ast}_\text{tail}$ increases), leading to lower curvature, as illustrated in \fig\ref{radii-1}(ii). The transition from squeezing to dripping regimes fundamentally alters both droplet length (\eqns\ref{eq:l_squeezing} and \ref{eq:l_dripping}) and topology.
%
%
\noindent
For instance, in the dripping regime  (Regime - ii), the droplet morphology exhibits a nonlinear dependence on both $\qr$ and $\cac$. Unlike the symmetric plug-shaped droplets observed in squeezing, the dominance of viscous shear forces in dripping causes the dispersed phase to elongate and deform asymmetrically, as depicted in \fig\ref{radii-1}(ii). The droplets adopt an asymmetric shape, with a more rounded front and a tapered rear, such that $r^{\ast}_\text{tail} > r^{\ast}_\text{front}$ and a finite asymmetry ($\Delta r^{\ast} > 0$) increases with $\cac$ and $\qr$. This asymmetry grows with increasing $\cac$, as higher shear stresses enhance droplet elongation, and is further amplified at larger $\qr$, where the increased dispersed phase flux intensifies pressure and velocity gradients. Thus, droplet radii ($r^{\ast}$) in the dripping regime are sensitive to both $\cac$ and $\qr$, with a more pronounced dependence on $\cac$ as highlighted in \fig\ref{radii-2}b (Regime-ii).
Furthermore, \figs\ref{dradii}c clearly display non-linear increase in the droplet asymmetry, i.e., difference in the front and back radii {($\Delta r^{\ast}$)} with the capillary number ($\cac$) and flow-rate ratios ($\qr$) in the  dripping regime. The numerical and experimental values for the droplet radii ($\Delta r^{\ast}$)  in the  dripping regime are empirically correlated (\eqn\ref{eq:deltar2}) as follows.
\begin{gather}
   \Delta r^{\ast} = a_0 + a_1 \cac + a_2 \qr + a_3 \cac^2 + a_4 (\cac \qr) + a_5  \qr^2 \qquad  \text{(Dripping regime)}
   \label{eq:deltar2}
\end{gather}
The regression coefficients corresponding to the empirical model, as listed in \tab\ref{tab:drcoeff}, yield prediction errors ranging from a minimum of 2.1\% to a maximum of 7.4\% across the entire range of flow-rate ratio ($0.1\le\qr\le 2$) in the dripping regime ($\cac > 0.01$).
\begin{table}[h]
	\centering\renewcommand{\arraystretch}{1.5}
	\caption{Statistical parameters for empirical model (\eqn\ref{eq:deltar2}) for $\cac > 10^{-2}$ and $0.1\le\qr\le 2$.}\label{tab:drcoeff}
	\begin{tabular}{|l|c|c|c|c|c|c|c|c|}
		\hline
		Coefficient &$a_0$	&$a_1$	&$a_2$	& $\cac \to$ &	 0.009& 0.029	& 0.06	&	0.1 	\\\hline
		Value & $-$0.0935 &	12.08&0.129	&Max. Error (\%) &7.4	&6.1	&6.8	&3.4	 \\\hline
		Coefficient &$a_3$	&$a_4$	&$a_5$	&Min. Error (\%)&3.1	&5.2	&4.6	&2.1	  \\\hline
		Value &$-$48.80	&$-$0.142	&$-$0.023	&	&	& &	 &\\\hline 
	\end{tabular}
\end{table}

\noindent
This transitional behaviour of the droplet morphology, reported above in this section, is clearly reflected in the  phase profiles  (\figs\ref{fig:Expregime} - \ref{fig:flowrate ratio}) and flow regime maps (\fig\ref{L/Qr}) , where the squeezing regime spans a broad range of $\qr$ at low $\cac$, producing long, symmetric plug-shaped droplets, while the dripping regime emerges at higher $\cac$, confined to narrower ranges of $\qr$, and is associated with smaller, asymmetric droplets. Furthermore, the present experimental and numerical predictions align closely with prior regime maps reported for rectangular T-junction devices \citep{Garstecki2006,Christopher2008}, thereby validating the applicability of the empirical correlations and confirming the robustness of the transition trends across different geometrical cross-sections.
\subsection{Droplet generation frequency}
\begin{figure}[!h]
	\centering
	\subfigure[$f^\prime$ =  $f(Q_r)$ in squeezing regime]{
		\includegraphics[width=0.48\linewidth]{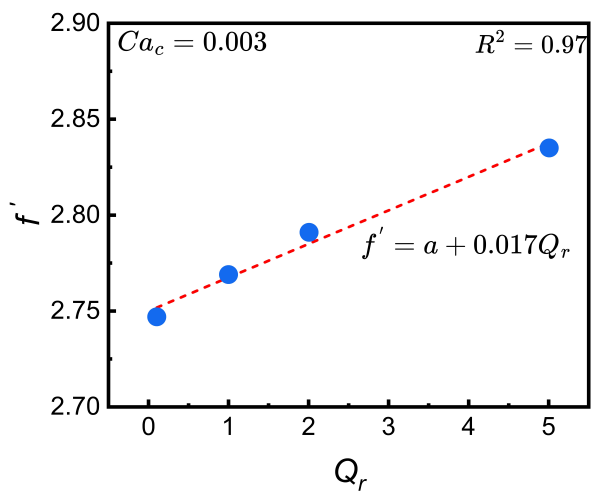} \label{frequency-1}
	}
	\subfigure[$f^\prime$ =  $f(Q_r)$ in squeezing regime]{
		\includegraphics[width=0.48\linewidth]{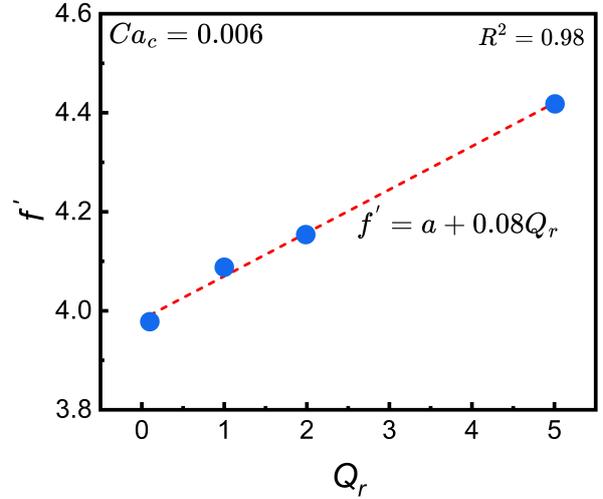} \label{frequency-2}
	}\\ 
	\subfigure[$f^\prime$ = $ f(Q_r)$ in dripping regime]{
		\includegraphics[width=0.48\linewidth]{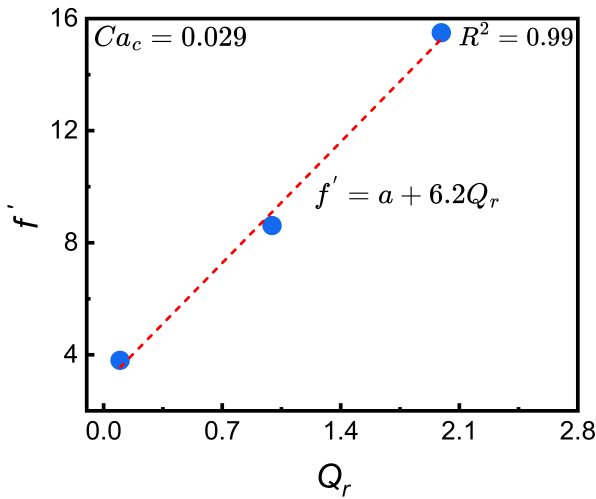} \label{frequency-3}
	}
	\subfigure[$f^\prime$ =  $f(Ca_c)$ in dripping regime]{
		\includegraphics[width=0.48\linewidth]{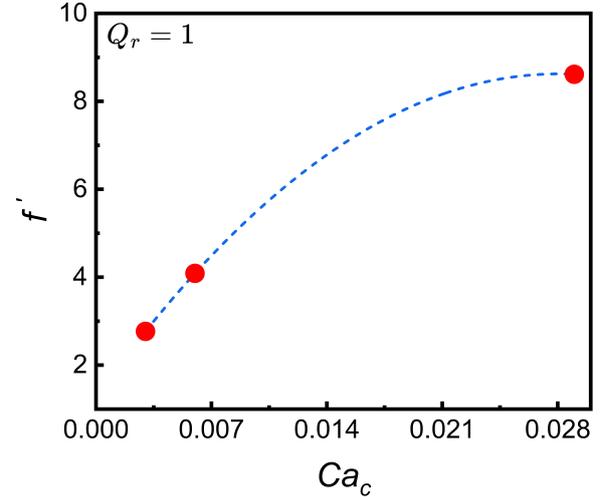} \label{frequency-4}
	}
	\caption{Droplet generation frequency ($f^\prime$) as a function of  flow rate ratios ($\qr$), and  capillary number ($\cac$) .}
	\label{frequency}
\end{figure}

The droplet generation frequency is calculated as the inverse of the total time required for a droplet breakup cycle, i.e., the time elapsed from the lag stage to the pinch-off stage  \citep{venkateshwarlu2021effects}. \fig\ref{frequency} presents the effect of the flow rate ratio ($0.1 \le \qr \le 5$) on droplet generation frequency ($f^\prime$, s$^{-1}$) for a wide range of capillary numbers ($0.001 \le \cac \le 0.1$) covering both the squeezing and dripping regimes.

\noindent
\figs\ref{frequency-1} and \ref{frequency-2} illustrate the variation in $f^\prime$ with $\qr$ in the squeezing regime. At low capillary numbers ($\cac = 0.003, 0.006$), the frequency increases  linearly with $\qr$, although the overall  increase remains small. For example, at $\cac = 0.003$, $f^\prime$ increases only from 2.75 to 2.85 s$^{-1}$, corresponding with  a slope of 0.017, as $\qr$ increases from 0.1 to 5. Similarly, at $\cac = 0.006$, the frequency rises modestly from 4 to 4.4 s$^{-1}$, with a slope of 0.08, over the same $\qr$ range. This can be attributed to the fact that, since the breakup process is dominated by geometric confinement and pressure-driven interface deformation, influence of $Q_r$ only marginally affect the overall droplet generation frequency.

\noindent
In contrast, under the dripping regime (($\cac = 0.029$), the influence of the flow-rate ratio ($\qr$) on the droplet generation frequency ($f^\prime$) becomes significantly more pronounced, as shown in \fig\ref{frequency-3}. Here, $f^\prime$ exhibits a strong linear dependence on $\qr$, increasing from approximately 3 to 15 s$^{-1}$ as $Q_r$ increases from 0.1 to 2. Compared to the squeezing regime ($\cac = 0.003$), the slope of the $f^\prime$–$Q_r$ relationship increases from 0.017 to 6.2, indicating an enhancement of nearly two orders of magnitude in the rate at which droplet frequency responds to variations in $Q_r$. This marked increase in sensitivity arises from the higher interfacial shear imposed by the continuous phase (CP) at elevated $\cac$, which accelerates interface deformation and neck thinning, thereby  resulting to higher droplet generation frequencies.
For low flow-rate ratios ($\qr < 1$), the droplet generation frequency remains comparable across $\cac$, as the CP flow rate is sufficient to dictate the breakup dynamics. However, when $\qr \ge 1$, the influence of $\cac$ on $f^\prime$ becomes significant. \fig\ref{frequency-4} demonstrates this exponential trend at $\qr = 1$: increasing $\cac$ from 0.003 to 0.029 results in a 22\% increase in $f^\prime$, while at $\qr = 2$ the corresponding increase is amplified to 43.5\%. Physically, this occurs because increasing both $\qr$ and $\cac$ amplifies the imbalance between the DP inflow and the CP shear stresses, causing faster pinch-off and higher droplet production rates.
\subsection{Film thickness}
\begin{figure}[!h]
	\centering
	\hspace*{0.9cm}
	\subfigure[Illustration of film thickness measurement for droplets formed under (i) Squeezing regime (ii) Dripping regime]{
		\includegraphics[width=0.9\linewidth]{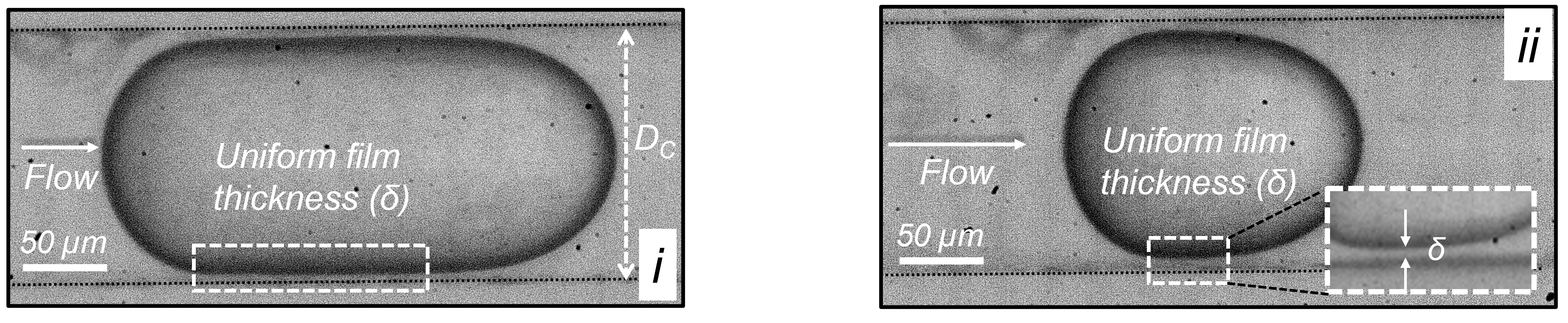}\label{film-1}
	}\\[0.2em] 
	
	\subfigure[$\delta^*$ as $f (\cac)$ for different $\qr$ ]{
		\includegraphics[width=0.48\linewidth]{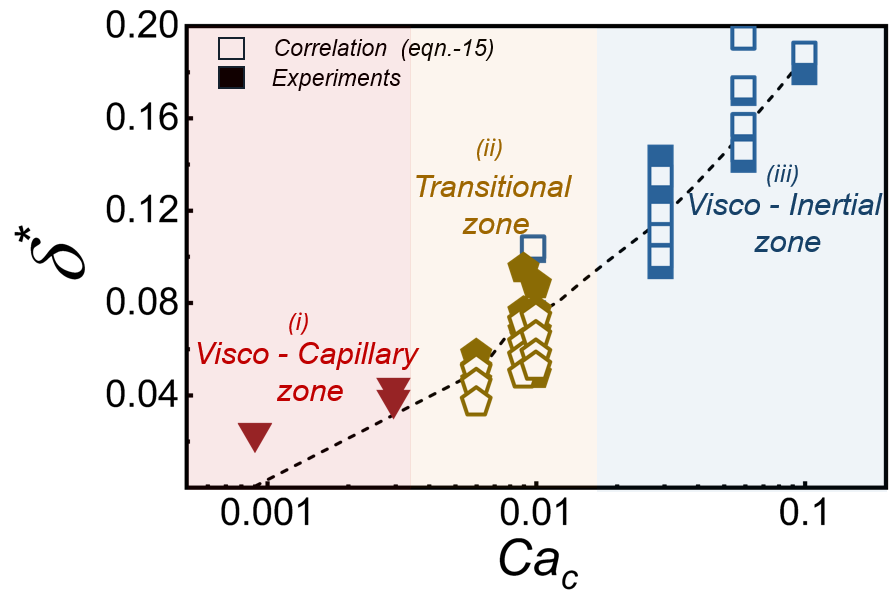}\label{film-2}
	}
	\hfill
	\subfigure[$\delta^*$ as $f (\qr)$ for different $\cac$ ]{
		\includegraphics[width=0.485\linewidth]{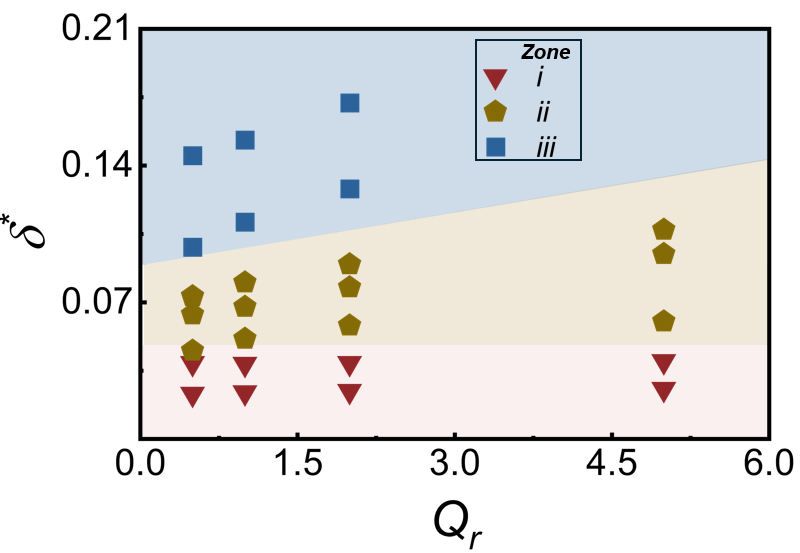}\label{film-3}
	}
	
	\caption{ (a) Illustration of film thickness measurement. Film thickness ($\delta^{\ast}=\delta/R_\text{c}$) between the wall and droplet as a function of (b) capillary number ($\cac$), and (c) flow rate ratio ($\qr$). The proposed correlation (\eqn\ref{eq:lfilm2}) in (b) is formulated by extending Taylor's law (\eqn\ref{eq:lfilm}).}
	\label{film}
\end{figure}

In two-phase flow through tubes, the dispersed phase (DP) occupies the central core region, while the continuous phase (CP) flows in the annular region, forming a `thin film' of CP between the DP and the tube walls \citep{Fairbrother_1935,Taylor_1961,Bretherton_1961}. This flow phenomenon is observed under the squeezing and dripping flow regimes, as shown in \fig\ref{film} which  illustrates for a thin CP film (thickness $\delta \lll R_\text{c}$, on the order of micrometers),  formed parallel to the tube wall and circumstantially surrounds the droplet/plug (i.e., DP) in both the squeezing (\fig\ref{film-1}i) and dripping  (\fig\ref{film-1}ii) regimes.

\noindent
In this study, the film thickness ($\delta$) is determined by averaging the local thickness ($\delta(x)$) at multiple axial positions under otherwise identical conditions. Subsequently, the normalized film thickness ($\delta^{\ast}$) as a function of the capillary number ($\cac$) is plotted in \fig\ref{film-2} as filled symbols (for the present data), dashed line (for the predictions based on Taylor’s law \citep{aussillous2000quick}, \eqn\ref{eq:lfilm}),  and unfilled symbols (for the data obtained using the proposed correlation, \eqn\ref{eq:lfilm2}) for the comparison purpose. The statistical correlation coefficients for empirical relation (\eqn\ref{eq:lfilm})  corresponding with the present data are summarized in \tab\ref{tab:Lfcoeff}. 
These figure show that the present measurements align qualitatively with the literature \citep{aussillous2000quick} in the squeezing regime, but deviations appear as the flow transitions into the dripping regime. Similar to the droplet morphology (i.e., shape and size) discussed in the previous sections, the thickness of the CP liquid film is also strongly influenced by the interplay of inertial, viscous, and surface tension forces, as quantified by the capillary number ($\cac$) and the flow-rate ratio ($\qr$) and discussed in the literature \citep{Fairbrother_1935,Taylor_1961,Bretherton_1961,aussillous2000quick,Mac_Giolla_Eain_2013,Youn_2021}. The balance among these forces governs the behavior of the liquid film under different flow conditions ($\cac$, $\qr$).

\noindent
Furthermore, it has been reported \citep{aussillous2000quick} that the geometry-dominated viscous force promotes the convergence of film thickness (i.e., Taylor convergence behavior), whereas inertial effects act in the opposite direction, leading to a thickening (divergence) of the film. Thus, the thin film behavior has been broadly categorized  \citep{Taylor_1961,Bretherton_1961,aussillous2000quick,Mac_Giolla_Eain_2013} into two regimes: (i) the visco-capillary regime, and (ii) the visco-inertial regime.

\noindent
The transition between these two regimes is characterized by a transitional capillary numbers ($Ca^{\ast}$ or $Ca^{\ast\ast}$), as defined in previous studies \citep{aussillous2000quick,Mac_Giolla_Eain_2013}. Physically, this capillary number represents the threshold at which inertial effects begin to compete with or dominate over capillary forces, thereby shifting the thin-film behavior from the visco-capillary to the visco-inertial regime.  This transitional threshold can be expressed \citep{aussillous2000quick} as follows.
\begin{gather}
	\cac^{\#} \sim (R_{\text{c}}/\ell_{\text{c}})^{3/2}
	\label{eq:cacss}
\end{gather}
where, $R_\text{c}$ is the capillary (tube) radius and $\ell_{\text{c}} = \sqrt{\sigma/(\rho_{\text{d}}\ g)}$ is the capillary length. 
%
%
An alternative form of transitional capillary number (\eqn\ref{eq:cacss}) is provided in subsequent refinements \citep{Mac_Giolla_Eain_2013} that account for geometry and flow-rate effects, but the underlying interpretation remains the same that it marks the onset of inertial influence on film thickness. Consequently, the transitional capillary numbers ($\cac^{\ast}$ and $\cac^{\ast\ast}$) are defined as follows.
\begin{gather}
	{\cac^{\ast} = (\cac^{\ast\ast})^{3/2}	
		\qquad
	\text{and}\qquad
	\cac^{\ast\ast} = X^{1/2} 
	\qquad\text{where}\qquad
	X =  \left(\frac{\mu_\text{d}^2}{\rho_{\text{d}} \sigma R_\text{c}}\right)}
	\label{eq:cacs}
\end{gather}
%

\noindent
Notably, $\cac^{\ast} < \cac^{\ast\ast}$ in the pre-Taylor convergence regime ($\cac < 1$). In general, the visco-capillary regime is observed for $\cac \leq \cac^{\ast}$, while the visco-inertial regime corresponds to $\cac \geq \cac^{\ast}$. 
\newline 
For $\cac < \cac^{\ast}$, the film thickness ($\delta^\ast$) is governed primarily by a balance between viscous and capillary forces, leading to the visco-capillary regime where Bretherton-type scaling laws apply. In contrast, for $\cac > \cac^{\ast}$, inertial effects become significant, resulting in the visco-inertial regime, leading to film thickening and deviation from the classical visco-capillary law (visco-inertial regime). However, since the transition between these regimes is gradual rather than abrupt, the limits are more precisely demarcated as: (i) visco-capillary regime ($\cac \leq \cac^{\ast}$), (ii) transitional regime ($\cac^{\ast} \leq \cac \leq \cac^{\ast\ast}$), and (iii) visco-inertial regime ($\cac \geq \cac^{\ast\ast}$), as demarcated in \figs\ref{film-2} and \ref{film-3}. 
\newline
In the present study, the critical values are determined as $\cac^{\ast} = 0.002$ and $\cac^{\ast\ast} = 0.014$, confirming that the proposed model is capable of accurately predicting the liquid film thickness across regimes. These values are consistent within the order of magnitude reported in earlier works \citep{aussillous2000quick,Mac_Giolla_Eain_2013}, thereby validating the applicability of the model within the established theoretical framework.

\noindent
In the visco-capillary regime, the normalized film thickness ($\delta^{\ast}$) depends primarily on the capillary number ($\cac$), governed by the balance between viscous and surface tension forces, and its estimation can be obtained from Taylor's law \citep{Taylor_1961,Bretherton_1961,aussillous2000quick,Mac_Giolla_Eain_2013} as follows.
\begin{equation}
	\delta^{\ast} =\frac{\delta}{R_\text{c}}=\frac{a_1 (\cac)^{n}}{a_2 + (a_3\times a_4) (\cac)^{m}}
	\label{eq:lfilm}
\end{equation}
{where, $a_{i}$'s, $n$ and $m$ are the theoretical/statistical correlation coefficients.
Various theoretical, numerical, and experimental studies have estimated the thin-film thickness in conventional (i.e., macro-scale) two-phase flow through tubes, and expressed their results in terms of Taylor's law (\eqn\ref{eq:lfilm}), with the corresponding statistical correlation coefficients summarized in \tab\ref{tab:Lfcoeff}.
{\begin{table}[!b]
		\caption{Correlation coefficient for the thin film thickness (\eqn\ref{eq:lfilm}). }\label{tab:Lfcoeff}
		\begin{center}\vspace{-1em}\renewcommand{\arraystretch}{1.5}
			\resizebox{\textwidth}{!}{\begin{tabular}{|l|c|c|c|c|c|c|c|p{0.25\linewidth}|}
				\hline %
				{Source} &  $a_1$ & $a_2$ & $a_3$ & $a_4$& $n$ & $m$ & $\cac$ & Remarks (LG: liquid-gas; LL: liquid-liquid)\\  \hline
				\citet{Fairbrother_1935} & 1& 0	&1 &1 & $(3/4)B$ & 0 & $7.5\times 10^{-5} - 0.014$ & LG flow; \newline Good approximation \citep{Taylor_1961,Bretherton_1961} for $\cac \le 0.09$ \\  \hline
				\citet{Bretherton_1961} & $A$	& 0	&1 &1 & $B$ & 0 & $\le 0.003$ & LG flow; \newline Bretherton law;\newline $A=(0.645\times 3^{n})=1.34$, $B=(2/3)$ \\  \hline
				\citet{aussillous2000quick} & $A$	& 1	& 2.5 & $A$ &$B$ & $B$& $0.015 - 2$ & LG flow;  \citet{Taylor_1961} data; Taylor law\\  \hline
				\citet{Mac_Giolla_Eain_2013}&$A$	&1	&1.6 &$A$ & $B$ & $B$& $0.002 - 0.119$ & LL flow, \newline Modified Taylor law\\  \hline
				\citet{Klaseboer_2014}&1.02 A&	1& 2.79 A &A& B& B & $\le 2$ & LG flow\\  \hline
			
				\citet{Datar_2025} & $(26/100)A$- $(42/100)A$	& 1	& -2.53 A & 1   & $B$ & $B$& $0.001 - 0.002 $& LG flow \\		  \cline{2-8}
				& $(30/100)A$- $(46/100)A$	& 1	& -2.53 A & 1   & $B$ & $B$& $0.014 - 0.1$ &  \\		  \hline
				\citet{Magnini_2022}& 1&1	&1 &1&B &B & 0.005-1 & LG flow\\  \hline
				\textbf{Present study} & $A$	& 1.14	& 1 & 1   & $(81/100)B$ & $(9/10)B$& $0.001 - 0.002 $& LL micro-tube flow \\		  \cline{2-8}
               	 (\eqns\ref{eq:lfilm} and \ref{eq:lfilm2}) & $(86/100)A$	& 1	& 1 & 1   & $(80/100)B$ & $(2/10)B$& $0.014 - 0.1$ &  \\		  \hline
			\end{tabular}}		
		\end{center}
        \label{corr}
\end{table}}

\noindent
Beyond the critical capillary number ($\cac^{\ast}$), a noticeable deviation is observed in \fig\ref{film-2}, marking the transition from the visco-capillary to the visco-inertial regime. In this transitional zone ($\cac^{\ast} \leq \cac \leq \cac^{\ast\ast}$), the film thickness ($\delta^{\ast}$) increases more rapidly with the capillary number ($\cac$) than predicted by Taylor's law \citep{aussillous2000quick}. During this transition, inertial forces become significant, and the film thickness is influenced by both the flow-rate ratio ($\qr$) and the capillary number ($\cac$). \fig\ref{film-3} shows the variation of film thickness  ($\delta^{\ast}$) with the flow-rate ratio ($\qr$) for different capillary number ($\cac$). In the visco-capillary regime, no distinct deviation is observed as $\qr$ varies from $0.1$ to $5$. However, upon transitioning to the visco-inertial regime, the effect of $\qr$ becomes pronounced, leading to distinct variations in film thickness. This trend clearly confirms the increasing dominance of inertia over surface-tension effects in governing the thin-film behavior.
\newline
To capture this observed behavior, a refined empirical correlation is proposed in this work for predicting the film thickness in the visco-inertial regime. The proposed correlation is formulated by extending Taylor’s law  (\eqn\ref{eq:lfilm}) to incorporate inertial effects, expressed as BD-Taylor's law (\eqn\ref{eq:lfilm2}) as follows.
\begin{equation}
	\delta^{\ast} 
	=\frac{a_1 (\cac)^{n}(\qr)^{m}}{a_1 + (\cac)^{n}(\qr)^{m}}
	\label{eq:lfilm2}
\end{equation}
The correlation coefficients ($a_1$, $m$, $n$) for \eqn\eqref{eq:lfilm2} are listed in \tab\ref{tab:Lfcoeff}.  
%
As demonstrated through unfilled symbols in \fig\ref{film-2}, the proposed model (BD-Taylor's law, \eqn\ref{eq:lfilm2}) shows excellent agreement with the experimental data, with deviations confined to the range of $\pm 2\%$ to $\pm 7\%$.
\newline
Furthermore, \tab\ref{tab:Lfcoeff} compares the present correlation coefficients for the continuous-phase film thickness ($\delta^\ast$) with existing empirical models. The results generally follow the Bretherton-type scaling ($\delta^\ast \propto \cac^{2/3}$) but with modified pre-factors reflecting the combined influence of liquid-liquid (LL) interfacial dynamics and cylindrical geometry. In the present case of 3D microtube, the smooth curved wall produces a more symmetric pressure distribution and promotes circumferential recirculation of the continuous phase, leading to slightly thicker and more uniform films than in planar geometries. Unlike rectangular microchannels, where liquid can become trapped in corner gutters and reduce the effective film thickness, the cylindrical surface allows continuous redistribution of the lubrication layer. At low $\cac\ (0.001- 0.002)$, the coefficients ($a_1 = A$, $n \approx 0.81B$) indicate curvature-induced flow enhancement, while at higher $\cac\ (0.014-0.1)$, the weaker $\text{Ca}$-dependence ($n \approx 0.8B$, $m \approx 0.2B$) suggests that film thinning slows despite increased viscous stress. Overall, the cylindrical geometry minimizes confinement-induced shear and modifies curvature-driven pressure gradients, yielding a thicker  film compared to planar and rectangular T-junction channels.
\begin{figure}[!b]%
	\centering{{\includegraphics[width=0.95\linewidth]{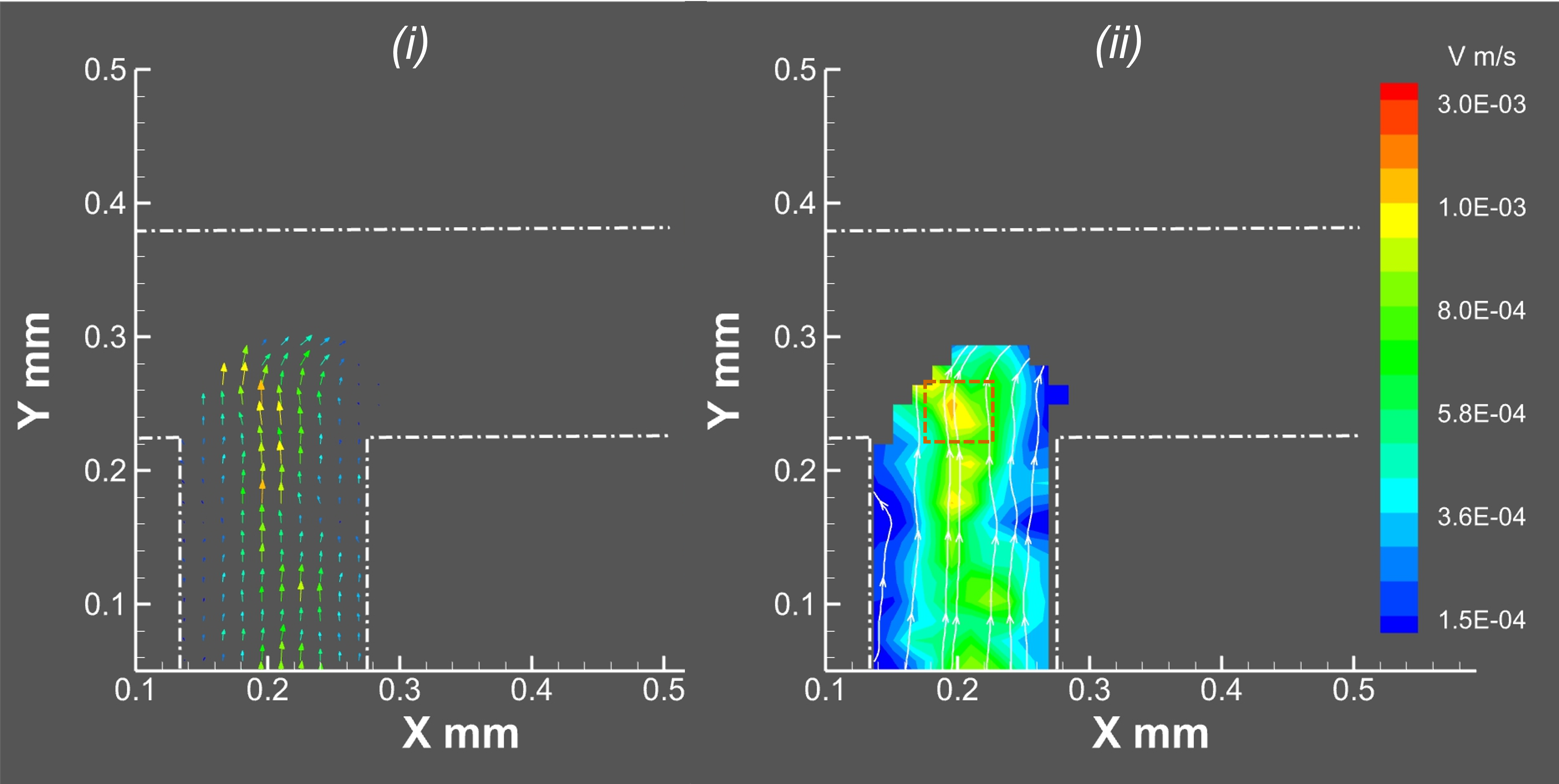}}}%
	\caption{PIV visualization of velocity vectors (i) and contours (ii) inside the droplet at the lag stage of breakup.}%
	\label{fig:vector6}%
\end{figure}
\begin{figure}[!b]%
	\centering{{\includegraphics[width=0.95\linewidth]{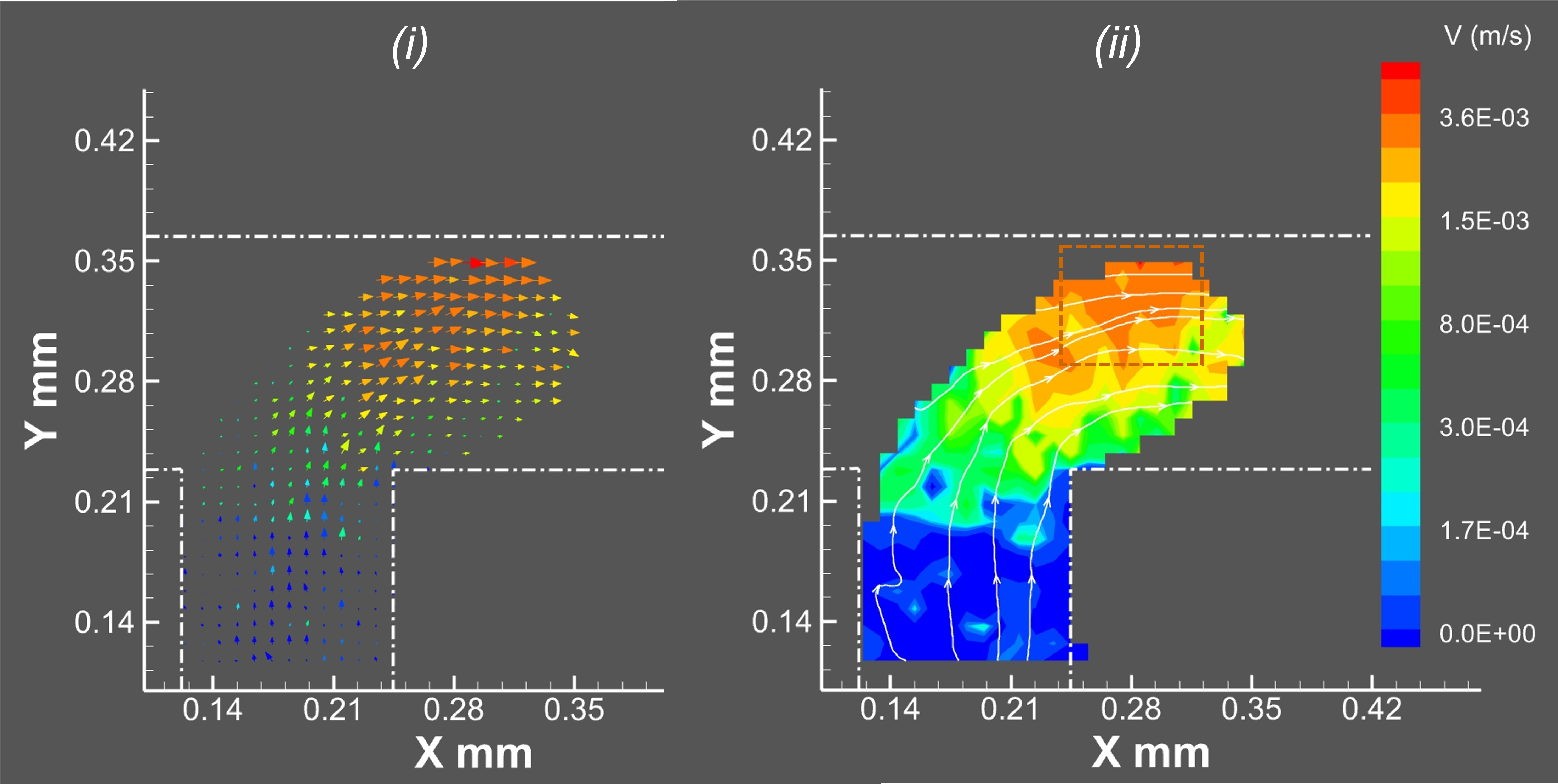}}}%
	\caption{PIV visualization of velocity vectors (i) and contours (ii) inside the droplet at the filling stage of breakup. }%
	\label{fig:vector1}%
\end{figure}
\begin{figure}[!b]%
	\centering{{\includegraphics[width=0.95\linewidth]{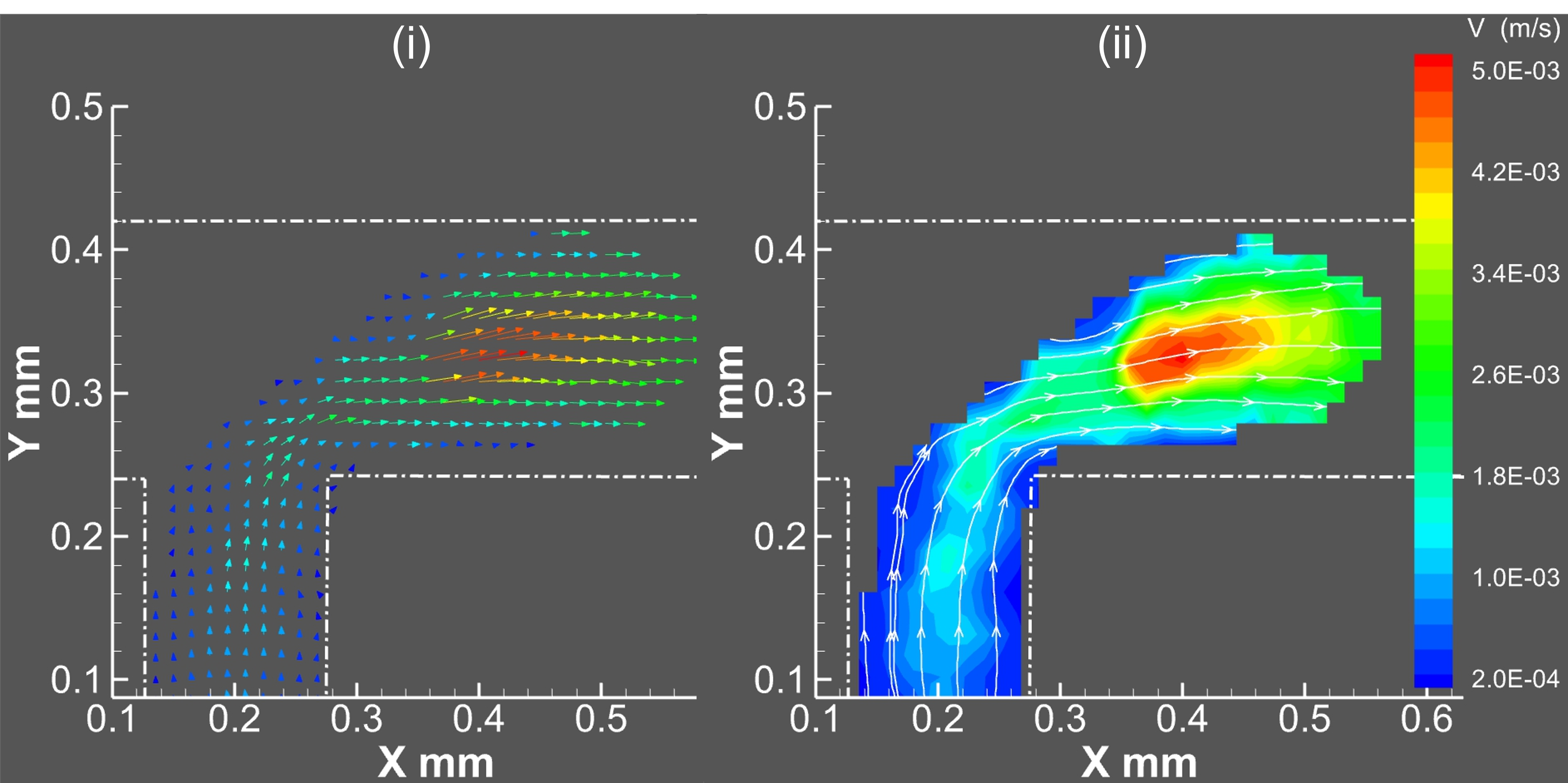}}}%
	\caption{PIV visualization of velocity vectors (i) and contours (ii) inside the droplet at the necking stage of breakup. }%
	\label{fig:vector2}%
\end{figure}
\begin{figure}[!b]%
	\centering{{\includegraphics[width=0.95\linewidth]{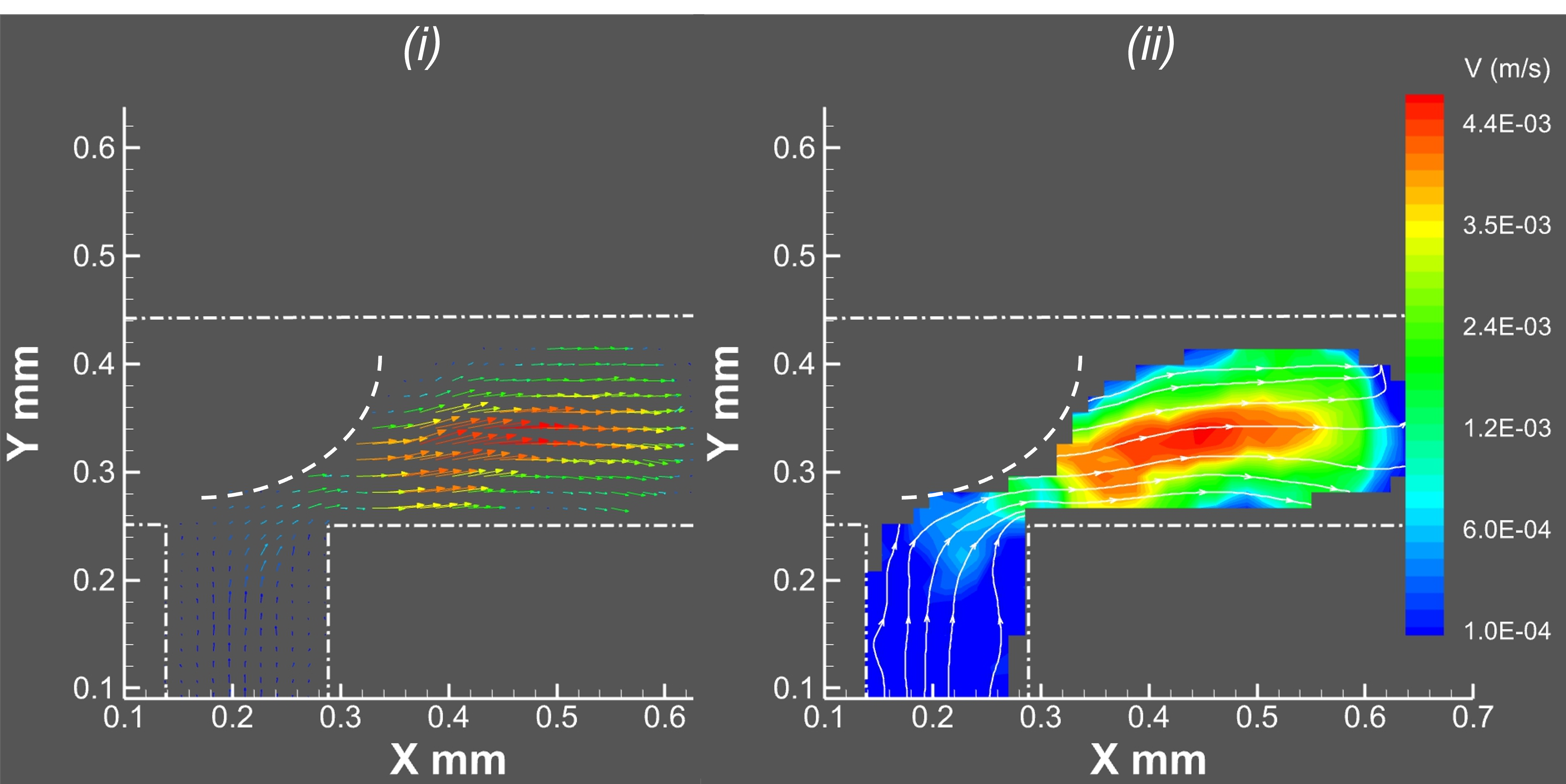}}}%
	\caption{PIV visualization of velocity vectors (i) and contours (ii) inside the droplet at the pinching-off stage of breakup. }%
	\label{fig:vector3}%
\end{figure}
\begin{figure}[!b]%
	\centering{{\includegraphics[width=0.95\linewidth]{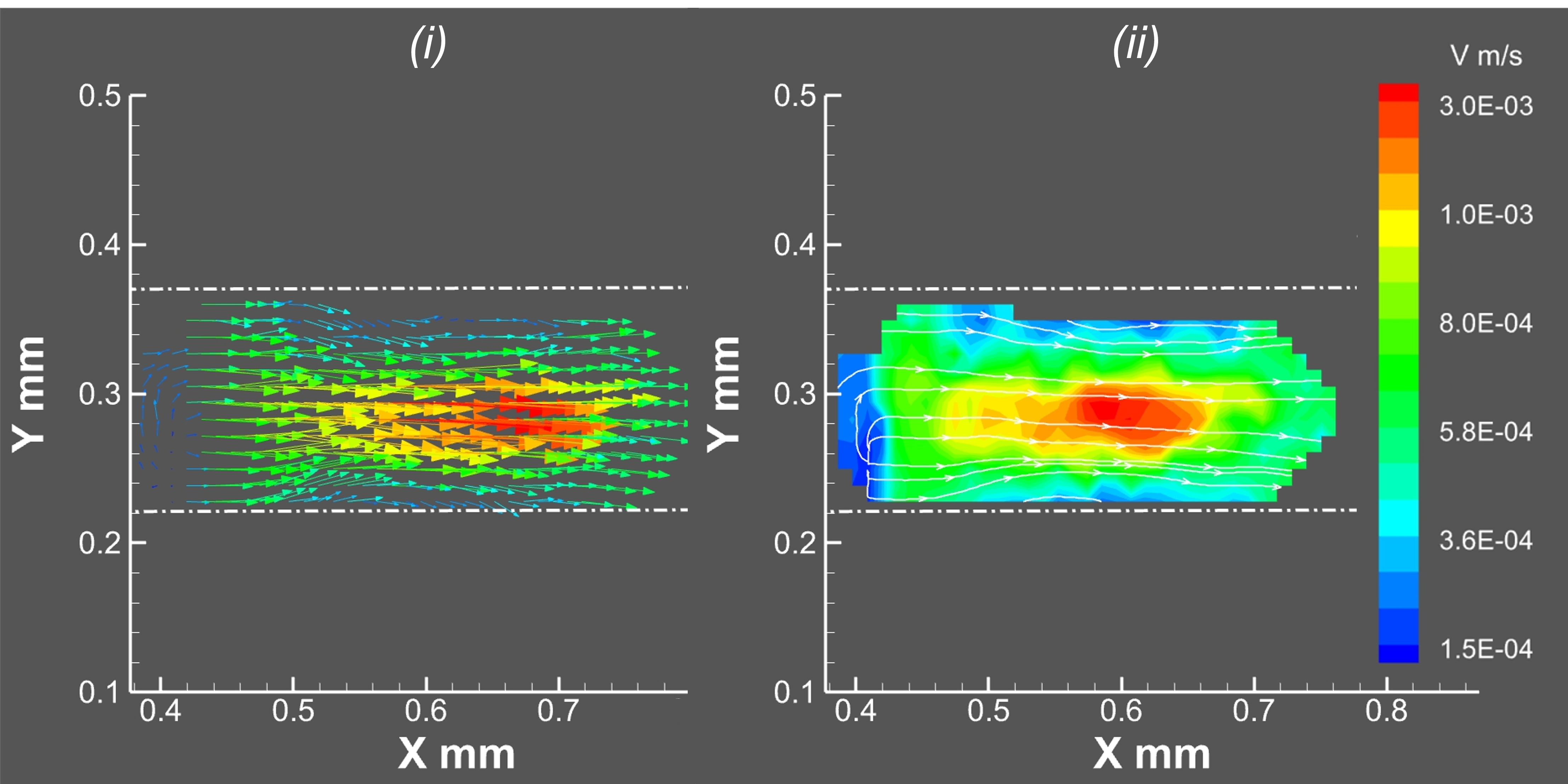}}}%
	\caption{Ensemble average velocity field inside the detached droplet under squeezing regime at $Q_\text{c}=Q_\text{d}= 0.03$ ml/hr (i) Velocity vectors (ii) Velocity contour. }%
	\label{fig:vector5}%
\end{figure}
\begin{figure}[!b]%
	\centering{{\includegraphics[width=0.95\linewidth]{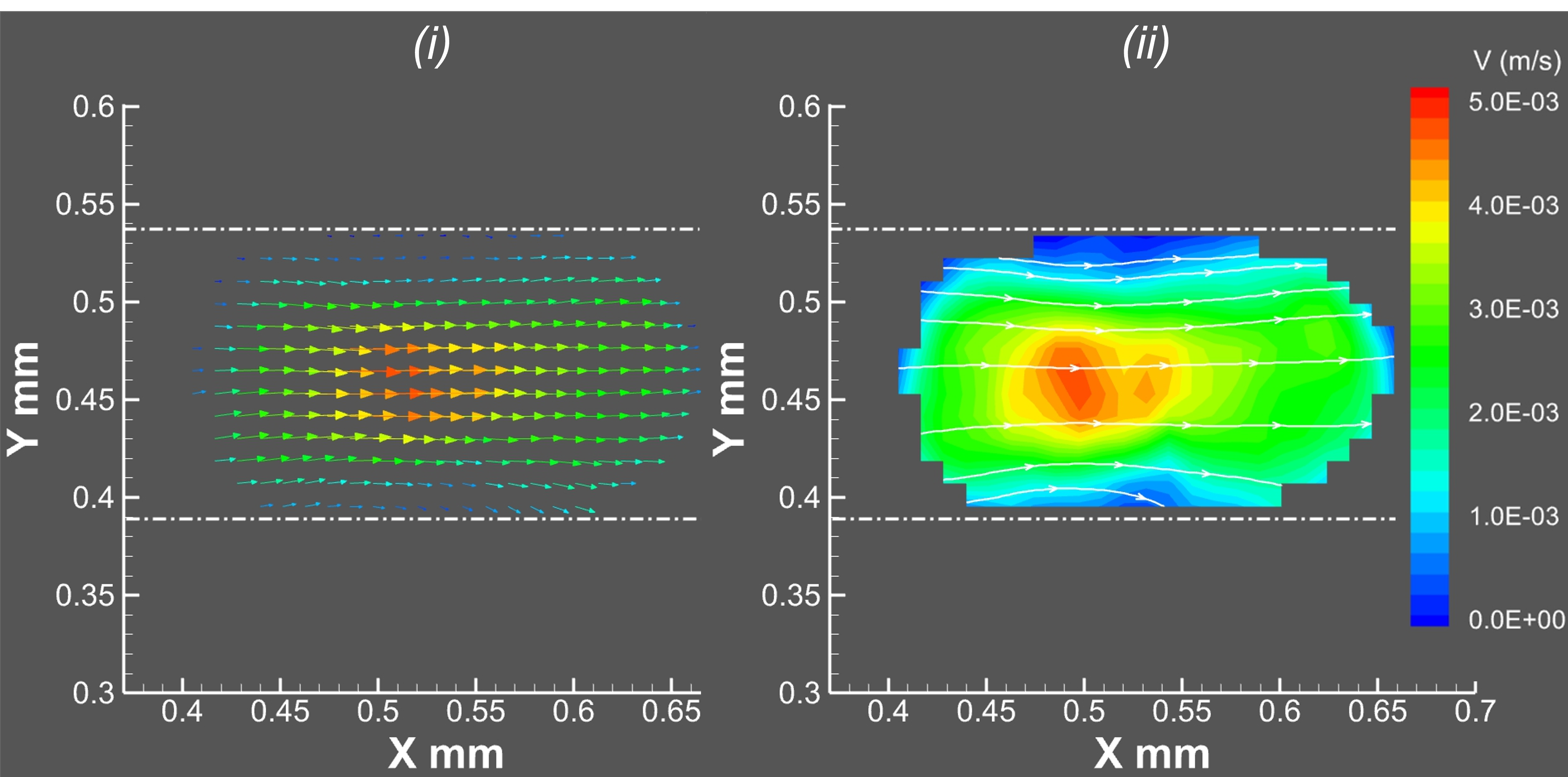}}}%
	\caption{Ensemble average velocity field inside the detached droplet under dripping regime at $Q_\text{c}=Q_\text{d}= 0.3$ ml/hr (i) Velocity vectors (ii) Velocity contour.}%
	\label{fig:vector7}%
\end{figure}
\begin{figure}[!h]
	\centering
	\hspace*{0.7cm}
	\subfigure[Illustration of different spatial locations  used for velocity magnitude extraction within the droplet under (i) Squeezing regime (ii) Dripping regime]{
		\includegraphics[width=0.95\linewidth]{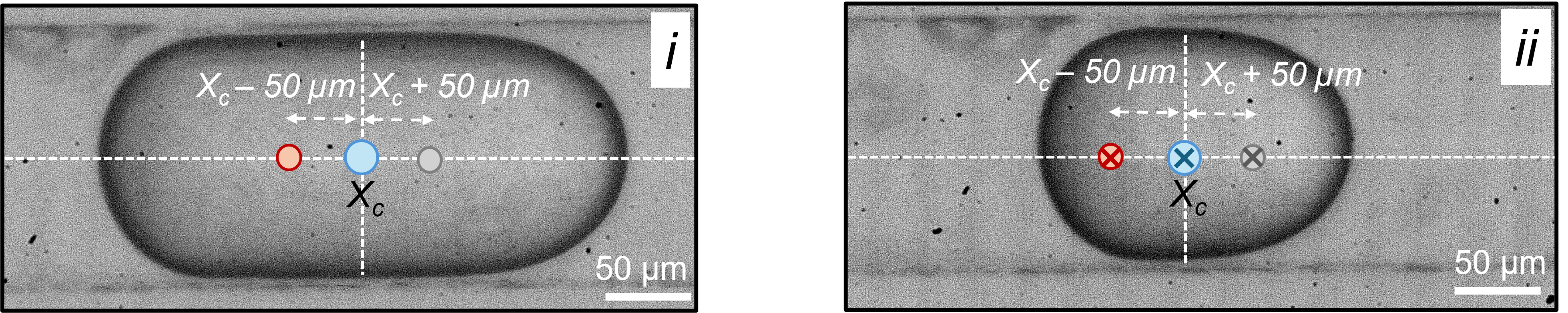}\label{Velocity profile-1}
	}\\[0.2em] 
	%
	\subfigure[Squeezing regime  for  $Q_r$ = 1]{
		\includegraphics[width=0.48\linewidth]{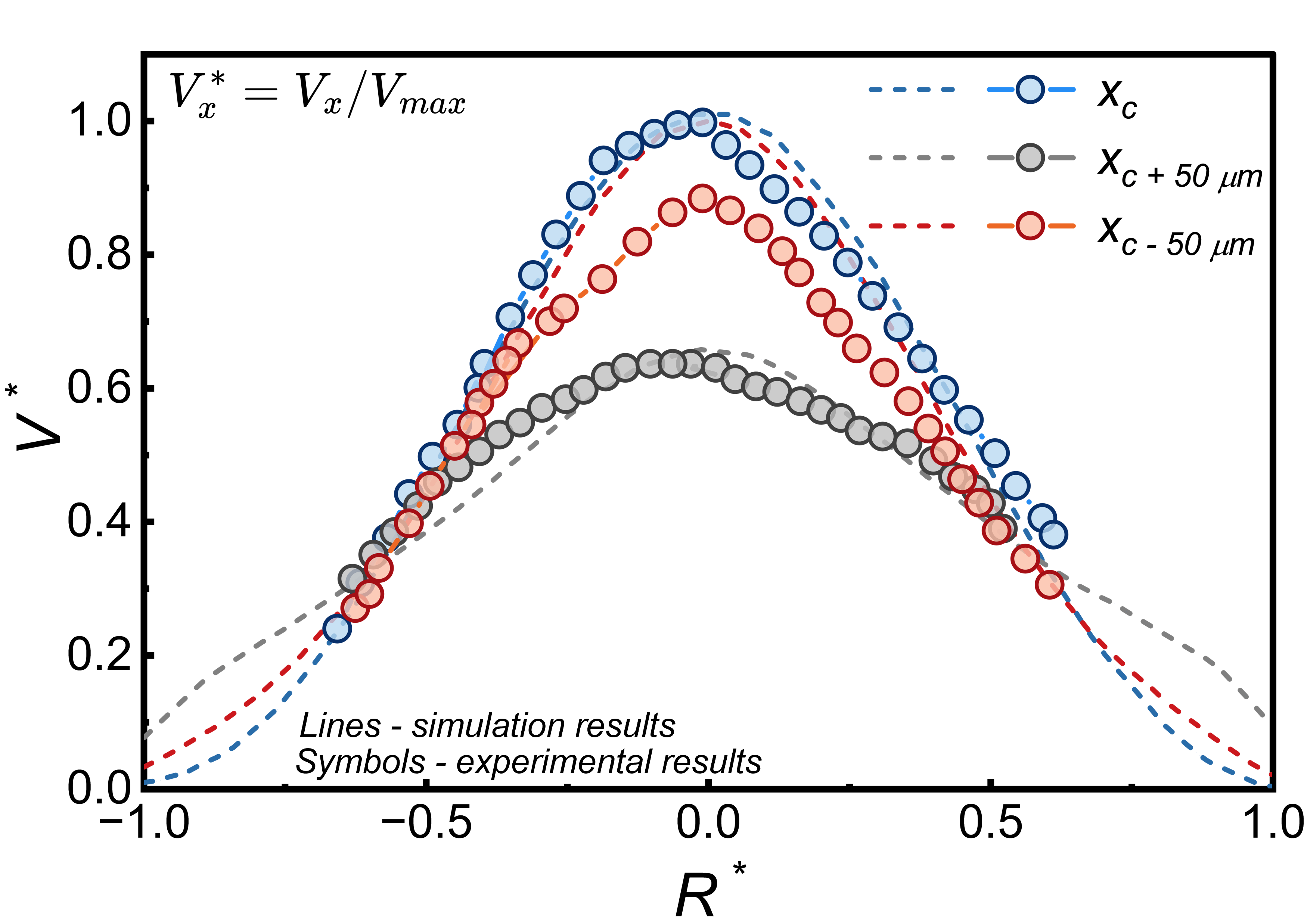}\label{Velocity profile-2}
	}
	\hfill
	\subfigure[Dripping regime  for $Q_r$ = 1]{
		\includegraphics[width=0.48\linewidth]{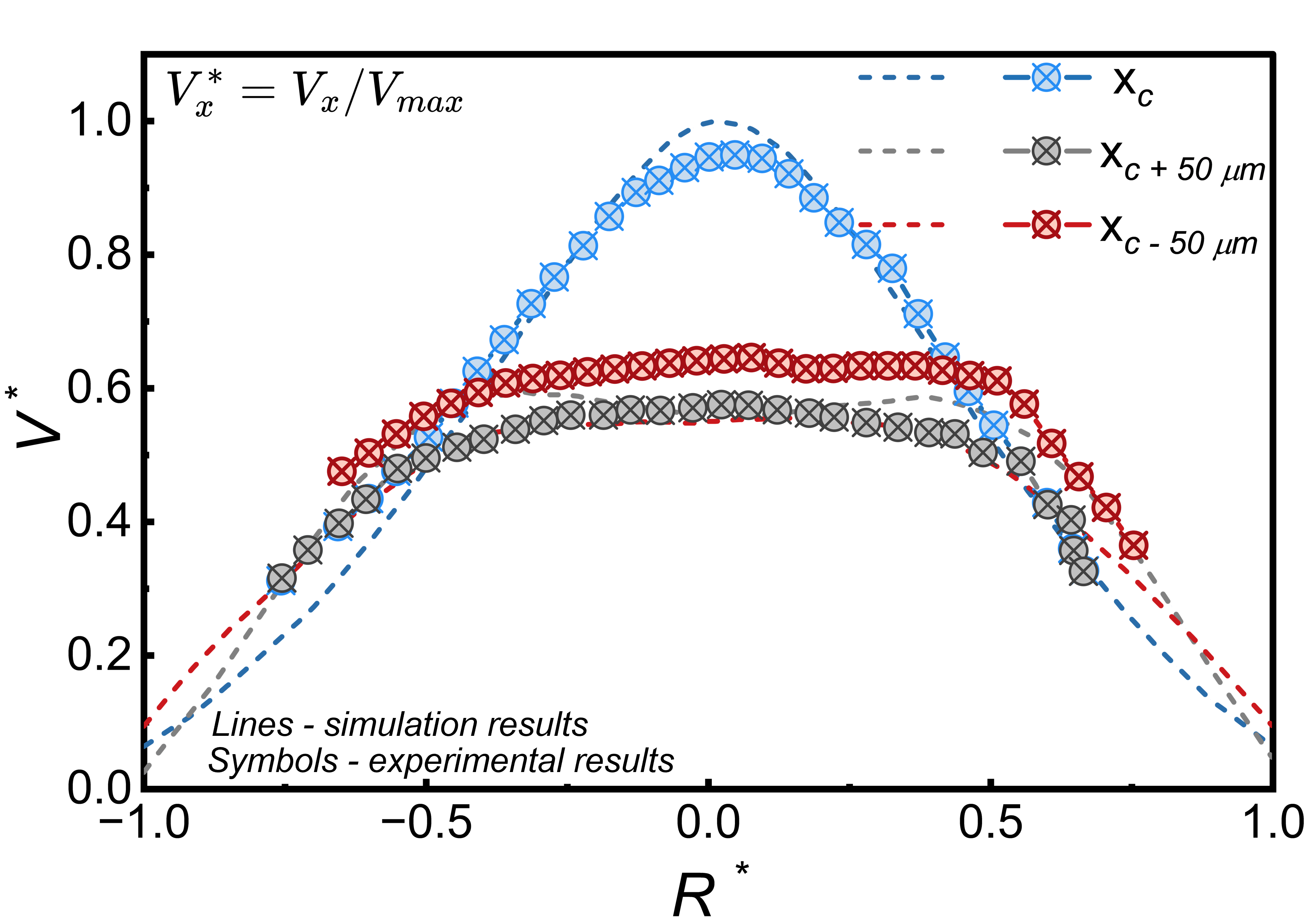}\label{Velocity profile-3}
	}
	\caption{Velocity profiles at different locations inside droplets for (a) squeezing regime $\cac=0.003$, and (b) dripping regime $\cac=0.06$. The lines and symbols indicate for the simulation and experimental results respectively.}
	\label{Velocity profile}
\end{figure}
\subsection{Velocity vectors}
\figs\ref{fig:vector6}-\ref{fig:vector5} illustrate the velocity vectors and contours in the dispersed phase at different stages of droplet formation under the squeezing regime at $Q_\text{d}=Q_\text{c}=0.3$ ml/hr observed from $\mu-$PIV experiments.
During the lag stage, the dispersed-phase (DP) flow remains steady and laminar with minimal interface instability. However, as a result of high pressure exerted by the continuous phase (CP) in the main channel, a single weak vortex forming near the advancing front is observed, as shown in \fig\ref{fig:vector6} (dashed region).  Because viscous forces are strongest at the wall due to the no-slip condition, we observe comparatively low velocity near the walls (indicated in blue color) than to the centreline (indicated in green and yellow colors). As DP evolves into the main channel, it experiences shear from CP. The continuous-phase flow around the forming droplet induces tangential shear at the top interface, increasing the local velocity magnitude near the top boundary (from 3 mm/s to 3.6 mm/s), as shown in \fig\ref{fig:vector1}. The velocity magnitude  further goes up to 5  mm/s as DP keeps progressing downstream of the main channel. It is to be noted that the local velocity during this stage  is approximately equal to the combined velocity of both DP and CP \fig\ref{fig:vector2}. As dispersed phase further progresses downstream and reaches the necking stage, the velocity gradients become very steep  depicted by red regions spreading across the pinching droplet and  the thinning neck. However the velocity magnitude at this stage decreases (from 5 mm/s to 4.4 mm/s), before it finally pinches off, as shown in \fig\ref{fig:vector3}.
\newline
Further, \figs\ref{fig:vector5} and \ref{fig:vector7} illustrate the ensemble-averaged velocity profiles of droplets generated under the squeezing (at $Q_\text{d}=Q_\text{c}=0.03$ ml/hr) and dripping (at $Q_\text{d}=Q_\text{c}=0.3$ ml/hr) regimes, respectively. In both cases, the profiles exhibit maximum velocity magnitudes near the droplet center and approach zero velocity at the interface. The pronounced velocity core reflects internal circulation driven by shear coupling between the dispersed and continuous phases. Additionally, in the squeezing regime, elongated circulatory regions appear at the interface; shorter but similar circulatory structures are also visible in the dripping regime.
\newline
The PIV measurements data extracted from the contours presented in \figs\ref{fig:vector5} and \ref{fig:vector7} is subsequently used to generate velocity profiles ($V_{x}$) inside the droplet at three positions ($X_c-50\ \mu$m, $X_c\ \mu$m, $X_c+50\ \mu$m; where $X_c$ is located over the axial centerline of the tube coinciding with center of droplet), as illustrated in \fig\ref{Velocity profile-1}. PIV measurements ensured that the profiles were measured away from regions influenced by end effects. 
The velocity profiles ($V_x$), obtained from PIV measurements, are normalized by the corresponding maximum velocity ($V_{\text{max}}$), while the transverse channel coordinate is normalized by the channel diameter. \figs\ref{Velocity profile-2} and \fig\ref{Velocity profile-3} shows the normalized velocity ($V^\ast = V_{x}/V_{\text{max}}$)  profiles inside the droplet as a function of radial distance ($-R_c^*\le R^*\le R_c^*$, where $R_c^*=2R_c/D_c=1)$) of tube under squeezing and dripping regimes, respectively.  The filled symbols  represent for the data obtained experimentally from PIV measurements, and dashed lines for data obtained using 3D simulations in this work. Experimentally, measurements are confined to the interior of the droplet; this is evident from the fact that the data symbols do not extend to the vicinity of the wall ($R^\ast\approx \pm 1$).
Broadly, both the squeezing and dripping regimes exhibit laminar velocity distributions; however, in the dripping regime, a Couette-type flow profile emerges in regions away from the droplet center, as depicted in \figs\ref{Velocity profile-2} and \ref{Velocity profile-3}. Both PIV measurements and CFD simulations reveal that the velocity profiles inside droplets are laminar and parabolic in the central region. In the squeezing regime, symmetric plugs ($L/D > 1$) display fully developed velocity profiles at both the front and rear. In contrast, in the dripping regime, although the central velocity profile remains parabolic, the front and rear sections are still in the developing stage. Furthermore, the experimental and numerical results show good agreement in the normalized velocity magnitude ($V_x^{*}$) profiles along the droplet length, with negligible deviations likely attributed to experimental uncertainties.
%
\section{Conclusions}
%
This study presented a comprehensive experimental (two-dimensional \micro-PIV) and numerical (three-dimensional CFD) investigation of droplet generation hydrodynamics in a T-type cylindrical microchannel with an inner diameter of 150 \micro m, using silicone oil and DI water as the continuous and dispersed phases, respectively, over a wide range of flow-rate ratios ($0.1 < \qr < 10$; $Q_\text{c} = 0.06$–$5$ ml/hr, $Q_\text{d} = 0.06$–$5$ ml/hr) and capillary numbers ($10^{-3} \leq \cac \leq 0.1$). The microfluidic device was in-house fabricated from PDMS using an economical embedded-template method. Based on the present investigation, the following conclusions can be drawn:  The phase flow field profiles reveal four distinct phases of droplet breakup, namely lag, filling, necking, and pinch-off. Four distinct droplet breakup regimes are identified as squeezing, dripping, parallel flow with tip streaming, and sausage flow. While the squeezing and dripping regimes have been experimentally observed, the parallel flow with tip streaming regime, though not captured experimentally due to the limited field of view and short detachment times at high flow rates, has been confirmed through simulations. Subsequently, a flow regime map is developed, wherein the squeezing regime transitions to parallel flow with tip streaming at high $\qr$, thereby delineating the operating boundaries of the breakup map between droplet-generating and non-droplet regimes. The droplet topology exhibits a complex dependence on the flow-rate ratio and capillary number. In the squeezing regime, the droplet length increases linearly with $\qr$, while the droplet curvature remains nearly independent of $\qr$. In contrast, in the dripping regime, both droplet length and curvature show strong dependence on $\cac$ and $\qr$, with the droplet length following a power-law relation. Furthermore, the variation in front and rear radii ($\Delta r^{\ast}$) increases non-linearly with $\cac$. To understand the behaviour of continuous phase film between channel walls and the droplets, a transition capillary number representing the threshold at which inertial effects begin to compete with or dominate over capillary forces, thereby shifting the thin-film behavior from the visco-capillary to the visco-inertial regime have been presented. A novel empirical correlation is proposed to determine the film thickness accounting the inertial and capillary effects. Furthermore, the flow field inside the forming droplet exhibits a complex dependence on the governing parameters. Both PIV measurements and CFD simulations reveal that the velocity profiles inside droplets are laminar and parabolic in the central region. In the squeezing regime, symmetric plugs ($L/D > 1$) display fully developed velocity profiles at both the front and rear. In contrast, in the dripping regime, although the central velocity profile remains parabolic, the front and rear sections are still in the developing stage. Overall, this study advances the understanding of droplet formation dynamics in confined cylindrical microchannels by integrating experimental observations with numerical simulations. The proposed correlations for droplet length, curvature, and film thickness provide predictive capability within the studied parameter range and offer valuable insights for the design and optimization of droplet-based microfluidic systems`	.
	%
\section*{Authors Contributions Statement}
%
\begin{table*}[!h]
	\begin{center}\renewcommand{\arraystretch}{1.5}		
			\resizebox{\textwidth}{!}{\small\begin{tabular}{|ll|p{0.75\linewidth}|} 
			\hline
			\multicolumn{2}{|l|}{Author(s)}  & Contribution(s) \\\hline
			1.& \ifblind\else Pratibha Dogra \fi	&
			Methodology, Software, Validation, Investigation, Data Curation, Visualization, Formal analysis, Writing - Original Draft\\\hline
			2. & \ifblind\else Ram Prakash Bharti	 \fi &
			Supervision, Conceptualization, Materials and Methodology, Software, Resources, Project administration, Funding acquisition, Formal analysis, Writing - Review \& Editing\\\hline
			3. & \ifblind\else Gaurav Sharma  \fi	& Supervision, Materials and Methodology, Resources, Funding acquisition, Writing - Review \& Editing\\\hline
		\end{tabular}}
	\end{center}
\end{table*}
\vspace{-2em}
%
\section*{Declaration of Competing Interest}
\noindent\small
The authors declare that they have no known competing financial interests or personal relationships that could have appeared to influence the work reported in this paper.
%
\section*{Acknowledgments}
\noindent
{\small 
	\ifblind
\else 
	\noindent The authors gratefully acknowledge the following support received from various funding agencies and individuals in completing this research.
	\begin{itemize}
		\item 
		Dr. Neha Jain for her valuable insights and hands-on support in the fabrication of the microchannels.		
		\item {Access to the \micro-PIV facility supported under  FIST (Fund for Improvement of S\&T Infrastructure) Grant (No. SR/FST/ET-II/2018/233(c)) by the Department of Science and Technology (DST), Government of India, and hosted at the Department of Chemical Engineering, IIT Roorkee.}
		\item 
		The high-end computing and experimental facilities supported under the Empowerment and Equity Opportunities for Excellence in Science (EMEQ) scheme (Grant No. EEQ/2022/001053) and Core Research Grant (No. CRG/2023/002759) by the Anusandhan National Research Foundation (ANRF), erstwhile Science and Engineering Research Board (SERB), Department of Science and Technology (DST), Government of India.
		\item Ministry of Education (MoE) for awarding the Prime Minister’s Research Fellowship (PMRF) to Pratibha Dogra (PMRF ID: 2801817, Cycle 8, Dec 2021).
		\item {Access to the COMSOL Multiphysics\textsuperscript{\textregistered} software suite under academic licenses (CKL and FNL) hosted at the Institute Computer Center (ICC), IIT Roorkee.}
		\item 
		The National Supercomputing Mission (NSM) for providing access to the `PARAM Ganga supercomputing facility' at the Indian Institute of Technology Roorkee, implemented by C-DAC and supported by the Ministry of Electronics and Information Technology (MeitY) and the Department of Science and Technology (DST), Government of India. 
	\end{itemize}
	\fi
}
%
%

\begin{spacing}{1.2}
\noindent 
\renewcommand{\nomgroup}[1]{%
  \ifthenelse{\equal{#1}{A}}{\item[\textbf{Abbreviations}]}{%
  \ifthenelse{\equal{#1}{G}}{\item[\textbf{Greek Symbols}]}{%
  \ifthenelse{\equal{#1}{D}}{\item[\textbf{Dimensionless Groups}]}{%
  \ifthenelse{\equal{#1}{S}}{\item[\textbf{Subcripts and Superscripts}]}
  {}}}}
}
 \nomenclature[A]{CFD}{computational fluid dynamics}
 \nomenclature[A]{CP}{continuous phase}
 \nomenclature[A]{DP}{dispersed phase}
 \nomenclature[A]{DAE}{differential algebraic equations}
 \nomenclature[A]{LSM}{level set method}
 \nomenclature[A]{FDM}{finite difference method}
 \nomenclature[A]{FEM}{finite element method}
 \nomenclature[A]{PARDISO}{parallel direct solver}
\nomenclature[A]{Nd:YAG}{Neodymium-doped:Yttrium-Aluminum-Garnet (Nd:Y$_3$Al$_5$O$_{12}$)}
%
%
\nomenclature[]{$\mathbf{D}$}{rate of strain tensor, s$^{-1}$}
\nomenclature[]{$\mathbf{F}_{\sigma}$}{interfacial force, N}
\nomenclature[]{$L_{\text{u}}$}{upstream length of the main channel, m} 
\nomenclature[]{$L^{\ast}_{\text{u}}$}{normalized upstream length of the main channel, --}
\nomenclature[]{$L_{\text{d}}$}{downstream length of the main channel, m}
\nomenclature[]{$L^{\ast}_{\text{d}}$}{normalized downstream length of the main channel, --}
\nomenclature[]{$L_{\text{v}}$}{length of the side channel, m}
\nomenclature[]{$L^{\ast}_{\text{v}}$}{normalized length of the side channel, --}
\nomenclature[]{$p^{\ast}$}{normalized point pressure, Pa} 
\nomenclature[]{$P^{\ast}$}{dimensionless point pressure, Pa} 
\nomenclature[]{$Q_{\text{c}}$}{flow rate of CP, ml/hr}
\nomenclature[]{$Q_{\text{d}}$}{flow rate of DP, ml/hr}
\nomenclature[]{$Q_{\text{r}}$}{flow rate ratio, dimensionless}
\nomenclature[]{$f^{\prime}$}{droplet detachment frequency, s$^{-1}$}
\nomenclature[]{$t$}{droplet detachment time, s}
\nomenclature[]{$r^{\ast}_\text{tail}$}{normalized radius of rear end of droplet}
\nomenclature[]{$r^{\ast}_\text{front}$}{normalized radius of front end of droplet}
\nomenclature[]{$Ca_{\text{c}}$}{capillary number for CP, --}
\nomenclature[]{$Ca^{\ast}$}{Critical capillary number for CP, --}
\nomenclature[]{$\mathbf{V}$}{normalized velocity magnitude, --}
\nomenclature[]{${V^{\ast}}$}{velocity magnitude, m/s}
\nomenclature[]{$D_{\text{c}}$}{Diameter of the main channel, mm}
\nomenclature[]{$D_{\text{d}}$}{Diameter of the side channel, mm}
\nomenclature[G]{$\delta$}{film thickness}
\nomenclature[G]{$\delta^{\ast}$}{normalized film thickness}
\nomenclature[G]{$\gamma$}{re-initialization or stabilization parameter, m/s}
\nomenclature[G]{$\epsilon_{\text{ls}}$}{interface thickness controlling parameter, m}
\nomenclature[G]{$\mu_{\text{c}}$}{viscosity of CP, Pa.s}
\nomenclature[G]{$\mu_{\text{d}}$}{viscosity of DP, Pa.s}
\nomenclature[G]{$\mu_{\text{r}}$}{viscosity ratio, dimensionless}
\nomenclature[G]{$\rho_{\text{c}}$}{density of CP, kg/m$^3$}
\nomenclature[G]{$\rho_{\text{d}}$}{density of DP, kg/m$^3$}
\nomenclature[G]{$\rho_{\text{r}}$}{density ratio, --}
\nomenclature[G]{$\sigma$}{interfacial tension, N/m}
\nomenclature[G]{$\tau$}{extra stress tensor,  N/m$^2$}
\nomenclature[G]{$\phi$}{level set function, --}
\nomenclature[G]{$\kappa$}{curvature of the interface, m}
\nomenclature[G]{$\theta$}{contact angle, degrees}
%
%
%
%
	\printnomenclature[5em]
\end{spacing}
%
%
%
%
\noindent
\bibliography{references}

%
%
%
%
%
%
%
%
%
%
\end{document}